\documentclass[12pt,preprint]{aastex}

\newcommand\diamondrule{\line{$\m@th
   \leaders\hrule\hfill\rlap{$\m@th\bracerd\braceld$}
   \braceru\bracelu\leaders\hrule\hfill$}}

\newcommand\bupskip{\vskip-17pt}
\newcommand\smallneg{\kern-.0800em}
\newcommand\negskip{\kern-.5em}

\newcommand\lsim{\rlap{\raise.4ex\hbox{$<$}}\lower.55ex\hbox{$\sim$}\,}
\newcommand\gsim{\rlap{\raise.4ex\hbox{$>$}}\lower.55ex\hbox{$\sim$}\,}
\newcommand\implies{{\bf=\kern-0.45em>}}

\newcommand\unit{\,\rm}
\newcommand\back{\negthinspace\negthinspace}
\newcommand\kms{\rm\, km\cdot s^{-1}}

\newcommand\um{\unit\mu m}

\newcommand\CO{\rm {}^{12}\smallneg CO}
\newcommand\COex{\rm {}^{12}\smallneg C{}^{16}\smallneg O}
\newcommand\CObf{\bf {}^{12}\smallneg CO}
\newcommand\COit{{}^{12}\smallneg\it CO}
\newcommand\cO{\rm {}^{13}\smallneg CO}
\newcommand\cOex{\rm {}^{13}\smallneg C{}^{16}\smallneg O}
\newcommand\cObf{\bf {}^{13}\smallneg CO}

\newcommand\cOit{{}^{13}\smallneg\it CO}
\newcommand\Co{\rm C{}^{18}\smallneg O}
\newcommand\Coex{\rm {}^{12}\smallneg C{}^{18}\smallneg O}

\newcommand\Jone{\rm J=1\rightarrow 0}
\newcommand\Joneit{\it J=1\rightarrow 0}
\newcommand\Jonebf{\bf J=1\rightarrow 0}

\newcommand\COone{\CO\ \Jone}
\newcommand\COoneit{\COit\ \Joneit}

\newcommand\Coone{\Co\ \Jone}
\newcommand\cOone{\cO\ \Jone}
\newcommand\cOoneit{\cOit\ \Joneit}
\newcommand\cOonebf{\cObf\ \Jonebf}
\newcommand\COonebf{\CObf\ \Jonebf}

\newcommand\NH{\rm N(H_2)}
\newcommand\nH{\rm n(H_2)}

\newcommand\mh{\rm m_{_{H2}}}

\newcommand\aef{{\hbox{$a_{_{eff}}$}}}
\newcommand\Tk{\rm T_{{}_K}}

\newcommand\Tr{\rm T_{{}_R}}
\newcommand\Tx{\rm T_{{}_X}}

\newcommand\Tbg{\rm T_{{}_{BG}}}

\newcommand\cKkms{\unit cm^{-2}\cdot (K\cdot km\cdot s^{-1})^{-1}}

\newcommand\Ia{\rm I_\nu(140\um)}
\newcommand\Ib{\rm I_\nu(240\um)}
\newcommand\Ic{\rm I(\cO)}

\newcommand\Ngas{\rm N(H\,I+2H_2)}

\newcommand\Dv{\rm\Delta v}
\newcommand\Dvc{\rm\Delta v_c}

\newcommand\rs{{\hbox{${\cal R}$}}}
\newcommand\ph{p_{1/2}}
\newcommand\Dh{d_{1/2}}
\newcommand\rh{r_{1/2}}
\newcommand\yh{y_{1/2}}
\newcommand\Nc{\rm N_c}
\newcommand\Ncs{{\hbox{${\cal N}_c$}}}
\newcommand\At{\rm A(\tau_0)}
\newcommand\Ath{\rm A(\tau_{0,13})}
\newcommand\Atw{\rm A(\tau_{0,12})}
\newcommand\Jnu{{\cal J}_\nu(\Tk)}
\newcommand\Jnux{{\cal J}_\nu(\Tx)}

\newcommand\nb{\hbox{\=n}}
\newcommand\rb{\hbox{\=\back\back{\hbox{$\rho$}}}}

\newcommand\Xf{\rm N(H_2)/I(CO)}
\newcommand\Xfbf{\bf N(H_2)/I(CO)}

\newcommand\degree{\rlap{$^\circ$}\kern.06em}
\newcommand\tef{\tau_{ef}}
\newcommand\thef{\tau_{ef,13}} 
\newcommand\twef{\tau_{ef,12}}  

\newcommand{\mymail}{wwall@inaoep.mx}

\slugcomment{
	     \hbox to \hsize {\hfil Version --   1/Mar/2007}}

\shorttitle{X-Factor}
\shortauthors{W. F. Wall}

\begin{document}


\title{Rethinking the $\Xfbf$ Conversion Factor}


\author{W. F. Wall}
\affil{Instituto Nacional de Astrof\'{\i}sica, \'Optica, y Electr\'onica,
Apdo. Postal 51 y 216, Puebla, Pue., M\'exico}
\email{\mymail}







\begin{abstract}

An improved formulation for the $\Xf$ conversion factor or X-factor is proposed.  The statement that the 
velocity-integrated radiation 
temperature of the $\COone$ line, $I(\CO)$, ``counts'' optically thick clumps is quantified using the formalism 
of \citet{Martin84} for line emission in a clumpy cloud.  Adopting the simplifying assumptions of thermalized 
$\COone$ line emission and isothermal gas, an effective optical depth, $\tef$, is defined as the product of the 
clump filling factor within each velocity interval and the clump effective optical depth as a function of the 
optical depth on the clump's central sightline, $\tau_0$.  The clump effective optical depth is well approximated 
as a power law in $\tau_0$ with power-law index, $\epsilon$, referred to here as the clump ``fluffiness,'' and has 
values between zero and unity.   While the $\COone$ line is optically thick within each clump (i.e., high 
$\tau_0$), it is optically thin ``to the clumps'' (i.e., low $\tef$).  Thus the dependence of $I(CO)$ on $\tef$ 
is linear, resulting in an X-factor that depends only on clump properties and {\it not\/} directly on the entire 
cloud.  Assuming virialization of the clumps yields an expression for the X-factor whose dependence on physical 
parameters like density and temperature is ``softened'' by power-law indices of less than unity that depend on 
the fluffiness parameter, $\epsilon$.  The X-factor provides estimates of gas column density because each 
sightline within the beam has optically thin gas within certain narrow velocity ranges.  Determining column 
density from the optically thin gas is straightforward and parameters like $\epsilon$ then allow extrapolation of 
the column density of the optically thin gas to that of all the gas.  Implicit in this formulation is the 
assumption that fluffiness is, on average, constant from one beam to the next.  This is also required to some 
extent for density and temperature, but the dependence of the X-factor, $X_f$, on these may be weaker.  

One important suggestion of this formulation is that {\it virialization of entire clouds is {\rm ir}relevant.\/}  
The densities required to give reasonable values of $X_f$ are consistent with those found in cloud clumps (i.e. 
$\sim 10^3\, H_2\unit cm^{-3}$).  Thus virialization of clumps, rather than of entire clouds, is consistent with 
the observed values of $X_f$.  And even virialization of clumps is not strictly required; only a relationship 
between clump velocity width and column density similar to that of virialization can still yield reasonable values 
of the X-factor.  The underlying physics is now at the scale of cloud clumps, implying that the X-factor can probe 
sub-cloud structure. 

The proposed formulation makes specific predictions of the dependence of $X_f$ on the CO abundance and
of the interpretation of line ratios.  In particular, the $\cOone/\COone$ line ratio values observed in the
Orion clouds suggest that $\epsilon\simeq 0.3\pm 0.1$.  If the
majority of the $\COone$ emission originates in structures with an $r^{-2}$ density variation, then the
constraints on $\epsilon$ also constrain the ratio of the outer-to-inner radii of the $r^{-2}$ region
within the clumps.  Specifically, this ratio for spherical clumps must be 2 to 9 and for cylindrical clumps
it must be 4 to 42.  This is apparently consistent with observations, but higher spatial resolution is
necessary to ensure that the observed ratios are not just lower limits.  This formulation also ties the
narrow range of the observed values of the $\cOone/\COone$ line ratio to the relative constancy of the 
X-factor.

The properties of real clumps in real molecular clouds can be used to estimate the X-factor within these
clouds and then be compared with the observationally determined X-factor.  This yields X-factor values 
that are within a factor of 2 of the observed values.  This is acceptable for the first attempt, but 
reducing this discrepancy will require improving the formulation.  While this formulation improves upon 
that of \citet{Dickman86}, it has shortcomings of its own.  These include uncertainties as to why $\epsilon$
seems to be constant from cloud to cloud, uncertainties in defining the average clump density and neglecting 
certain complications, such as non-LTE effects, magnetic fields, turbulence, etc. 

Despite these shortcomings, the proposed formulation represents the first major improvement in 
understanding the X-factor because it is the first formulation to include radiative transfer.

\end{abstract}




\section{Introduction\label{sec1}}

One of the most basic questions that can be asked in any field of research
is about the quantity of the material under investigation.  In studies of the 
interstellar medium (ISM), for instance, the amount of gas and dust bears on 
questions of the physical mechanisms that effect and control the ISM and, 
consequently, how that ISM evolves and can affect the evolution of an entire 
galaxy.  In particular, the amount of molecular gas in a cloud, cloud complex,
spiral arm, or galaxy constrains the number of stars that form and the way that 
they form.  The workhorse molecule for estimating molecular gas masses has been, 
and still is, CO \citep[e.g., see][and references therein]{iau170}.  Specifically, 
observations of the $\Jone$ rotational line of the isotopologue, $\COex$ (just 
CO for short), permit simple, but crude, estimates of the mass of molecular 
hydrogen in an astronomical source.  The velocity-integrated radiation temperature, 
I(CO), often called the {\it integrated intensity\/}, is multiplied by a standard 
conversion factor, $\Xf$, to yield the molecular hydrogen column density, $\rm 
N(H_2)$, which gives the H$_2$ mass of the source after integrating over the source's
projected area.  The most current value of this conversion factor is about $2\times 
10^{20}\, H_2\cKkms$ for the molecular gas in the disk of our Galaxy \citep{Dame01}.  

Why the CO $\Jone$ line should yield an estimate of column density is far
from clear.  Even if this line were optically thin, the conversion factor
would depend in a simple way on the physical conditions in the molecular gas.  
Given that CO has many rotational levels and that the spacings of these levels
(in temperature units) are comparable to the temperatures found in molecular gas, 
the $\Xf$ conversion factor would depend, at the very least, on the gas kinetic 
temperature.  Given that the densities inferred with molecular clouds (of at 
least $\sim 10^3\, H_2\unit cm^{-3}$) are comparable to, and not much higher than, 
the critical densities of the observed rotational transitions, then it is obvious 
that $\Xf$ should also have at least a weak dependence on molecular gas density.
Another obvious dependence would be on the abundance of CO relative to
$\rm H_2$, X(CO).  If these physical conditions were known, then the
molecular hydrogen column density could be recovered easily \citep[see 
Appendix~A of][]{W06}.  Even if these conditions are not known exactly, 
observations of molecular clouds on galactic scales would yield reasonable 
estimates of molecular gas column densities, because reasonable values for 
the relevant physical parameters are well known and are relatively constant 
from source to source.  For example, estimating the mass of the molecular 
medium of an entire galaxy using an optically thin molecular line would be, 
on average, more reliable than estimating the mass of a molecular cloud core 
{\it of unknown physical conditions,\/} because the physical conditions in the 
molecular gas averaged over the scale of a galaxy are less extreme and vary 
much less from galaxy to galaxy than they do from cloud core to cloud core.  
In any event, the conversion from integrated intensity to column density is 
very straightforward and relatively simple in the optically thin case.  However, 
as is well known \citep[e.g., see][]{Evans80, Kutner84, Evans99}, the CO $\Jone$ 
line is optically thick, obfuscating any simple explanation as to why it should 
probe molecular gas column densities.    

Other tracers of molecular gas mass exist, tracers that do not possess the 
potentially serious uncertainties posed by CO $\Jone$.  The rotational lines 
of the isotopologues $\cOex$ and $\Coex$ (just $\cO$ and $\Co$ in short form), for 
example, can be optically thin and, consequently, their integrated intensities 
have a straightforward relationship with the molecular gas column density 
(provided the physical conditions are known).  While potentially simpler to use 
for determining column densities, these optically thin lines are normally factors 
of about 3 to 50 weaker than the CO $\Jone$ line \citep[e.g.,][]{Kutner84, 
Langer90, Nagahama98, Maddalena86}; the $\CO$ lines are better for mapping large 
areas of molecular gas or for detecting weak sources, such as high-redshift 
galaxies \citep[e.g.,][]{Brown92, Barvainis97, Barvainis98, Alloin00, Carilli02, 
Carilli02a}.  This makes the $\Jone$ line of CO, and the $\Xf$ factor, more useful or 
even essential in estimating the total molecular gas mass in some sources, 
resulting in a strong incentive for understanding the $\Xf$ factor's behavior.   

The usual attempts at accounting for why the $\Xf$ factor, or X-factor (or $X_f$), 
is relatively constant on multi-parsec scales are variations of the explanation
given by \citet{Dickman86}, hereafter DSS86 \citep[e.g.,][]{Sakamoto96}.  A 
summary of the DSS86 explanation follows.  If $\Tr$ is the peak radiation 
temperature of the CO $\Jone$ line and $\Dv$ is the appropriately defined 
velocity width of this line, then $\rm I(CO) = \Tr\Dv$.  If the molecular gas 
under observation is virialized, then the observed velocity width is related 
to the mass of this gas and, therefore, the gas column density averaged over 
the solid angle subtended by the observed gas.  It was then easy to show that 
$\Xf\propto\ $$n$$^{0.5}/\Tr$.  The $n$ was the gas density {\it averaged over 
the virialized volume of gas.\/}  DSS86 found that $n$ had to be $\sim few\times
10^2\, H_2\unit cm^{-3}$ to give the observed value of N(H$_2$)/I(CO); therefore
it was assumed that this volume included entire clouds.  Because the CO $\Jone$ 
line is optically thick, relatively easily thermalized (compared to higher 
rotational lines of CO), and may almost fill the radio telescope's beam at the 
line peak, it has been assumed that $\Tr\simeq\Tk$, where $\Tk$ is the gas
kinetic temperature, $\Tk$ \citep[e.g.,][]{Kutner85, Weiss01}.  Even if the gas 
does not fill the beam (a point to which we will return later), we would have 
$\Tr$ roughly proportional to $\Tk$ and we would still have $\Xf\propto $
$n$$^{0.5}/\Tk$.  The basic argument is that the quantity $n^{0.5}/\Tk$ does not 
strongly vary on multi-parsec scales, especially due to the weak dependence on 
density, resulting in a fairly stable value of X.  Observational evidence does 
indeed seem to support a roughly constant value of the X-factor to within a factor 
of about 2 for the disk of our Galaxy, where $\rm X\simeq 2\times 10^{20}\cKkms$ 
\citep[see, e.g.,][and references therein]{Dame01, Strong88}, although the 
observations of \citet{Sodroski94} and \citet{Strong04} suggest a higher value 
of $X_f$ in the outer disk \citep[a claim that is at odds with][]{Carpenter90}.
The values of the X-factor 
that apply to the disks of other spiral galaxies are often within factors of
about 3 of that of the Galactic disk X-factor \citep[e.g.,][]{Young82, Adler92,
Guelin95, Nakai95, Brouillet98, Rand99, Meier00, Meier01, Boselli02, Rosolowski03}.  

Nonetheless, there is good observational evidence that the usual value of the 
X-factor does {\it not always\/} apply.   In the centers of 
external galaxies, the X-factor is factors of about 5 or more lower than the 
standard value \citep[e.g.,][]{Rickard85, Israel88, W93, Regan00, Paglione01}, 
as well as in the central region of our own Galaxy \citep{Sodroski95, Dahmen97, 
Dahmen98}.  In contrast, clouds in the central region of the galaxy M$\,$31 have an 
X-factor that is an order of magnitude larger than that for the disk of our Galaxy 
\citep{Sofue93, Loinard98}.  The high-latitude translucent clouds in our Galaxy show 
X-factor variations with a total range of an order of magnitude \citep[see][]
{Magnani98}.  In some infrared luminous galaxies there is evidence that the X-factor 
can be roughly an order of magnitude lower than the standard value \citep{Yao03}.  
In irregular galaxies, the X-factor can be more than an order of magnitude higher 
than the standard value \citep[e.g.,][]{Israel88, Dettmar89, Israel97, Israel97a, 
Madden97, Fukui99}.  A detailed discussion of the shortcomings of the X-factor 
can be found in \citet{Maloney88}\citep[also see][]{Israel88, Kutner85}.  

Hence, any complete explanation or theory of the X-factor must allow for and account 
for inferred variations of $X_f$ in some cases and, at the same time, relative 
stability of $X_f$ in other cases.  This is a difficult balancing act, but achieving 
such a theory is important for the very basic reason that scientific tools must be 
thoroughly understood.   While there are a few articles that explore the underlying 
physics of $X_f$ \citep[e.g., DSS86][]{Maloney88, Israel88, Kutner85}, there are 
literally {\it hundreds\/} of articles in the literature that use or mention the 
X-factor without a {\it detailed} examination of its physical properties 
\citep[including even][]{W96}.  This is in stark contrast to the situation with 
supernovae type~Ia (SNe~Ia), for example.  SNe~Ia can be used as standard candles 
\citep{Phillips93, Reiss95, Reiss96, Hamuy96, Hamuy96a} and can constrain cosmological 
models \citep{Reiss98, Perlmutter99}.  While these supernovae are often used as 
standard candles, there is also much theoretical and observational work to
understand SNe~Ia \citep[e.g.][]{Sauer06, Wang07, Reiss06, Garg07, James06, Neill06,
Borkowski06}.  Such work will elucidate why SNe~Ia are standard candles, or why they 
might not be in some cases \citep[see][]{Howell06}.  In a sense, the X-factor is 
{\it almost\/} a standard candle that relates surface brightness to surface density 
(i.e., $I(CO)$ to $N(H_2)$).  And yet there is comparatively little effort to shed
light on the physics underpinning the X-factor. 

Furthermore, there are a number of more specific reasons for understanding $X_f$:
\begin{itemize}
\item[] {\it Reliable molecular gas mass estimates.\/} Among other things, this
can refine our understanding of star formation yields (i.e., what fraction
of the gas goes into star formation). 
\item[] {\it Improved knowledge of molecular cloud structure and physics.\/}
Better comprehension of the X-factor can constrain estimates of molecular
cloud physical parameters and can supply new insights into star formation 
processes. 
\item[] {\it Improved radiative transfer theory.\/} While it is unlikely that
there would be fundamental improvements of radiative transfer theory, such 
improvements are still possible, benefiting astrophysical theory in general.
\end{itemize}
These represent possible long-term goals of research into the physics 
underlying the X-factor. 

The goals of the current paper are considerably more modest: addressing the
deficiencies of the DSS86 explanation of the X-factor, improving upon this
explanation, and examining a few consequences of the formulation proposed here.  
Improvements are necessary because DSS86 has the following problems:
\bupskip
\bupskip
\begin{enumerate}
\item {\it No treatment of radiative transfer.\/}  This is a fundamental problem 
with DSS86.  At first glance, it might seem superfluous to treat radiative transfer
in the optically thick case.  However, if we consider a clumpy medium, where the 
clumps can have optically thin edges and optically thin frequencies in their line
profiles, then treating radiative transfer is essential for understanding the
X-factor.  In particular, {\it the optically thin limit of CO $\Joneit$ must also
be included.\/} Any {\it complete\/} treatment must include the optically thin case, 
whether this case is observed in nature or not.  This case cannot be included easily 
in the DSS86 explanation because it includes the virial theorem with{\it out\/} 
including radiative transfer --- virialization by itself says nothing about the
optical depth of the emission. 
\bupskip
\item {\it Sensitivity to T$_{_K}$ and n(H$_2$).\/}  As discussed in \citet{W06}, 
I(CO) and the X-factor estimate the molecular hydrogen column densities to within 
factors of about 2 of the values determined from optically thin tracers for the 
majority of positions in the Orion clouds.  And yet we know from \citet{W06} 
(looking at two-component model temperatures), the range of gas kinetic temperatures 
can be an order of magnitude.  In general, we know that molecular cloud kinetic 
temperatures and densities have a full range of an order of magnitude on multi-parsec 
scales \citep[cf.][]{Sanders85, Sakamoto94, Helfer97, Plume00}.  Since the X-factor 
supposedly varies as $n^{0.5}/\Tk$, the temperature and density variations can 
{\it each\/} change $X_f$ by factors of 3 to 10 (unless $n^{0.5}$ were to vary like 
$\Tk$, but this is unlikely to be true in general, especially if there is pressure 
equilibrium).  Thus the $X_f$ of DSS86 is too sensitive to the density and kinetic 
temperature.  Having a weaker dependence of $X_f$ on $n$ and $\Tk$, like $\rm 
X\propto$$(n/\Tk)^{0.3}$, would resolve this sensitivity problem; variations of
an order of magnitude in either $n$ or $\Tk$ would allow $\rm X$ to vary by less
than a factor of 2.
\bupskip
\item {\it Virialization of entire clouds.\/} DSS86 require low densities 
(i.e. $\nH\sim$$few\times 10^2\unit cm^{-3}$) to obtain the observed value of the 
X-factor.  Given that the critical density of the CO $\Jone$ transition is
$\sim 3\times 10^3\, H_2\unit cm^{-3}$, the densities of the CO-emitting 
structures are about an order of magnitude higher \citep[also see the average
densities of the filaments found by][]{Nagahama98}.  The low density required by 
DSS86 may represent the density averaged over an entire cloud.  For the Orion$\,$A 
and B clouds, this volume-averaged density is between about 200 and 600$\, 
H_2\unit cm^{-3}$, depending on the precise assumptions used.  Therefore, DSS86 are 
assuming that entire clouds are stable and virialized.  Some evidence suggests that 
larger molecular clouds are indeed virialized \citep[i.e. for masses $\gsim 
10^4\unit M_\odot$, e.g., see][]{Heyer01, Simon01}, while other evidence suggests 
that many molecular clouds may not be \citep{Pringle01, Clark04, Vazquez07}, especially in 
extragalactic systems \citep[e.g., see][]{Israel00}.  Also, the X-factor seems to 
yield reasonable column density estimates for gas on scales smaller than  entire 
clouds \citep[e.g.,][]{W06}.
\bupskip 
\item {\it Stronger dependence of peak $T_{_{R}}$ on N(H$_2$) than of $\Delta$v on 
N(H$_2$) is not explained.\/}  DSS86 require that the observed velocity width of the 
line depends on the gas column density.  However, there is evidence that it is the peak 
radiation temperature, $\Tr$, that depends on N(H$_2$) and that $\Dv$ has only a weak 
dependence on N(H$_2$) \citep[see Figure~\ref{fig01} and][]{W06, Heyer96, Pichardo00, 
Ostriker01, Balle02}.
\end{enumerate}
\bupskip
The purpose of the current paper is to propose an improved approach for understanding
the X-factor that will resolve, or at least mitigate, the problems with DSS86.  For 
example, the explanation proposed here includes radiative transfer in a clumpy medium 
and shows how the optically thick CO~$\Jone$ emission of a cloud can be sensitive to 
the optical depths of the individual clumps.  As a result, this explanation will permit, 
in some circumstances, a very weak dependence on $\Tk$ and $\nH$.  Also, even though we 
will also use the virial theorem (except in one case), we can apply it to scales smaller 
than entire clouds.  And the X-factor in the current proposed explanation will lose its 
dependence on virialization in the optically thin case.  In addition, the proposed 
approach will naturally explain the dependence of the peak $\Tr$ on N(H$_2$).  This 
improved approach has shortcomings of its own, but nevertheless {\it represents the 
first major improvement in understanding the X-factor since DSS86,\/} because it is
the first formulation to include radiative transfer.  \citep[The reader may also 
consult][for a very brief description of the method.]{W06c}

\section{A Formulation for the X-Factor\label{sec2}}

\subsection{Radiative Transfer in a Clumpy Cloud\label{ssec21}}

The X-factor may yield a reasonable estimate of the molecular gas column density,
because the integrated intensity of the CO $\Jone$ line is essentially counting 
optically thick clumps in the gas in the beam \citep[e.g., see][]{Evans99}.  Two
clumps on the same line of sight within the beam will be, on average, separated 
in velocity by more than the velocity widths of the individual clumps, thereby
allowing the clumps to contribute their intensities to separate velocities within
the line profile without absorption of the emission from the more distant clump.  And 
clumps at the same velocity within the line profile will be, on average, at different 
locations within the beam, their intensities simply added together at that velocity 
within the profile.  This explanation does not, by itself, directly relate the masses 
of individual clumps to the observed integrated intensity, because, again, the clumps 
are optically thick in the CO $\Jone$ line.  Applying {\it only\/} the DSS86 approach 
to the clumps will not work, because, as discussed in the introduction, DSS86 and the 
observed value of the X-factor together require densities an order of magnitude lower 
than is found in the clumps of real clouds.  The DSS86 derivation of the X-factor depends 
on the beam-averaged column density, $N$, and the observed velocity width, $\Dv$.  We 
need a treatment of the problem in which the beam-averaged quantities, $N$ and $\Dv$, 
are cancelled out in favor of the corresponding quantities for an individual clump, 
i.e., $N_c$ and $\Dvc$.  And we need a treatment of the radiative transfer in a clumpy 
medium.

\citet{Martin84} (hereafter MSH84) developed a method for describing radiative
transfer through a clumpy medium in a highly simplified case: they assumed that
each clump was homogeneous and in LTE.  For additional simplicity, they also
assumed that the clumps were identical, although they pointed out that their 
method could be easily generalized to clumps with a spectrum of properties
(see the Appendix of MSH84).  The assumption of LTE was necessary because the
implicit assumption is that the excitation temperature {\it of the transition\/}
is constant throughout each clump, which is easily attained if the density is
high enough for LTE.  If the density is {\it not\/} high enough for LTE, then the 
populations of the rotational levels are affected by the ambient radiation
field at the frequency of the line: all other things being equal, the molecules on 
the surface of a clump will be less excited than those in the clump center, because
the former only see the radiation from roughly 2$\pi$ steradians of solid
angle, while the latter see it from the full 4$\pi$ steradians.  This results in a 
spatial gradient in the line's excitation temperature. If the density {\it is\/} high 
enough for the transition to be in LTE, then this excitation temperature will be equal 
to the kinetic temperature of the gas throughout the clump.  Since the clump is assumed
to be homogeneous, this kinetic temperature is constant throughout the clump,
thereby ensuring that the line's excitation temperature is also constant.  These
assumptions are particularly appropriate for the CO $\Jone$ line: because of its
high optical depth (i.e., $\tau\sim\ few$) and low critical density (i.e.,
$n_{_{crit}}\simeq 3\times 10^3\unit cm^{-3}$), this line is largely thermalized
(i.e., close to LTE).  Hence the method of MSH84 is appropriate here.

MSH84 used a statistical approach to find the appropriately averaged optical depth
on a sightline through a cloud with clumps in a vacuum.  The effective optical depth on 
a given sightline was expressed in terms of the individual clump opacities and the mean
number of clumps on a sightline with velocities within a clump's velocity width for the case 
of identical clumps.  This effective optical depth is the expectation value of the total 
optical depth of the clumps on a given sightline, considering the probability of a given 
clump impact parameter with respect to the sightline (i.e., perpendicular displacement of 
the clump center from the sightline) and, accordingly, of a given line-of-sight optical depth 
through each clump.  Computing this expectation value then depends on an average opacity over 
all impact parameters for each clump, which is the appropriately determined average opacity 
over the clump's surface area projected in the sightline's direction.   The appropriate average
of the optical depth is determined from the average over values of $[1-exp(-\tau)]$ and 
{\it not\/} over $\tau$ itself because the observed emission depends directly on the former 
and only indirectly on the latter --- and the relationship between the two is non-linear.  
Accordingly, the effective optical depth, $\tef$, is given by $1-exp(-\tef) = 
\langle 1-exp(-\tau)\rangle$, which implies $exp(-\tef) = \langle exp(-\tau)\rangle$, where 
$\langle\rangle$ indicates expectation value.  In the approximation of the spectral line
width, $\Dv$, being much larger than the velocity width of an individual clump, $\Dvc$, $\tef$ 
can be expressed as the product of the number of clumps per clump velocity width on a sightline
and the effective optical depth of an individual clump.  If $N$ is the beam-averaged
gas column density and $\Nc$ is the gas column density averaged over the projected area
of a single clump, then $(N/\Nc)(\Dvc/\Dv)$ is the number of clumps per sightline averaged
over the beam per clump velocity width at the line central velocity. (Note that MSH84 defined 
$N$ as the number of clumps per unit projected cloud area rather than the column density of 
gas. Note also that, below, $\Nc$ is actually the column density on the central sightline through 
the clump, but this change in definition accords with the definition of the clump average optical 
depth; see Appendix~\ref{appa} for details.)  If $\tau_0$ is the optical depth on a sightline 
through the center of a single clump, then, following MSH84, $\At$ is the clump effective optical 
depth.  (Note that MSH84 called $\At$ the effective optically thick area of the clump.  Even though 
their term is more accurate, the simpler ``clump effective optical depth'' is adopted here.) 
Consequently, 
\begin{equation}
\tef(v_z) = {N\over\Nc}{\Dvc\over\Dv}\,\At\, 
exp\left(-{\hbox{$v_z$}^2\over 2\,\Dv^2}\right)\quad ,
\label{bi01}
\end{equation}
where $v_z$ is the velocity component along the sightline and where a Gaussian
line profile has been assumed.  If $\tau(x,y)$ is the clump optical depth on the
sightline at position (x,y) with respect to a sightline through the clump center, then 
the clump effective optical depth is given by
\begin{equation}
\At = {1\over\sqrt{2\pi}\;\Dvc\;\aef}\int dv\int dx\int dy \left\{1-exp\left[-\tau(x,y)
\, exp\left(-{\hbox{$v$}^2\over 2\,\Dvc^2}\right)\right]\right\} \quad ,
\label{bi02}
\end{equation}
where $\aef$ is clump's effective projected area defined in terms of its optical depth:
\begin{equation}
\aef \equiv {1\over\tau_0}\int dx\int dy\ \tau(x,y) \quad .
\label{bi03}
\end{equation}
The $\tau_0$ is simply $\tau(x=0,y=0)$, the optical depth through the clump's center and at the 
center of the clump's velocity profile.  The integrals are over the projected area of 
the clump and over the clump's velocity profile.  For more details, see MSH84 and 
Appendix~\ref{appa}.  Figure~\ref{fig02} shows the variation of $\At$ as a function of
$\tau_0$ for two types of clumps: cylindrical (seen orthogonally to the axis of symmetry) 
and spherical.     

     The observed line radiation temperature, $\Tr$, is then related to $\tef$ by the 
usual expression
\begin{equation}
\rm\Tr(\nu) = \Jnu\,\,\left[1-exp(-\tef)\right]\qquad ,
\label{bi04}
\end{equation}
where
\begin{equation}
\rm\Jnu\equiv {\strut h\nu\over
\strut k}\Biggl\{\,\Biggl[exp\Biggl({\strut h\nu\over
\strut k\Tk}\Biggr) - 1\Biggr]^{-1} - 
\ \Biggl[exp\Biggl({\strut h\nu\over\strut 
k\Tbg}\Biggr) - 1\Biggr]^{-1}\Biggr\}\quad .
\label{bi05}
\end{equation}
$\Tk$ and $\Tbg$ are the gas kinetic and cosmic background temperatures, respectively.  As 
stated earlier, LTE is assumed for the emission of the spectral line at frequency, $\nu$.  
The $\Jnu$ is the source function in temperature units and is the correction for the cosmic 
microwave background emission and for the failure of the Rayleigh-Jeans approximation.  Of 
course when $\tef\ll 1$, we have the simplified form of equation~(\ref{bi04}):
\begin{equation}
\rm\Tr(\nu) = \Jnu\,\tef \quad .
\label{bi06}
\end{equation}
This is often called the ``optically thin limit'' for the equation of radiative 
transfer.  A very interesting and important point here is that the effective optical
depth is in the optically thin limit {\it even though the individual clumps can still be
quite optically thick.\/}  And this, of course, will provide a partial explanation for the 
X-factor.  Substituting equation~(\ref{bi01}) into equation~(\ref{bi06}) yields 
\begin{equation}
\Tr(\hbox{$v_z$}) = \Jnu\, {N\over\Nc}{\Dvc\over\Dv}\,\At\, 
exp\left(-{\hbox{$v_z$}^2\over 2\,\Dv^2}\right)\quad ,
\label{bi07}
\end{equation}
As mentioned previously, the quantity $(N/\Nc)(\Dvc/\Dv)$ is the number of clumps
per sightline averaged over the beam within a clump velocity width at the line central velocity. 
(Note that this is not exactly correct.  See the last paragraph of Appendix~\ref{appa} for 
an explanation.)  This quantity can be much less than unity, thereby permitting $\tef\ll 1$ 
even for $\At\gg 1$.  (When $(N/\Nc)(\Dvc/\Dv) < 1$, it is similar to the geometric area 
filling factor within a narrow velocity interval, although it is not {\it necessarily\/} 
equivalent.)  Since $\At$ is roughly equivalent to $[1-exp(-\tau)]$ for a single clump, 
the meaning of expression~(\ref{bi07}) is clear: it is the specific intensity of a single 
clump at velocity $v_z$ --- $\sim\Jnu\, [1-exp(-\tau)]$ --- multiplied by the number of 
clumps at that velocity within a clump velocity width --- 
$(N/\Nc)(\Dvc/\Dv)\, exp[-\hbox{$v_z$}^2/(2\,\Dv^2)]$.  Simply multiplying the intensity 
of a single clump by the number of clumps gives the observed intensity {\it if\/} the 
clumps are radiatively de-coupled, and this is ensured if $\tef\ll 1$.  As $\tef$ increases 
and becomes optically thick, the different clumps within each velocity interval start 
absorbing each other's emission and the radiation temperature approaches the source function 
$\Jnu$ asymptotically. 

      We are now better equipped to understand the behavior of the curves in 
Figure~\ref{fig02}.  $\At$ represents the level of emission from a single clump 
averaged over the clump's projected area.  When $\tau_0\ll 1$, $\At\simeq\tau_0$
because all lines of sight through the clump and the line's profile at all the
clump's internal velocities are optically thin.  As $\tau_0$ increases past unity, 
the line profiles on sightlines passing near the clump's center start saturating in
their cores and $\At$ starts deviating noticeably from the $\At=\tau_0$ line.  Nevertheless,
$\At$ continues rising with increasing $\tau_0$ because sightlines away from the
clump's center are still optically thin.  Even for sightlines near the clump's
center, the line remains optically thin at velocities outside the line's core.
As $\tau_0$ continues increasing, $\At$ deviates further and further from the $\At=\tau_0$ 
line because the area of optically thick emission slowly increases outwards from the
sightline through the clump center, covering more and more of the clump's projected
area.  The optically thick portion of the line profile on each sightline increases
as well.  Nevertheless, $\At$ continues growing because of those lines of sight
and those velocities at which the emission is still optically thin.  Figure~\ref{fig02}
shows two curves: one for a cylindrical clump of gas (i.e. a filament) and one for a 
spherical clump.  The cylinder is viewed side-on (i.e. with its symmetry axis perpendicular
to the sightline) and has length $h$.  If the symmetry axis is the $x$-axis, then a Gaussian
variation of the optical depth with $y$ was adopted:
\begin{equation}
\tau(x,y) =  \tau_0\, exp\left(-\pi h^2 {y^2\over\aef^2}\right)\quad .
\label{bi08}
\end{equation}
The spherical clump also has Gaussian spatial variation with optical depth, but with
radial distance, $p$, from the central sightline through the clump:
\begin{equation}
\tau(x,y) =  \tau_0\, exp\left(-\pi {p^2\over\aef}\right)\quad ,
\label{bi09}
\end{equation}
where $p=\sqrt{x^2+y^2}$.  This case was also treated by MSH84, and it is included 
here for comparison. (Note that the $\aef$ used here corresponds to the $r_o^2$ of
MSH84.)  The effective optical depth of the spherical clump grows faster with $\tau_0$
for $\tau_0\gsim 1$ than that of the cylindrical clump because the former's optically
thick area is growing simultaneously in two dimensions, whereas the latter's grows only
in one.  While $\At$ can grow without bound in these idealized cases, $\Tr$ cannot. 
Eventually, $\At$ will grow large enough that $\tef\ll 1$ is no longer valid and $\Tr$
asymptotically approaches $\Jnu$.  The growing $\tau_0$ causes this to happen because
the clumps start crowding each other spatially and in velocity, due to their increasing
optically thick areas and their increasingly saturated line profiles.

The curves of Figure~\ref{fig02} demonstrate that we can represent them as power-laws
in $\tau_0$ for $\tau_0\leq 1$ or $\tau_0\geq 3$:
\begin{equation}
\At\simeq k_A\, \tau_0^\epsilon\quad .
\label{bi09a}
\end{equation}
The values of $k_A$ and $\epsilon$ obviously depend on the specific $\tau(x,y)$ --- the
opacity structure of the clump, except in the optically thin case.  When $\tau_0<1$, we
have $k_A = 1$ and $\epsilon=1$, regardless of the specific variation of $\tau(x,y)$. 
A lower value of $\epsilon$, i.e. closer to zero, indicates a clump with a better defined 
outer edge like a hard sphere.  Conversely, aside from the optically thin case, a higher 
value of $\epsilon$, i.e. closer to unity, indicates a clump with a more tenuous, or 
fluffier, outer region.  Accordingly, $\epsilon$ will be called the ``fluffiness" of
the clump.


\subsection{Relating Clump Velocity Width with Column Density\label{ssec22}}

DSS86 required virialization in order to relate the line velocity width to the gas
column density.  That is also required here, but it will be combined with the 
radiative transfer in a clumpy cloud discussed in the previous subsection.  The virial 
theorem in its simplest form neglects the effects of surface pressure and magnetic
fields, yielding
\begin{equation}
2\,T+W = 0 \quad ,
\label{bi10}
\end{equation}
with $T$ as the total internal kinetic energy of the cloud and $W$ as its total
internal potential energy.  Assuming a spherical clump of uniform density gives
\begin{equation}
W = -{3\over 5}\,{G\,M_c^2\over R} \quad ,
\label{bi11}
\end{equation}
where $M_c$ and $R$ are the clump mass and radius, respectively.  The total kinetic
energy is
\begin{equation}
T = {1\over 2}\, M_c\,\Delta v_{3d}^2 \quad .
\label{bi12}
\end{equation}
The $\Delta v_{3d}$ is the three-dimensional velocity dispersion of the gas.
The exact kind of velocity dispersion this is depends on a number of factors,
including the radiative transfer through the gas (E.~Vazquez Semadeni, priv.
comm.).  Nevertheless, this velocity width most closely resembles an rms width.
The $\Delta v_{3d}$ is related to the one-dimensional velocity dispersion $\Dvc$
by $\Delta v_{3d}^2 = 3\,\Dvc^2$.  If we assume that molecular hydrogen is the
only form of hydrogen in the gas with number density, $n_c$, then the mass of
the spherical clump is
\begin{equation}
M_c = {4\pi\over 3} n_c\mu\mh R^3 \quad .
\label{bi12a}
\end{equation}
$\mh$ is the mass of the hydrogen molecule, $\mu$ is the helium correction, and
$R$ is the clump radius.  The density $n_c$ is related to the column density through 
the clump center by $n_c=\Nc/$$L_c$, where $L_c$ is the path length on the central 
sightline through the clump is equal to $2R$.  
Substituting equations~(\ref{bi12}), (\ref{bi12a}), and (\ref{bi11}) into (\ref{bi10}) gives 
us
\begin{eqnarray}
\Dvc &=& k_v\, \Nc^{0.5}\, L_c^{0.5} 
\label{bi13}\\
\noalign{\noindent and} 
k_v&\equiv&\left({\pi\over 15}\, G\,\mu\mh\right)^{0.5} \quad . 
\label{bi14}\\
\noalign{\noindent Numerically in $cgs$ units, this is}
k_v &=& 2.47\times 10^{-16}\quad ,
\nonumber
\end{eqnarray}
where $\mu = 1.3$ was used.  For a more detailed treatment of spherical clumps,
see Appendix~\ref{appb}.  (Also, see Appendix~\ref{appc} for a treatment of
cylindrical clumps.) 


\subsection{Relating Clump Optical Depth with Column Density\label{ssec23}}

The clump optical depth on the sightline through the clump's center,
$\tau_0$, can be written in terms of the column density of CO in level $J$, $N_J$:
\begin{equation}
N_{_{J}} = {\strut 8\pi\over\strut A_{_{J,J-1}}\lambda_{_{J,J-1}}^3}
\,\Biggl[exp\Biggl({\strut T_{_{J,J-1}}\over\strut T_{_K}}\Biggr)-1
\Biggr]^{-1}\tau_0\sqrt{2\pi}\Dvc\quad .
\label{bi18}
\end{equation}
This comes from equation~(A9) of \citet{W06} after applying the Boltzmann
factor to change $N_{J-1}$ to $N_J$.  The velocity integral was replaced by
$\tau_0\sqrt{2\pi}\Dvc$, where $\tau_0$ is the optical depth at the center
of the clump's velocity profile and on the sightline through the clump's
center.  $T_{_{J,J-1}}$ is the energy of the $J\rightarrow J-1$ transition
in units of temperature: i.e., $T_{_{J,J-1}} = h\,\nu_{_{J,J-1}}/k$ with
$\nu_{_{J,J-1}}$ as the frequency of the transition.  $A_{_{J,J-1}}$ and
$\lambda_{_{J,J-1}}$ are the spontaneous transition rate and the wavelength 
of the transition, respectively.  LTE is assumed, so $\Tk$ applies in place
of $\Tx(J\rightarrow J-1)$.  We can determine the total column density of
CO, $N(CO)$, by substituting equation~(\ref{bi18}) into equation~(A22) of 
\citet{W06}:  
\begin{equation}
N(CO) = {\strut 8\pi\over\strut (2J+1) A_{_{J,J-1}}\lambda_{_{J,J-1}}^3}
\, Q(T_{_K})\, exp\left({T_{_{J,0}}\over T_{_{_K}}}\right)\,
\Biggl[exp\Biggl({\strut T_{_{J,J-1}}\over\strut T_{_K}}\Biggr)-1
\Biggr]^{-1}\tau_0\sqrt{2\pi}\Dvc\quad ,
\label{bi19}
\end{equation}
where $Q(T_{_K})$ is the partition function of CO.  Setting J to 1 and rearranging 
for $\tau_0$ results in
\begin{equation}
\tau_0 = {\strut 3 A_{10}\lambda_{10}^3\over
\strut 8\sqrt{2}\,\pi^{3\over 2}\Dvc Q(T_{_K})}\, 
\Biggl[1-exp\Biggl(-{\strut T_{10}\over\strut T_{_K}}\Biggr)
\Biggr]\hbox{$\Nc$}X(CO)\quad .
\label{bi20}
\end{equation}
The $N(CO)$ was replaced by $\Nc$$X(CO)$, where $X(CO)$ is the abundance of
CO relative to H$_2$.  The following values are used \citep[see][and references
therein]{W06}: $A_{10}=7.19\times 10^{-8}\unit s^{-1}$, $T_{10}=5.54\unit K$, 
$\lambda_{10}=0.2601\unit cm$, and $X(CO) = 8\times 10^{-5}$.  Accordingly,
\begin{equation}
\tau_0 = {1.21\times 10^{-14}\over\sqrt{2\pi}\Dvc Q(T_{_K})}\, 
\Biggl[1-exp\Biggl(-{\strut 5.54\over\strut T_{_K}}\Biggr)
\Biggr]\hbox{$\Nc$}\quad .
\label{bi21}
\end{equation}
The above expression can be represented more simply as a power-law in $\Tk$:
\begin{equation}
\tau_0 = {k_\tau\over\sqrt{2\pi}\Dvc}\,\Nc\Tk^{-\gamma}\quad .
\label{bi22}
\end{equation}
The exact values of $k_\tau$ and $\gamma$ depend on the temperature range
and can be computed by numerically comparing expressions~(\ref{bi22}) and 
(\ref{bi21}).  In the high-temperature limit, however, an analytical solution
is possible.  This limit means that $\Tk\gg$$T_{10}$ and $Q(\Tk)\to 
2$$\Tk$$/T_{10}$ and $[1-exp(-T_{10}/T_{_K})]\to T_{10}/T_{_K}$.  This results 
in $k_\tau = 1.85\times 10^{-13}$ in $cgs$ units and $\gamma = 2$.  But we 
will be interested in the temperature range $\Tk=10$ to 20$\,$K.   The 
necessary numerical comparison gives us
\begin{eqnarray}
k_\tau &=& 7.23\times 10^{-14}\ (cgs\ units)
\nonumber\\
\noalign{\noindent and} 
\gamma &=& 1.75 
\nonumber
\end{eqnarray} 
for that range.  This approximation is accurate to within 1-2\% on the above
specified range. Equation~(\ref{bi22}) simplifies further by using expression~(\ref{bi13})
for $\Dvc$ and $n_c=\Nc/$$L_c$:
\begin{equation}
\tau_0 = {k_\tau\over k_v\sqrt{2\pi}}\, n_c^{0.5}\Tk^{-\gamma}\quad .
\label{bi22x}
\end{equation}
This interesting result suggests that the optical depth for this simplified case (i.e., the velocity 
profile of the optical depth is a simple Gaussian) of a virialized clump does not explicitly depend 
on the sightline pathlength nor the velocity width, but on their ratio.  This is related to the Sobolev
approximation \citep[e.g., see][]{Shu91} in which the optical depth is dependent on the velocity gradient 
within a given region and not explicitly on the region's size.   The pathlength-to-velocity-width ratio 
($L_c/\Dvc$) in a virialized clump is determined by the average density, the spatial variation of the 
density, and the geometry.  Therefore, the optical depth depends on those things and the gas temperature, 
but with {\it no\/} dependence on the clump size or velocity width (at least for this simplified case).

\subsection{The X-Factor\label{ssec24}}

Understanding how to combine the results of the previous subsections to derive an
expression for the X-factor requires examining the observational data that inspired 
the current paper in the first place. Figure~\ref{fig03} shows the Orion data discussed 
in \citet{W06}: the peak radiation temperature of the $\COone$ line (i.e., $\Tr$) for 
various positions in the Orion clouds normalized to the source function at each position 
(i.e., $\Jnu$) versus the gas column density (i.e., N(H$_2$)) as determined from $\cOone$.   
The plots demonstrate a clear correlation between $\Tr/\Jnu$ and N(H$_2$).   The Spearman 
rank-order correlation test indicates that the correlation exists at better than the 99.99\% 
confidence level.  (In fact, the confidence level of the null proposition of {\it no\/} 
correlation is zero to within the machine precision --- $10^{-38}$.)  This is more than just 
the expected correlation between the $\Jone$ lines of $\CO$ and $\cO$, because the $\Jnu$ is 
determined from the dust temperature \citep[see][for details]{W06}.  This suggests that the 
dust temperature really is a reliable measure of the kinetic temperature of the molecular gas, 
at least for the Orion clouds on the scales of parsecs \citep[see][for more discussion of 
this]{W06, W06a, W06b}.  One way of explaining the correlation visible in Figure~\ref{fig03} 
is that the area filling factor of the clump in each clump velocity interval is less than 
unity.  A rising beam-averaged column density, $N$, could mean that the filling factor is 
rising as more and more clumps fill the beam in each velocity interval.   Eventually the 
clumps start crowding each other within the beam and within the line velocity profile and 
the $\Tr/\Jnu$ ratio starts to saturate and asymptotically approaches unity. (Another possibility
is that $N$ rises because of rising $N_c$ within each clump, thereby increasing the 
clumps' optical depths.  This would also produce the observed saturation effect without 
increasing the number of clumps within each velocity interval.)  Obviously, 
the goal here is to be very specific about the relationship between $\Tr/\Jnu$ and N(H$_2$).  
There is sufficient scatter and uncertainty in the data that it is not possible to rule out {\it a 
priori\/} a number of such relationships.  

Nevertheless, from simple radiative transfer 
theory, we know that the specific intensity of a source normalized to its source function,
usually written $I_\nu/S_\nu$, will vary like $1-exp(-\tau)$ when plotted against
the optical depth through the source, $\tau$.  Given that the column density, $N$, 
is proportional to $\tau$ for constant kinetic temperature and density, the data in
Figure~\ref{fig03} mimic a curve with the form $1-exp(-aN)$ (see the plotted curves), 
where the $aN$ probably represents some kind of optical depth.   The majority of the 
data points are on the roughly linearly rising portion of the curve.  This represents 
the optically thin region of the curve, {\it but the $\COone$ is known to be optically 
thick\/} from comparisons with the optically thin isotopologue $\cO$.  Therefore, 
{\it a clue to understanding the X-factor is realizing that CO~$\Jone$ emission  
behaves like it is optically thin, despite being optically thick.\/}  This apparent
contradiction is resolved when we consider the effective optical depth as described
previously.  While the individual clumps are themselves optically thick in  
CO~$\Jone$, the cloud is optically thin ``to the clumps.''  In other words, the 
emission from every clump in the telescope's beam through the cloud reaches the observer.  
An analogy would be observing the H$\,$I 21-cm line from an atomic cloud.  In this case, 
the cloud is optically thin ``to the atoms'' in the sense that the emission from every 
atom in the telescope's beam through the cloud reaches the observer.  And since every
hydrogen atom is nearly identical in its 21-cm line emission properties, the conversion
from I(H$\,$I) to N(H$\,$I) is physically straightforward and undisputed.  For converting
from I(CO) to N(H$_2$), assuming absolutely identical clumps would give a constant value 
of the X-factor, relatable to the clumps' properties.  However, assuming identical clumps 
contradicts observational evidence \citep[see, e.g.,][]{Tachihara00, Nagahama98, Kawamura98,
Onishi96}.  But we need not restrict the clump properties so severely to explain the X-factor.  
All we need is to have the clumps similar {\it on average\/} from one beam to the next for 
the X-factor to stay relatively constant.  And, of course, we must also allow the X-factor to 
vary in some cases (see Introduction); the clumps' average properties must vary from the 
``norm'' in some clouds and locations. 

We now need to quantify this picture, so that we might better understand it and
its limitations.  As Figure~\ref{fig03} clearly shows, the $\Tr/\Jnu\propto N$
for $N\lsim 1$ to $2\times 10^{22}\, H_2\unit cm^{-2}$.  This is in the $\tef\ll 1$
limit, so equation~(\ref{bi07}) applies and it has the desired proportionality.
Of course, this proportionality is only visible if the clump properties --- $\Nc$,
$\Dvc$, and $\At$ --- and the observed line width, $\Dv$, do not vary strongly with $N$. 
In fact, the scatter visible in the plots of Figure~(\ref{fig03}) is probably due
to variations in all four of these quantities.  Integrating equation~(\ref{bi07})
over velocity, $v_z$, gives
\begin{eqnarray}
I(CO) &=& \sqrt{2\pi}\,\Tr(0)\,\Dv
\nonumber\\
      &=& \sqrt{2\pi}\,\Jnu\tef(0)\,\Dv\quad .
\label{bi23}
\end{eqnarray}
$\Tr(0)$ is the radiation temperature of the CO~$\Jone$ line at $v_z=0$ and is
also the peak radiation temperature of this line.  Similarly, $\tef(0)$ is the
effective optical depth at $v_z=0$.  The X-factor is then given by
\begin{eqnarray}
X_f &=& \left[\sqrt{2\pi}\,\Jnu\,\tef(0)\,\Dv\, N^{-1}\right]^{-1}\quad .
\label{bi24}\\
\noalign{\noindent If we now substitute equation~(\ref{bi01}) evaluated at
$v_z=0$ into the above, then}
X_f &=& \left[\sqrt{2\pi}\,\Jnu\,\At\,\Dvc\,\Nc^{-1}\right]^{-1}\quad .
\label{bi25}
\end{eqnarray}
The important thing to notice here is that the directly observed quantities,
$N$ and $\Dv$, have been replaced by the corresponding clump properties,
$N_c$ and $\Dvc$.  In fact, all the parameters in expression~(\ref{bi25})
are clump parameters, as desired.  It is convenient to define
\begin{eqnarray}
C_T &\equiv& {\Tk\over\Jnu}\quad .
\label{bi26}\\
\noalign{\noindent We now use the approximation $\Jnu\simeq\Tk-3.4\unit K$ for 
$\Tk\gsim 10\,$K and the frequency of the $\COone$ line, $\nu = 115.271\,$GHz.
This is good to within 0.4\% of $\Tk$ (and within 0.6\% of $\Jnu$).  Consequently,}
C_T &=& {\Tk\over\Tk-3.4\unit K}\quad ,
\label{bi27}
\end{eqnarray}
which approaches unity as $\Tk$ grows large.   Now we substitute the results of the 
previous subsections into equation~(\ref{bi25}): equation~(\ref{bi09a}) for $\At$, 
(\ref{bi22x}) for $\tau_0$, (\ref{bi13}) for $\Dvc$, and $\Tk$$/C_T$ for $\Jnu$. 
Except in the case of the end-on cylinder, where we defined the relationship between 
$N_c$ and $n_c$ differently, we also use $n_c=\Nc$$/L_c$.  These substitutions yield
\begin{equation}
X_f = (2\pi)^{{1\over 2}(\epsilon - 1)}\, C_T\, k_A^{-1}\,k_\tau^{-\epsilon}
\,k_v^{\epsilon - 1}\,\hbox{$\Tk^{\gamma\epsilon - 1}$}\,n_c^{{1\over 2}(1-\epsilon)}
\quad .
\label{bi29}
\end{equation}
The expression~(\ref{bi29}) and its variants (e.g., see Appendices) will be examined 
in detail.  It should be mentioned that expression~(\ref{bi29}) is more general than
for just a clumpy medium and can also apply to a uniform-density cloud (see 
Appendix~\ref{appab}).

The above formulation for $X_f$ obviously accomplishes the goal of insensitivity
to the parameters $\Tk$ and $n_c$ that we have sought for $X_f$.  The fluffiness
parameter, $\epsilon$, is in the range 0 to 1;  any value in that range that
is greater than 0 will confer a greater {\it in}sensitivity than occurs for the DSS86 
explanation.  A particularly interesting example is that value of $\epsilon$ for 
which $\gamma\epsilon - 1=0$.   In the high temperature limit, $C_T\to 1$ and 
$\gamma\to 2$ and, if $\epsilon=0.5$, then $X_f$ {\it has no dependence on 
temperature.\/}  (Notice that the density dependence is also very weak in this case: 
$X_f\propto n_c^{0.25}$.)  Given that the CO~$\Jone$ line is optically thick, this is 
counterintuitive; raising the temperature by some factor should simply increase 
$I(CO)$ by the same factor (in the high-$\Tk$ limit), thereby decreasing $X_f$ by 
that factor.  That is not the case here.   Here we are dealing with a clumpy medium 
where the optical depth varies across the projected area of each clump.  There will 
always be some sightlines through a clump that will still be optically thin.  There
will also be some velocities in the clump's spectral line profile where the line
emission is still optically thin.  As $\Tk$ increases, $\tau_0$ goes like
$\Tk^{-2}$, so that $\At$ goes like $\Tk^{-1}$ (see equations~\ref{bi22} and 
\ref{bi09a}).  But $\Jnu$ goes like $\Tk^1$ (in this high-$\Tk$ limit), meaning
that the observed $\Tr$ stays constant.  The effect of the increasing kinetic 
temperature of the gas is cancelled by the decreasing effective optical depth
of the clumps.  Another way of saying this is that the effect of the rising
temperature is cancelled by the shrinking effective optically thick areas of the 
clumps;  the filling factors of the clumps decline as the temperature rises.   (Note 
that this special case also occurs for lower temperatures.  For $\Tk = 10$ to 20$\,$K, 
for example, $C_T\propto\Tk^{-0.32}$ and $\gamma=1.75$.  The value of $\epsilon$ for 
which $X_f$ is independent of temperature would be 0.75.)  Therefore, {\it despite 
the optical thickness of the spectral line in the emitting clumps, changing the 
optical depths of the individual clumps will still have an appreciable effect on the 
line strength.\/}   And this will reduce the dependence of the X-factor on the 
temperature and the density of the gas within the clumps.

In general, the X-factor provides estimates of gas column density because each
sightline within the beam has some optically thin gas within certain narrow velocity 
ranges.  Parameters like $\epsilon$ then allow extrapolation from the optically thin
gas to all the gas. 

One problem with the above analysis is that it assumes clumps of homogeneous 
density.  This is inconsistent with most of the $\tau(x,y)$ functions that will 
be discussed in Section~\ref{sec30}.  The full analyses of these cases are
given in the appendices.

\section{Examination of the Properties of the X-Factor\label{sec30}}

In this section we examine the properties of the X-factor as formulated in the
previous section.

\subsection{Dependence on Clump Type\label{sec3}}

Here we examine how the X-factor relates to properties 
of the individual clumps, such as velocity width, optical depth, and mass.  
Accordingly, a ``standard'' clump --- or, rather, a {\it set\/} of standard clumps 
--- with specific input parameters must be adopted: geometry, dimensions, and density.  
These input parameters are based on the filamentary clumps as identified by 
\citet{Nagahama98} with their $\cOone$ map of the Orion$\,$A molecular cloud.  (It
should be mentioned here that the units responsible for the CO-line and dust-continuum
emission may be subfragments within the filaments identified by \citet{Nagahama98}.  
The filaments identified by that paper are nonetheless used as a first test of the X-factor's 
properties.)  Table~2 of that paper lists the Orion$\,$A clumps and their characteristics, 
including dimensions and masses.   Based on this table, the roughly cylindrical clumps have 
lengths ranging from 1.7 to 21$\,$pc with a mean length of 6.2$\,$pc, and diameters ranging 
from 0.7 to 3.5$\,$pc with and a mean diameter of 1.8$\,$pc.  Using the mass-to-volume ratios, 
the average densities range from about 300 to 4000$\, H_2\cdot molecules\cdot\rm cm^{-3}$.  
\citet{Nagahama98} also used $\COone$ data to estimate the gas kinetic temperature for 
each filament.  Their Table~2 shows values ranging from about 10 to 40$\,$K.  

The adopted kinetic temperature and density values for the standard clumps use the above 
numbers in combination with other considerations.   The adopted kinetic temperatures, for 
instance, also consider the numbers found for the Galactic disk at many-parsec scales.  
For example, the maximum temperature expected for the molecular gas and its dust on such 
scales in the Galaxy is about 20$\,$K \citep[e.g., see][]{Sanders85, Sodroski94}.  Therefore, 
the standard clumps have adopted kinetic temperatures of either 10$\,$K or 20$\,$K, and not
temperatures as high as the 30 to $\sim$40$\,$K values in Table~2 of \citet{Nagahama98}. 
The adopted densities for the standard clumps are loosely related to the range of densities 
mentioned in the previous paragraph, but also expand the range to better explore the effect 
(or to demonstrate the lack of it) on the derived X-factor:  the adopted densities for the 
standard clumps are 200, 2000, and 20000$\, H_2\cdot\rm cm^{-3}$. 

The adopted dimensions for the standard clumps have no effect on the derived X-factor or clump 
optical depth (see equations~\ref{bi29} and \ref{bi22x}), except in the case of a cylindrical 
filament viewed end-on.  And even in that case, the derived X-factor still does not depend on the 
{\it individual\/} values of the diameter or length, but on their ratio (see Appendix~\ref{appc}).  
Both the X-factor and clump optical depth depend on the clump temperature and average density, but 
{\it not\/} on clump size.  Nevertheless, the clump dimensions still affect the derived clump mass 
and clump velocity width.  For the spherical clumps, the adopted diameter is 1.8$\,$pc --- the same 
as the mean diameter of the observed Orion filaments.  For the cylindrical filaments, the adopted 
length and diameter are 6.2$\,$pc $\times$ 1.8$\,$pc --- again the same as the mean values of the 
observed Orion filaments.

Given that three kinds of density variations are considered for the spherical clumps and one for
the cylindrical filamentary clumps, there are twenty-four standard clumps in the set.  For each
type of spherical clump, each with diameter 1.8$\,$pc, there are six $\Tk$,$n_c$ combinations:
$\Tk = 10\,$K, and $n_c = 200$, 2000, and 2000$\, H_2\unit cm^{-3}$ and $\Tk = 20\,$K with those 
same three densities.  For the single type of cylindrical clump, with $length$~$\times$~$diameter$ 
$=$ 6.2$\,$pc $\times$ 1.8$\,$pc, there are the same six $\Tk$,$n_c$ combinations as for each type 
of spherical clump.  The different kinds of density variations considered for the spherical clumps 
are as follows: uniform, Gaussian, and squared Lorentzian.  For the cylindrical filamentary clump, 
only the Gaussian radial density variation is considered, viewed from the side (perpendicular to 
the symmetry axis) and from the end (along that axis).  All these cases are listed in 
Table~\ref{tbl-1}, along with two extra cases discussed in the next two subsections.  

The following subsections and the appendices examine these cases in detail, the results of which 
are summarized in Tables~\ref{tbl-1} -- \ref{tbl-5}.

\subsubsection{Completely Optically Thick Case: Optically Thick Disks with Flat-Topped 
Velocity Profiles\label{ssec31}}

For this case we will examine face-on optically thick disks with completely
flat-topped velocity profiles.  This means that the optical depth of the
$\COone$ line will be equal to $\tau_0$ for all lines of sight through the
clump and at every point in the velocity profile for each sightline.   $\At$ 
would then have the thoroughly familiar form, 
\begin{equation}
\At = 1- exp(-\tau_0)\quad .
\label{xx1}
\end{equation}
In the optically thick case, i.e. $\tau_0\gsim 2$,
\begin{equation}
k_A = 1. \quad {\rm and}\quad\epsilon = 0. \quad .
\label{xx2}
\end{equation}
With fluffiness and $k_A$ set to zero and unity, respectively, all dependence on optical depth 
parameters $k_A$, $\epsilon$, $k_\tau$, and $\gamma$ disappears from expression~(\ref{bi29}), 
resulting in an expression for $X_f$ that is devoid of radiative transfer: 
\begin{eqnarray}
X_f &=& (2\pi)^{-{1\over 2}}\, C_T\,\,k_v^{- 1}\,\hbox{$\Tk^{- 1}$}\,n_c^{{1\over 2}}\quad .
\label{xx3}\\
\noalign{\noindent Numerically, this is}
X_f(cgs) &=& 1.62\times 10^{15}\ C_T\,\hbox{$\Tk^{- 1}$}\,n_c^{{1\over 2}}\quad ,
\nonumber\\
\noalign{\noindent or}
X_f(X_{20}) &=& 1.62\ C_T\,\hbox{$\Tk^{- 1}$}\,n_c^{{1\over 2}}\quad ,
\label{xx4}
\end{eqnarray}
where the X-factor in (\ref{xx4}) is in units of $X_{20}$ or $10^{20}\, H_2\cdot molecules\cdot\cKkms$.
(Because the expressions~\ref{xx3} and \ref{xx4} come from \ref{bi29}, we have actually assumed that 
the clumps are spheres rather than disks; the result is almost the same so long as $\At$ is given by 
\ref{xx1}.)

This of course is the DSS86 result with the $n_c^{0.5}/\Tk$ dependence for $X_f$.  The resultant
numerical values for $X_f$ are listed in the first row of Table~\ref{tbl-1}.  Notice that these $X_f$ 
values are higher than those for the other cases.  Also, notice that the only reasonable values for 
$X_f$ occur for a density of $2\times 10^2\, H_2\unit cm^{-3}$.     Given that this result only occurs 
for this highly contrived case of optically thick disks with flat-topped velocity profiles, the DSS86 
result is unlikely.  Nevertheless, DSS86 represents a useful limit.

\subsubsection{Completely Optically Thin Case\label{ssec32}}

This case assumes that the line emission is optically thin on all sightlines through the
clump and at all velocities within the line profile.  Consequently, $\At=\tau_0$ and 
\begin{equation}
k_A = 1. \quad {\rm and}\quad\epsilon = 1. \quad .
\label{xx5}
\end{equation}
Now with fluffiness set to unity, dependence of $X_f$ on $k_v$ and $n_c$ disappears.  So
expression~(\ref{bi29}) is free of the dependence on virialization; this is the exact opposite
of the optically thick case discussed in the previous subsection.  Accordingly, 
\begin{eqnarray}
X_f &=& C_T\,\,k_\tau^{- 1}\,\hbox{$\Tk^{\gamma -1}$}\quad .
\label{xx6}\\
\noalign{\noindent For $\Tk = 10$ to 20$\,$K,}
X_f(X_{20}) &=& 1.38\times 10^{-2}\ C_T\,\hbox{$\Tk^{0.75}$}\quad .
\label{xx7}\\
\noalign{\noindent These values of $X_f$ are listed in the second row of Table~\ref{tbl-1}.
Notice that the $X_f$ values are lower than those for the other cases.   
Equation~\ref{xx7} can also be written as, } 
N(H_2\cdot cm^{-2}) &=& 1.38\times 10^{18}\ C_T\,\hbox{$\Tk^{0.75}$}\, I(CO)(K\cdot km\cdot s^{-1})\quad .
\label{xx8}
\end{eqnarray}
Therefore, the relationship between column density and integrated intensity in the optically 
thin case is recovered.  (This is easily verified by starting with expression~\ref{bi23} for
$\At = \tau_0$ and combining this with expression~\ref{bi22}.)

\subsubsection{Spherically Symmetric Clumps\label{ssec33}}

Here we examine cases examined by MSH84: the hard sphere, the Gaussian sphere, and the 
squared-Lorentzian sphere.   These examples do not necessarily represent the real clumps in real 
molecular clouds, but simply represent an interesting exploration of the parameter space that 
determines the $X_f$ values.  In all the cases considered in this paper, the clumps are isothermal 
and in LTE, implying that the clump optical depth as a function of impact parameter, or projected 
radius $p$, is proportional to that for the column density: $\tau_c(p) \propto N_c(p)$.   See 
Appendix~\ref{appb} for a detailed treatment of the spherically symmetric examples.  As stated 
at the beginning of Section~\ref{sec3}, the spherical clumps considered here have radii equal to 
1.8$\,$pc. 

First, the hard sphere example is examined.  This is a uniform-density sphere with a well-defined
edge at radius, R.  Given that the density is uniform, the optical depth profile over the projected
area of the clump is simply the sightline path-length for impact parameter, $p$:
\begin{equation}
\tau(p) = \tau_0\,\left[1 - \left({p^2\over R^2}\right)\right]^{0.5} \quad ,
\label{xx9}
\end{equation}   
the same as equation~(12) of MSH84, with $R$ in place of $r_0$.   Using (\ref{xx9}) along with
(\ref{bi01}), (\ref{bi02}), and (\ref{bi03}) yields the curve depicted with plus signs in Figure~5
of MSH84.  Measuring the curve for $log\,\tau_0\geq 0.5$ results in $k_A = 1.7$ and $\epsilon=0.14$. 
Following the derivation in Subsections~\ref{ssec22}, \ref{ssec23}, and \ref{ssec24} or in 
Appendix~\ref{appb} yields expression~(\ref{apb12o}) for $X_f$.   The numerical results for this
case are listed in Tables~\ref{tbl-1} to \ref{tbl-5} inclusive.  The hard sphere is the case closest
to the completely optically thick case and yields results within factors of a few of those of DSS86
(see Table~\ref{tbl-1}).  The hard sphere has the advantage of giving reasonable $X_f$ values (i.e.
$\sim 2\, X_{20}$ for reasonable densities (i.e. $\sim 10^3\, H_2\unit cm^{-3}$) and temperatures
(i.e. 20$\,$K), but has the disadvantage of sensitivity to density and temperature (i.e. $X_f\propto
C_T\, n^{0.43}\Tk^{-0.76}$). 

The Gaussian sphere has an optical depth profile given by expression~(\ref{apb13}), with $\tau(p)$ and 
$\tau_0$ in place of $N_c(p)$ and $n_0\,\sqrt{\pi}$, respectively.   Using this $\tau(p)$ and the 
equations~(\ref{bi01}), (\ref{bi02}), and (\ref{bi03}) yields $\At$ versus $\tau_0$: the solid curve in 
Figure~5 of MSH84 and the dashed curve in Figure~\ref{fig02} of the current paper.   Measurement of this 
curve gives $k_A = 1.6$ and $\epsilon=0.36$ for $\tau_0\geq 3$.  The derivations in Subsection~\ref{appb2}
of Appendix~\ref{appb} give the numerical results listed in Tables~\ref{tbl-1} to \ref{tbl-5}.  As seen in
Table~\ref{tbl-1}, the expected $X_f$ values for $\nb=2\times 10^3 H_2\unit cm^{-3}$ are within about a 
factor of 2 of the observed value for the Galactic disk.  Also, expression~(\ref{apb16}) shows that 
$X_f\propto C_T\,\nb^{0.32}\Tk^{-0.38}$, which is a greater {\it in}sensitivity to physical conditions 
than occurs for the hard sphere case.

The squared-Lorentzian sphere's optical depth profile is given by equation~(\ref{apb18}), with a similar 
expression (i.e. \ref{apb19}) for the column density profile.  Following the usual procedure gives the 
$\At$ versus $\tau_0$ curve: the ``$\times$'' symbols in Figure~5 of MSH84, for which we find that 
$k_A=1.5$ and $\epsilon=0.57$.  This particular optical depth profile has the advantage of low sensitivity
to density and temperature (i.e. $X_f\propto C_T\, n^{0.22}\Tk^{-0.003}$), but the disadvantage of $X_f$
values of roughly factors of 4 too low for a density of $\sim 10^3\unit cm^{-3}$. 

In addition to the $X_f$ values (i.e. Table~\ref{tbl-1}), the tables also have the $\Dvc$, $\tau_0$, 
$\At$, and $M_c$ values (i.e. Tables~\ref{tbl-2} through \ref{tbl-5} inclusive).  The $\Dvc$ and $M_c$
in Tables~\ref{tbl-2} and \ref{tbl-5}, respectively, can be compared with the velocity widths and masses
of the clumps in the Orion~A cloud, as listed in Table~2 of \citet{Nagahama98}.  Given that spherical
clumps with a diameter of 1.8$\,$pc were adopted here, we need the roughly analogous clumps in 
\citet{Nagahama98}.  These are the clumps numbered 14 and 23 in their Table~2, with dimensions 
$2.3\unit pc\times 1.8\unit pc$ and $1.7\unit pc\times 1.5\unit pc$, respectively.  The observed masses
of these clumps, 330$\unit M_\odot$ and 220$\unit M_\odot$, suggest that they most closely resemble 
either the hard sphere of density $2000\unit cm^{-3}$ or the Gaussian sphere of average density
$200\unit cm^{-3}$.  However, their observed velocity width of 1.1$\kms$ suggests that they more 
closely resemble the hard sphere, but with about 2/3 the density, or $\sim 1300\unit cm^{-3}$. Here
we have also considered that the masses given in Table~2 of \citet{Nagahama98} do not include the
correction for helium.  This lower density would imply $\Dvc\simeq 1.2\kms$, close to that observed.  This 
implies that these clumps are virialized, as is required for a roughly constant X-factor to apply.   However, 
hard spheres with $\Tk\simeq 20\unit K$ \citep[see column~5 of Table~2 of][]{Nagahama98} and $n_c=1300\unit 
cm^{-3}$ would result in an $X_f$ of about 1.3$\, X_{20}$ or about 2/3 that observed for the Orion clouds 
\citep[see][]{W06, Dame01}.  Consequently, clumps~14 and 23 are not entirely representative of the rest of 
the clumps of the Orion clouds. 

Tables~\ref{tbl-3} and \ref{tbl-4} give the optical depths on the central sightline, $\tau_0$, and
the clump effective optical depths, $\At$.  The $\tau_0$ values listed in Table~\ref{tbl-3} seem to
be too high.  The observed I($\cO$)/I($\CO$) values for the Orion clouds (see Figure~\ref{fig05}) are 
mostly in the range 0.1 to 0.4, implying $\tau(\CO)$ of 6 to 24.  However, Table~\ref{tbl-3} lists 
values that are an order of magnitude larger.   Nevertheless, given that a roughly constant X-factor 
only occurs on large scales (i.e. many parsecs) and that the data for which the X-factor was estimated 
had a spatial resolution of about 1$^\circ$ \citep[see][]{W06}, the relevant optical depth to use would 
be $\At$ and not $\tau_0$.  The former is listed in Table~\ref{tbl-4} and has values in the desired 
range and lower. (Although interpretation of the I($\cO$)/I($\CO$) ratio is a little more complicated 
than in the case of homogeneous gas. See Section~\ref{ssec39} for more details.) 


\subsubsection{Cylindrically Symmetric Clumps\label{ssec34}}

Filaments or cylindrical clumps are discussed in detail in Appendix~\ref{appc}.  The behavior of these
clumps --- e.g., the derived $\tau_0$ and $X_f$ values --- depend on whether these filaments are observed 
side-on (i.e., perpendicularly to their symmetry axes) or end-on (i.e., parallel to these axes).  The
filament considered here has a Gaussian density variation with distance from the central axis, viewed
side-on and viewed end-on.   As mentioned near the beginning of Section~\ref{sec3}, the adopted 
dimensions of the filament are $6.2\unit pc\times 1.8\unit pc$. 

For the case of the Gaussian cylinder viewed side-on, the optical depth, $\tau$, as a function of
projected distance from the central axis, $y$, is similar to equation~(\ref{apc08}), but with $\tau(y)$
in place of $N_c(y)$ and $\tau_0$ in place of $N_c(0)$.  Using $\tau(y)$ with (\ref{bi01}), (\ref{bi02}), 
and (\ref{bi03}) yields the solid curve in Figure~\ref{fig02}.  Measuring the curve for $log\,\tau_0\geq
0.5$ gives us $k_A=1.5$ and $\epsilon=0.25$.  Side-on cylinders are expected to be less fluffy than spheres 
with the same optical depth profile, given that the former have optical depth variations in only one projected 
dimension and the latter have those variations in two projected dimensions.  Here we see that the Gaussian
side-on cylinder has a fluffiness that is only about 2/3 that of the Gaussian sphere (i.e. 0.25 versus 0.36). 
The derivation in Appendix~\ref{appc}, Subsection~\ref{apppc1}, gives the numerical results in Tables~\ref{tbl-1} 
to \ref{tbl-5}.  As in the case of the Gaussian sphere, the resultant $X_f$ values for an average density of 
$2\times 10^3 H_2\cdot\unit cm^{-3}$ are within a factor of 2 of the value inferred for our Galaxy.  Like the 
hard sphere case, this $X_f$ is a little too sensitive to $\Tk$ and $\nb$: $X_f\propto C_T\, \Tk^{-0.56}\, 
\nb^{0.37}$.

The Gaussian cylinder viewed end-on is similar to the Gaussian sphere in its optical depth profile.  However,
the central sightline optical depth, $\tau_0$, depends on the cylinder length-to-diameter ratio or aspect ratio,
$h/\Dh$.  Using the same $k_A$ and $\epsilon$ as the Gaussian sphere, the derivation in Appendix~\ref{appc}, 
Subsection~\ref{apppc2}, gives the numerical results in Tables~\ref{tbl-1} to \ref{tbl-5}.  The $X_f$ value
for an average density of $2\times 10^3 H_2\cdot\unit cm^{-3}$ and $\Tk =20\,$K is very close to that observed
for our Galaxy.  Also like the Gaussian sphere, this $X_f$ also has the advantage of relative insensitivity to 
$\Tk$ and $\nb$: $X_f\propto C_T\,\nb^{0.32}\Tk^{-0.38}$.  Unfortunately, these advantages are merely lucky 
coincidences, because the $X_f$ is dependent on the filament's aspect ratio ($X_f\propto (h/\Dh)^{0.64}$),
which can vary by an order of magnitude or more.  For example, the filaments of Orion$\,$A listed by 
\citet{Nagahama98} have aspect ratios that vary from about 1 to 10.  Another coincidence is that the end-on
case implicitly assumes that the cylindrical filament is absolutely straight and pointing along the 
sightline --- a low probability event.  

Comparison of the results in Tables~\ref{tbl-1} to \ref{tbl-5} with the clump properties listed in Table~2
of \citet{Nagahama98} suggests that the Orion$\,$A filaments behave more or less like Gaussian cylinders, 
except for the inferred $X_f$ values.  Clumps~12, 22 and 33 have dimensions similar to the standard dimensions 
adopted here: $6.2\unit pc\times 1.8\unit pc$.  The average densities of these filamentary clumps, $\nb$, can 
be inferred from the column densities divided by the product of the observed thicknesses and the $k_N$ for a
Gaussian cylinder (see Appendix~\ref{appc}).  These average densities and the dimensions can be combined to 
give the masses and velocity widths.  These are found to agree with those in Table~2 of \citet{Nagahama98} to 
within 7\%, except for the velocity width of clump~33.  The observed value for this clump is about double 
the virialized velocity width, strong evidence that this clump is not virialized.  Another problem is the 
inferred $X_f$ values.  Table~2 of \citet{Nagahama98} provides enough information to roughly estimate the $X_f$ 
that would correspond to each of the filaments listed.  Assuming side-on Gaussian cylinders for all 39 filaments 
listed gives $X_f=0.6$ to 1.2$\,X_{20}$, factors of roughly 2 to 4 lower than the estimated value for the Orion 
clouds \citep[see][]{W06, Dame01}.  One way that the $X_f$ can be raised to that estimated from observations is 
to consider uniform-density filaments.   As seen in Table~\ref{tbl-1} for the spheres for $\nb = 2\times 10^3
\unit cm^{-3}$, the uniform-density case (i.e., hard sphere) has $X_f$ about a factor of 2 larger than for
the Gaussian sphere.  Accordingly, uniform-density filaments would correspond to higher X-factors, closer
to that observed for Orion.  One slight problem with uniform-density filaments is that the inferred virialized
velocity widths are lower than those for the Gaussian filaments, resulting in a slightly greater number of
non-virialized filaments.  For uniform-density filaments, roughly half have velocity widths within 40\%
of the virial velocity widths.  For Gaussian filaments, roughly half have velocity widths within 30\% of the
virial velocity widths. 

\subsubsection{Gravitationally Collapsing, Magnetized Filaments\label{ssec35}}

The examples of clumps that have been examined up till now have all been virialized clumps.  Strictly 
speaking, virialization is not required for determining $X_f$; all that is required is a relationship 
between the clump velocity width and the clump's density.  \citet{Tilley03}, for example, examine 
gravitationally collapsing, magnetized filaments.  They investigate filaments with constant toroidal 
flux-to-mass ratio and those with constant thermal gas pressure to magnetic pressure.  In the current 
paper only the former is considered (see Appendix~\ref{appc}, Subsection~\ref{appc3}).   Their 
equation~(35) is rearranged to give the clump velocity width in terms of the density; see 
equation~(\ref{apc19}).  \citet{Tilley03} find that the density goes like $r^{-\alpha}$ at large $r$,
where $\alpha =2$ in the case of strong magnetic fields and $\alpha = 4$ for weak magnetic fields.  
Both of these cases are examined here.

For the case of $\rho(r)\propto r^{-2}$ (see Appendix~\ref{appc}, Subsubsection~\ref{apppc31}), the
total clump mass, $M_c(r_1)$, depends on the ratio, $r_1/r_0$, where $r_1$ is the clump's outer
radius and $r_0$ is the radius of the $\rho = constant$ core.  In contrast, all the previous clump
examples had finite $M_c(\infty)$.  This dependence on $r_1/r_0$ also extends to $\At$.  Numerically
integrating (\ref{bi02}) and (\ref{bi03}) shows that $\epsilon$ is an increasing function of this
ratio, asymptoting out at about 0.9 for $r_1/r_0 \gsim 5\times 10^4$.  It is clear then that the
fluffiness of the filament depends not only on the density exponent, $\alpha$, but also on the
ratio $r_1/r_0$.  As was found previously, increased fluffiness had the advantage of reduced 
sensitivity to $\Tk$ and $\nb$, but the disadvantage of unrealistically low $X_f$.   For example,
for $r_1/r_0 = 1000$, we have $k_A = 1.46$, $\epsilon = 0.69$, and $X_f\propto C_T\, {\Tk^{0.21}}\, 
\nb^{0.16}$.  To determine a value for $X_f$, we must specify a reasonable number for the fragmentation 
wavelength, $\lambda_{frag}$.  The simplest way to specify this is to adopt the average cylinder length, 
6.2$\,$pc, that has been used up till now.   Equation~(35) of \citet{Tilley03} allows us to estimate
this number.  The observed $\Dvc(FWHM)$ in most of the filaments in Orion$\,$A is 1-2$\kms$ or velocity 
dispersions of 0.4-0.8$\kms$ (after dividing by $\sqrt{8\, ln\, 2}$).  The conversions from the central 
density to the average density are found in Appendix~\ref{appc}.  These together yield $\lambda_{frag}
= 4.4$ to 8.8$\,$pc for $\alpha = 2$ (and 3.5 to 7.0$\,$pc for $\alpha = 4$).  These numbers suggest
that choosing $\lambda_{frag} = 6.2\unit pc$ is acceptable.   For reasonable temperature and density, 
$\Tk =20\,$K and $\nb = 2000\, H_2\cdot\unit cm^{-3}$, $X_f = 0.36\, X_{20}$, or a factors of $\sim$5 
smaller than that for the Galaxy.  On the other hand, if $r_1/r_0 = 10$, we have $k_A = 1.74$, 
$\epsilon = 0.27$, and $X_f\propto C_T\, {\Tk^{-0.53}}\,\nb^{0.37}$.  For $\Tk =20\,$K and $\nb = 
2000\, H_2\cdot\unit cm^{-3}$, $X_f = 1.6\, X_{20}$, which is close to the Galactic value.  One problem 
with the $r_1/r_0 = 10$ case is that the power-law approximation for $\At$ is noticeably poorer than 
for the othercases studied in the current paper.   For these other cases, $\At\simeq k_A\, \tau_0^\epsilon$ 
is good to within about 13\%.  For $r_1/r_0 = 10$, it is only good to within about 26\%. 

For the case of $\rho(r)\propto r^{-4}$ (see Appendix~\ref{appc}, Subsubsection~\ref{apppc32}), 
$M_c(\infty)$ is finite and the power-law approximation for $\At$ is as good as that for the majority 
of cases discussed in the current paper.  Numerical integration of (\ref{bi02}) and (\ref{bi03}) yields 
$k_A = 1.15$ and $\epsilon = 0.37$, which results in  $X_f\propto C_T\, {\Tk^{-0.36}}\,\nb^{0.32}$. 
For reasonable temperature and density, $\Tk =20\,$K and $\nb = 2000\, H_2\cdot\unit cm^{-3}$, $X_f = 
1.9\, X_{20}$, which is approximately the value for molecular clouds in the Galactic disk. 

One complication with the above analysis is that magnetized filaments have velocity dispersions that 
vary spatially within the filaments.  The central velocity dispersion used in equation~(35) of 
\citet{Tilley03} may differ from the observed velocity dispersion by a factor of about 2, as suggested 
by Figure~4 of \citet{Fiege00}.  If so, then $\lambda_{frag}$ is a factor of two lower than the numbers 
used above.  Of course, if this is true, then the value used for $k_v$ must be increased by this factor
of 2.  Since $X_f$ is a function of the product, $k_v\ \lambda_{frag}$, the factor of two decrease in
$\lambda_{frag}$ is cancelled by this factor of increase in $k_v$, leaving $X_f$ unchanged.  

The density dependences found for the collapsing, magnetized filament are those expected for an
isothermal filament in hydrostatic equilibrium \citep[][]{Tilley03}.  \citet{Ostriker64} solved
for the density distribution in an isothermal filament and found the $\rho(r)\propto r^{-4}$
at large $r$.  This solution would also permit $\rho(r)\sim r^{-2}$ over a range of intermediate
$r$.  Since this is similar to the case of the collapsing, magnetized filament, how do the $X_f$
values compare?  The $k_M$ values were computed in Appendix~\ref{appc}, Subsection~\ref{appc3},
allowing determination of $k_v$ for a virialized filament.   Using the virialized $k_v$ values
and removing the $[\Dh/(\lambda_{frag}\ k_{max})]^{(1-\epsilon)}$ factor will give the corresponding
virialized $X_f$ values.  For $\rho(r)\propto r^{-2}$, $X_f$ for the virialized, non-magnetic filament
is up to nearly 20\% smaller than that for the collapsing, magnetized filament.   For $\rho(r)\propto 
r^{-4}$, $X_f$ for the virialized, non-magnetic filament is almost 50\% larger than that for the 
collapsing, magnetized filament.  The X-factor values derived in the isothermal, virialized case
are then easily within a factor of 2 of those for the gravitationally collapsing, magnetized 
filament.

\subsection{Other Effects on the X-Factor\label{ssec320}}

\subsubsection{Fluffiness\label{ssec36}}

One advantage of the current, proposed formulation of the X-factor is that $X_f$ can be insensitive to 
the kinetic temperature and density.  But this advantage is coupled to an important disadvantage: $X_f$ 
is sensitive to the fluffiness, $\epsilon$.  Another way of saying this is that the problem of 
sensitivity to temperature and density has been recast as one of sensitivity to another physical
parameter.  If real clumps in real clouds can indeed be characterized, at least approximately, with
a parameter $\epsilon$ and if the proposed formulation of the X-factor is roughly correct, then a
relatively constant $X_f$ requires constant $\epsilon$.  Understanding the X-factor would then provide 
insights into molecular cloud structure.  More specifically, the X-factor would itself be a probe of the 
opacity structure in CO-emitting clumps. 

One obvious way to do this is to plot $X_f$ as a function of $\epsilon$ and see what range of $\epsilon$ values 
gives realistic $X_f$ values.   This must be done for specific and realistic values of $\Tk$ and $\nb$, but the 
parameters $k_A$, $k_v$, and $k_N$ have implicit dependences on $\epsilon$.   These dependences are not simple, 
given that they are affected by many things, such as geometry and density variations.  However, these dependences 
are not strong and a crude variation of $X_f$ versus $\epsilon$ can be plotted if we adopt ``typical'' values for 
these quantities: $k_A = 1.4$, $k_v = 3.4\times 10^{-16}\, cgs$, and $k_N = 1.5$.   These values along with $\nb 
= 2000\, H_2\unit cm^{-3}$ and $\Tk = 10$ and 20$\,$K were substituted into equation~(\ref{apb11}) to yield the 
curves of Figure~\ref{fig04}.  Note that the curve is about 30\% lower than the level expected for $\epsilon = 0$ 
and $\epsilon = 1$ (cf. Table~\ref{tbl-1}). This occurs because $k_v$, $k_N$, and especially $k_A$ were fixed. 
Another point to consider is that the positions of these curves also depend on exactly how the ``average''
densities are defined.  Nevertheless, the curves of Figure~\ref{fig04} still permit approximate estimates of 
$\epsilon$ in molecular clouds.

Figure~\ref{fig04} can estimate the range of $\epsilon$ that results in $X_f$ within a factor of 2 of that for 
Galactic disk molecular clouds.   Given that Galactic disk molecular clouds have $X_f\simeq 2\, X_{20}$ 
\citep[][]{Dame01}, what range of $\epsilon$ permits $X_f$ to be in the range 1 to 4$\, X_{20}$?  For $\Tk =
20\,$K and $\nb = 2000\, H_2\unit cm^{-3}$, $\epsilon = 0.0$ to 0.35.  (Taking into account the underestimate
of $X_f$ for $\epsilon = 0$, this lower limit is more like 0.05.)   For $\Tk = 10\,$K and $\nb = 2000\, H_2\unit 
cm^{-3}$, $\epsilon = 0.17$ to 0.46.  For $X_f = 2\, X_{20}$, $\epsilon = 0.15$ when $\Tk = 20\,$K and $\nb = 
2000\, H_2\unit cm^{-3}$ and $\epsilon = 0.31$ when $\Tk = 10\,$K and $\nb = 2000\, H_2\unit cm^{-3}$.  So for
the adopted density, the clumps in Galactic disk molecular clouds behave most like hard spheres if $\Tk = 20\,$K
and roughly like Gaussian spheres or filaments if $\Tk = 10\,$K.  If instead the clumps are filaments with
$\rho\propto r^{-2}$, then the estimated $\epsilon$ values constrain the ratio of the maximum-to-minimum radii
of the $r^{-2}$ region, $r_1/r_0$.   The $\epsilon = 0.15$ for $\Tk = 20\,$K and $\nb = 2000\, H_2\unit cm^{-3}$
requires $r_1/r_0 = 2$.  The $\epsilon = 0.31$ for $\Tk = 10\,$K and $\nb = 2000\, H_2\unit cm^{-3}$ requires 
$r_1/r_0 = 16$.

In short, the likely range of $\epsilon$ is subject to a number of assumptions, but is probably $\epsilon=0.05$ 
to 0.46.  One point that should be emphasized is that the curves of Figure~\ref{fig04}, for the given densities
and temperatures, are {\it not\/} universal.  These curves are dependent on the definition of average density
(see Section~\ref{sssec471}).

\subsubsection{CO Abundance\label{ssec37}}

The effect of the CO abundance on the X-factor has been discussed in the literature 
\citep[e.g., see][]{Kutner85, Maloney88}.   One might expect that $X_f$ would depend only weakly on the 
CO abundance given that the $\COone$ line is optically thick.  However, as discussed in \citet{Maloney88}, 
a reduced CO abundance means less self-shielding from the interstellar radiation field and a smaller 
CO-emitting volume, resulting in a lower area filling factor.   The current proposed formulation would
suggest a very specific dependence on the CO abundance, $X(CO)$.  From equations~(\ref{bi20}) and 
(\ref{bi22}) in Section~\ref{ssec23}, we see that $k_\tau\propto X(CO)$.  Equation~(\ref{bi29}) tells 
us that $X_f\propto k_\tau^{-\epsilon}$.  Accordingly, $X_f\propto X(CO)^{-\epsilon}$.  Given that the 
most likely values for $\epsilon$ are between 0.05 and 0.46 (see Section~\ref{ssec36}), this is a weak
dependence on the CO abundance. 

However, a more relevant approach is to determine the dependence on the volume-averaged abundance. In 
irregular galaxies it has been found that CO is virtually absent over large volumes of molecular clouds, 
while having roughly Galactic abundance of CO within small central regions within those clouds 
\citep[see][and references therein]{Israel97a, Israel00}.   Accordingly, an {\it effective\/} CO 
abundance, $X(CO)$, is defined which is related to the Galactic CO abundance, $X_G(CO)$, within 
a spherical central region of radius, $r_{co}$, within a spherical clump of radius, $r_1$, by
\begin{equation}
X(CO) = X_G(CO)\ \left({r_{co}\over r_1}\right)^3 \qquad .
\label{xx10}
\end{equation}
It is easy to show that the optical depth on the central sightline through the clump, $\tau_0$,
is proportional to $r_{co}/r_1$.  From equations~({\ref{bi07}) and ({\ref{bi09a}) we know that,
\begin{equation}
\Tr(0)\propto {N\over\Nc}\ \tau_0^{\epsilon}\qquad .
\label{xx11}
\end{equation}
The quantity $N/\Nc$ is the number of clumps per sightline averaged over the beam.  It had been
assumed up till this point that the entire volume of each clump was emitting in the CO line.  If
we now only consider the case where the projected emitting area is less than the entire projected area
of each clump, then it is easy to show that $N/\Nc\propto (r_{co}/r_1)^2$.  Putting all of this
together yields,
\begin{eqnarray}
X_f&\propto& X(CO)^{-{1\over 3}(\epsilon + 2)} \qquad\qquad .
\label{xx12}\\
\noalign{\noindent Given that $\epsilon = 0.05$ to 0.46,}
X_f&\propto& X(CO)^{-0.7\ to\ -0.8} \qquad .
\label{xx13}
\end{eqnarray}
This is a considerably stronger dependence on the CO abundance than was derived in the previous
paragraph.  In fact, the derived dependence is nearly as strong as that expected for optically thin
$\COone$: $X_f\propto X(CO)^{-1}$.  Notice that (\ref{xx12}), in fact, has this dependence in the
optically thin case: i.e., when $\epsilon = 1$.   

Filamentary clumps may imply a different dependence of $X_f$ on $X(CO)$ from that of spherical clumps.
If the filaments are much longer than their diameters (i.e., high aspect ratio), then the dependence
of $X(CO)$ on $r_{co}/r_1$ may be two-dimensional rather than three-dimensional.  This is only an
approximation because the strong radiation fields or low metallicities that can lead to low effective
$X(CO)$ \citep[see][and references therein]{Israel97a, Israel00} would reduce the length of a filament's 
CO-emitting volume as well as its diameter.  However, for a filament with a high aspect ratio, the {\it 
fractional\/} change would be larger for the diameter than for the length.  As long as the effective 
$X(CO)$ is not extremely low, we can roughly assume that $X(CO)\propto (r_{co}/r_1)^2$.  If the 
filaments are viewed side-on, $N/\Nc\propto r_{co}/r_1$.  Using a similar approach to that for the
spherical clumps,
\begin{eqnarray}
X_f&\propto& X(CO)^{-{1\over 2}(\epsilon + 1)} \qquad\qquad .
\label{xx14}\\
\noalign{\noindent Given that $\epsilon = 0.05$ to 0.46,}
X_f&\propto& X(CO)^{-0.5\ to\ -0.7} \qquad .
\label{xx15}\\
\noalign{\noindent For end-on filaments it is simple to show that,\smallskip}
X_f&\propto& X(CO)^{-1} \qquad\qquad ,
\label{xx16}
\end{eqnarray}
regardless of the value of $\epsilon$.  Since filaments, on average, are viewed neither completely
side-on nor completely end-on, the expected power-law index would be between $-$0.5 and $-$1.0.  Again,
notice that (\ref{xx14}) gives the dependence expected in the optically thin case for $\epsilon=1$.
For (\ref{xx16}) this optically thin dependence holds regardless of the value of $\epsilon$, given that
the effective $X(CO)$ is affected only by the {\it projected\/} area of the CO-emitting volume. 

\subsubsection{Effective Optical Depth\label{ssec38}}

The proposed formulation for the X-factor assumes that the $\COone$ line is optically thin ``to the
clumps'': i.e., $\tef\ll 1$.  But this is not necessarily the case.  The panels of Figure~\ref{fig03} 
suggest that saturation starts for $N(H_2)\gsim 1$ to $2\times 10^{22}\unit cm^{-2}$.  Finding the
$\COone$ brightness that corresponds to this surface density is {\it not\/} straightforward; the
inferred $\tef$ also depends on the source function, $\Jnu$.  Another consideration is that the 
conditions for saturation that are inferred here really only apply to the Orion clouds and not 
necessarily to other clouds.  Still another point is that the peak $\Tr$ is more relevant than the
I(CO), because the latter includes the velocity width of the line.  The I(CO) of an external galaxy,
for instance, can have a very large velocity width often dominated by large-scale systematic motions, 
such as the large-scale rotation of the galaxy, rather than the smaller scale virialized velocity 
dispersions within individual clumps or clouds.

With these caveats in mind, it is at least possible to specify a rough minimum peak $\Tr$ as a {\it
necessary\/} (and certainly {\it not\/} sufficient) condition for the start of saturation.  Within
the Orion clouds, this is $\Tr\gsim 6$ to 8$\,$K, although there are still positions for which 
$\Tr\simeq 10$ to 12$\,$K and $\tef\ll 1$ still applies.  So some measure of the source function is
necessary to be certain that saturation is occurring. 

If saturation is indeed a problem for some sources (or some positions within a source), then the actual 
X-factor necessary for determining the gas surface density could be factors of roughly two or more higher
than the ``standard'' value applicable to most other sources (or to most other positions within a source).  
This saturation behavior of the X-factor can be understood better by examining either panel of Figure~\ref{fig03}.  
The X-factor of a given point on the curve is proportional to the reciprocal of the slope of a line-segment 
joining the origin to that point.  As the column density increases, the point moves to the right and the 
line-segment to the origin has an ever decreasing slope; the X-factor increases without bound.  Therefore,
in {\it this\/} optically thick limit, the $\COone$ line loses its sensitivity to column density, as expected. 

\subsubsection{Interclump Gas\label{ssec38a}}

Including the effects of a more or less continuous low-density medium between the clumps --- an
interclump gas --- in the current treatment of the X-factor is beyond the scope of the current
paper.  Nonetheless, the effects of such gas on $X_f$ can be crudely estimated.  From 
expressions~(\ref{bi23}), (\ref{bi09a}), and (\ref{bi01}), we have
\begin{equation}
I(CO) = \sqrt{2\,\pi}\,\Jnu\,{N\over\Nc}\,\Dvc\, k_A\,\tau_0^\epsilon\qquad ,
\label{xx161}
\end{equation}
which assumes $\tef\ll 1$, as used throughout most of this paper.  Of course, the beam-averaged 
column density, $N$, is a measure of the mass of CO-emitting gas within the beam.  If CO-containing 
gas is added to the beam, then $I(CO)$ increases.  How $I(CO)$ increases depends on how this gas is 
added:
\bupskip
\bupskip
\begin{enumerate}
\item{If the gas is added in the form of more clumps to each velocity interval, without changing 
the clumps' properties, then $I(CO)\propto N$ and $X_f=N/I(CO)$ remains constant.}
\bupskip
\item{If the added gas increases $\Nc$ of each clump, without changing the clumps' dimensions, then 
$N/\Nc$ remains constant and $\Dvc\propto\Nc^{0.5}$. Given that $\tau_0\propto\Nc/\Dvc$, $I(CO)\propto 
N^{0.5(1+\epsilon)}$, requiring that $X_f\propto N^{0.5(1-\epsilon)}$.} 
\bupskip
\item{If a very distended envelope is added to each clump, then $\Nc$ changes very little and $\Dvc
\propto N^{-0.5}$.  This is a dependence of $\Dvc$ on $N$ because the total mass in the beam has 
increased, which will change the velocity widths of the clumps, even though $\Nc$ has changed little.
This is an inverse dependence of $\Dvc$ on $N$ because the mass is added at large distances from each
clump center (see more discussion of this below).  Again, it results that $I(CO)\propto 
N^{0.5(1+\epsilon)}$ and $X_f\propto N^{0.5(1-\epsilon)}$.}
\bupskip
\item{There are other cases where $\Nc$ is increased only a small amount, but $\Dvc$ is 
increased by a bit more, resulting in $I(CO)\propto N^\beta$, where $\beta>1$.  (Note that if
$\tau_0 < 1$, then $\beta=1$, because $\epsilon=1$ and $\tau_0\,(\Dvc/\Nc) = 1$.)  Hence, $X_f
\propto N^{1-\beta}$ and $X_f$ decreases with increasing $N$.} 
\end{enumerate}
\bupskip
One thing to notice about all these cases is that, if the gas is optically thin in the $\COone$ line
(i.e. $\tau_0< 1$ and $\epsilon=1$), then $I(CO)\propto N$ and $X_f$ is constant.  Except for point~\#4 
above, the addition of extra CO-emitting gas will increase, or leave unchanged, the X-factor.   And 
point~\#4 does not necessarily represent a very common case.
 
A quantitative estimate of how strongly the interclump gas could affect $X_f$ can come from assuming
that such gas behaves like an envelope for each clump.  The self-potential energy of such a 
clump-envelope system would be
\begin{equation}
W = -k_W\, y_W\ {G\, M_c^2\over\ph}\qquad ,
\label{xx162}
\end{equation}
where $k_W$ and $M_c$ are the quantities that apply to the clump without the envelope and $y_W$
is the correction due to the envelope.  The internal kinetic energy is 
\begin{equation}
T = {1\over 2}\, M_c\,\Delta v_{3d}^2\ (1\ +\ r_M)\qquad ,
\label{xx163}
\end{equation}
where 
\begin{eqnarray}
r_M &\equiv& {M_e\over M_c}\qquad 
\label{xx164}\\
\noalign{\medskip\noindent is the mass ratio of the envelope to the clump without the envelope.
The $r_M$ is also\medskip}
&=& r_\rho\, r_\sigma^3\qquad ,
\label{xx165}\\
\noalign{\medskip\noindent with \medskip}
r_\rho &\equiv& {\rho_{e0}\over\rho_{c0}}
\label{xx166}
\end{eqnarray}
as the ratio of the central densities of the envelope to bare clump.  The $r_\sigma$ is the ratio of 
sizes of envelope to bare clump.   Employing the Virial theorem in its simplest form and $\Delta v_{3d}^2
=3\,\Dvc^2$ yields
\begin{equation}
\Dvc = k_v\, y_v^{0.5}\,\nb\,\Dh\qquad ,
\label{xx167}
\end{equation}
with 
\begin{equation}
y_v\equiv {y_W\over 1\, +\, r_M}\qquad .
\label{xx168}
\end{equation}
The central column density of the clump-envelope system, $\Nc$, is
\begin{equation}
{\Nc} = 2\, k_N\, y_N\,\nb\,\ph\qquad .
\label{xx169}
\end{equation}
The $k_N$ is the quantity for the bare clump and 
\begin{eqnarray}
y_N&\equiv& 1\ +\ r_\rho\, r_\sigma
\label{xx170}\\
&=& 1\ +\ r_M\, r_\sigma^{-2}
\label{xx171}
\end{eqnarray}
From expressions~(\ref{xx167}) and (\ref{xx169}) we see that the $k_v$ and $k_N$ in expression~(\ref{apb11}) 
for $X_f$ must be replaced by $k_v\, y_v^{0.5}$ and $k_N\, y_N$, respectively.  The formula for $X_f$ ends
up with an extra factor of $y_X^{1-\epsilon}$ in which
\begin{equation}
y_X\equiv y_N\, y_v^{-0.5}\qquad .
\label{xx172}
\end{equation}

We are now ready to estimate the effects of the interclump gas on the X-factor.  To effectively be the
interclump gas the envelope must extend far from the clump; i.e., $r_\sigma\to\infty$.   Obviously, this 
yields $y_W\to 1$ and $y_v\to (1\, +\, r_M)^{-1}$.  Hence the addition of the envelope to the bare clump 
actually reduces the clump velocity width (see point~\#3 two paragraphs back).  Because this limit implies
that $y_N\to 1$, we have $y_X\to (1\, +\, r_M)^{0.5}$.  It is clear then that the interclump gas has only
a small effect on $X_f$.  
If the interclump gas mass does not dominate the mass of the cloud,
then $r_M<1$ and $X_f$ would increase by less than about 40\%.  For $r_\sigma$ not large, these numbers would 
be smaller.  Also, the envelope would increase the effective $\epsilon$ and $k_A$ of the clumps, thereby further 
decreasing the effect of the interclump gas on the X-factor.  (A numerical test of this found that $\epsilon$ can 
increase by about 50\%, requiring $r_\sigma=10$ and the extreme $r_M=10$.  A less extreme $r_M=1$ produced the
usual $\epsilon$ but a 25\% increase in $k_A$.)

Knowing the size of the effect on $X_f$ for finite $r_\sigma$ requires knowing the specific density variation 
within the clump and envelope.  It is easy to show that for $r_\sigma=1$ that $y_X=(1\, +\, r_M)^{0.5}$ as it
does for $r_\sigma\to\infty$.  (This comes from $y_W=(1\, +\, r_M)^2$ when $r_\sigma = 1$.)  Therefore, for
$r_\sigma$ between 1 and $\infty$, $y_X$ must achieve a minimum or a maximum.  (And, since the envelope must
surround the clump and not vice-versa, $r_\sigma$ must be greater than unity.)  Examining the cases of a
uniform-density spherical clump and envelope and a Gaussian spherical clump and envelope shows that both these 
cases have a minimum for $y_X$ in the desired range for $r_\sigma$.  In the former case, this minimum is greater 
than unity but less than $(1\, +\, r_M)^{0.5}$.  In the latter case, the minimum of $y_X$ is less than unity 
but greater than $(1\, +\, r_M)^{-0.5}$.  Therefore, the largest deviations from unity for $y_X$ occur for
$r_\sigma$ large.  Accordingly, only for $r_\sigma$ large is the effect on the X-factor maximized; and even
in this large-$r_\sigma$ limit the effect is no more than 40\% if $r_M\le 1$.  (Note that $r_\sigma$ large 
means that $r_\sigma\gg r_M^{1\over 3}$ when $r_M<1$ and $r_\sigma\gg r_M$ when $r_M>1$ for the uniform-density 
case and $r_\sigma\gg r_M^{1\over 2}$ when $r_M<1$ and $r_\sigma\gg r_M^2$ when $r_M>1$ for the Gaussian case.)

Another point to consider is if the interclump gas has a different temperature from that of the clumps.
Since this interclump gas is at a lower density than that of the clumps and, if there is even very crude
pressure equilibrium between the clumps and interclump medium, then $\Tk(interclump)>\Tk(clump)$.  
The warmer interclump gas could then possibly have appreciable emission compared to that of the clumps
and reduce $X_f$ appreciably.  However, if there is true clump/interclump pressure equilibrium and if 
the density of the interclump gas is lower than that of the clump gas by an order of magnitude, then the
temperature would be higher by an order of magnitude.  This would bring down the optical depth of the
$\COone$ line in the interclump medium by {\it two\/} orders of magnitude and could weaken its emission 
below what it would have been if $\Tk(interclump)=\Tk(clump)$.  If so, then the interclump gas would have 
an even weaker effect on the X-factor than has been estimated above. 

Of course, the above analysis assumes that the dominant CO-emitting gas is in the clumps, so that the presence
of the interclump gas only increases $X_f$.  However, if the interclump gas were to dominate the mass of 
the cloud and its $\COone$ emission, then the appropriate average density to use in the expression for $X_f$ 
would be considerably lower, thereby lowering $X_f$. 

In summary, interclump gas would likely increase the X-factor by less than about 40\%, so long as this
gas did not dominate the mass of the cloud.  If the interclump gas did dominate the mass, then the X-factor
would be decreased, because the average density of the dominate CO-emitting gas would decrease.  One other 
important effect of the interclump medium would be to increase the effective $\epsilon$.  Estimates of 
$\epsilon$ from inferred $X_f$ values (and assumed or estimated clump densities and temperatures, see 
Section~\ref{ssec36}) or from $\cOone/\COone$ line ratios (see Sections~\ref{ssec39} and \ref{ssec44}) 
could be higher than those expected from just clumps alone.

\subsection{$\cOonebf/\COonebf$\label{ssec39}}

One implication of the proposed formulation is that there is an additional layer of complication
in interpreting line ratios.  For example, if two spectral lines have $\tef\ll 1$, then the ratio of
their intensities depends on the ratio their respective $\Jnux\,\At$ values (where $\Tx$ is the 
excitation temperature of the transition), instead of depending on the ratio of their $\Jnux\,
[1 - exp(-\tau_0)]$ values.  If the two spectral lines have similar optical depths (i.e. similar $\At$ 
values) and $\tef\ll 1$, then this complication is minimized.   Given that the $\CO$ to $\cO$ abundance 
ratio is about 60 \citep[e.g.,][]{Langer90}, the ratios of $\cO$ lines to those of 
$\CO$ still have this complication.

To appreciate this, we start with a uniform slab of gas in which the $\Jone$ lines of $\CO$ and $\cO$
are in LTE:
\begin{eqnarray}
{I({\cO})\over I(\CO)}&\simeq& {\strut 1\ -\ exp(-x_r\tau_{12})\over\strut 1\ -\ exp(-\tau_{12})} \qquad ,
\label{xx17}\\
\noalign{\noindent 
where $x_r$ is the $\cO/\CO$ abundance ratio, $X(\cO)/X(\CO)$.  $\tau_{12}$ is the optical depth
of $\COone$ and, in LTE, the optical depth of $\cOone$, $\tau_{13}$, is related to $\tau_{12}$
by $\tau_{13} = x_r\cdot\tau_{12}$.  Expression~(\ref{xx17}) is an approximation because the
integrated intensities are used here instead of the radiation temperatures at a particular velocity
within the line profile.  (It is also an approximation because the temperature corrected for the
cosmic background and for failure of the R-J approximation, $\Jnu$, is slightly different for the
the two $\Jone$ lines.)  This line ratio has the following limiting cases:\medskip}
{I({\cO})\over I(\CO)}\simeq&\rlap{$x_r$}
\phantom{1\ -\ exp(-x_r\tau_{12})}
&,\ for\ \tau_{12}\ll 1\ and\ \tau_{13}\ll 1,
\label{xx17a}\\
\simeq&1\ -\ exp(-x_r\tau_{12})&,\ for\ \tau_{12}\gg 1,
\label{xx18}\\
\simeq& \rlap{$x_r\tau_{12}$} 
\phantom{1\ -\ exp(-x_r\tau_{12})}
&,\ for\ \tau_{12}\gg 1\ and\ \tau_{13}\ll 1,
\label{xx19}\\
\simeq& \rlap{$1$}
\phantom{1\ -\ exp(-x_r\tau_{12})}
&,\ for\ \tau_{12}\gg 1\ and\ \tau_{13}\gg 1 .
\label{xx20}
\end{eqnarray}
Equation~(\ref{xx17a}) represents the uncommon case of optically thin $\COone$ and has the
unsurprising result that the line ratio is the abundance ratio when both lines are optically
thin.  Equation~(\ref{xx19}) shows a linear relationship between the line ratio and the 
abundance ratio.  When both lines are optically thick, as represented by equation~(\ref{xx20}), 
the line ratio is completely insensitive to the abundance ratio.  

If we now consider a clumpy medium and use the MSH84 formalism, then
\begin{equation}
{I({\cO})\over I(\CO)}\simeq {1\ -\ exp(-a\twef)\over 1\ -\ exp(-\twef)} \qquad ,
\label{xx21}
\end{equation}
where $\twef$ is the effective optical depth of the $\COone$ line and where,
\begin{equation}
a\equiv {\thef\over\twef} = {\Ath\over\Atw} \qquad .
\label{xx22}
\end{equation} 
Here it is implicitly assumed that $\At$ has the same functional form for both $\COone$ and
$\cOone$.  This is not necessarily the case given that $\cO$ can be selectively photodissociated
more easily than $\CO$ \citep{Warin96}.  Nevertheless, for simplicity, the same $\At$ for both of the 
$\Jone$ lines will be assumed here.  Even for the same $\At$, the $k_A$ and $\epsilon$ values can
be different if $\cOone$ and $\COone$ are in different optical depth regimes (i.e. $\tau_0\lsim 1$
versus $\tau_0\gsim 3$):
\begin{equation}
a = {k_{A,13}\ \tau_{0,13}^{\epsilon_{13}}\over k_{A,12}\ \tau_{0,12}^{\epsilon_{12}}}\qquad .
\label{xx23}
\end{equation} 
When $\cOone$ is optically thin in the clump (i.e. $\tau_{0,13}\ll 1$), $k_{A,13} = 1$ and 
$\epsilon_{13} = 1$.   If, at the same time, $\COone$ is optically thick in the clump (i.e. 
$\tau_{0,12}\gsim 3$), then it is easy to see that
\begin{equation}
a = x_r\ k_A^{-1}\ \tau_0^{1-\epsilon}\qquad ,
\label{xx24}
\end{equation} 
where the subscript ``12'' was omitted from the $k_A$ and $\epsilon$.  In the other limit, when
$\cOone$ is optically thick in the clump, $k_{A,13} = k_{A,12}$ and $\epsilon_{13} = \epsilon_{12}$
and
\begin{equation}
a = x_r^\epsilon \qquad ,
\label{xx25}
\end{equation}
where $\epsilon$ is both $\epsilon_{13}$ and $\epsilon_{12}$.  With these expressions in mind, the
following limiting cases result:
\begin{eqnarray}
{I({\cO})\over I(\CO)}\simeq&\rlap{$x_r$}
\phantom{x_r^\epsilon\, k_A\,\tau_{0,12}^\epsilon\ {N\over\Nc}\ {\Dvc\over \Dv}}
&,\ for\ \twef\ll 1,\ \thef\ll 1,\ \tau_{0,12}\lsim 1,\ and\ \tau_{0,13}\lsim 1,
\label{xx25a}\\
\simeq&\rlap{$x_r\, k_A^{-1}\, \tau_{0,12}^{(1-\epsilon)}$}
\phantom{x_r^\epsilon\, k_A\,\tau_{0,12}^\epsilon\ {N\over\Nc}\ {\Dvc\over \Dv}}
&,\ for\ \twef\ll 1,\ \thef\ll 1,\ \tau_{0,12}\gsim 3,\ and\ \tau_{0,13}\lsim 1,
\label{xx26}\\
\simeq&\rlap{$x_r^\epsilon$}
\phantom{x_r^\epsilon\, k_A\,\tau_{0,12}^\epsilon\ {N\over\Nc}\ {\Dvc\over \Dv}}
&,\ for\ \twef\ll 1,\ \thef\ll 1,\ and\ \tau_{0,13}\gsim 3,
\label{xx27}\\
\simeq&\rlap{$x_r\,\tau_{0,12}\ {N\over\Nc}\ {\Dvc\over \Dv}$}
\phantom{x_r^\epsilon\, k_A\,\tau_{0,12}^\epsilon\ {N\over\Nc}\ {\Dvc\over \Dv}}
&,\ for\ \twef\gg 1,\ \thef\ll 1,\ and\ \tau_{0,13}\lsim 1,
\label{xx28}\\
\simeq&x_r^\epsilon\, k_A\,\tau_{0,12}^\epsilon\ {N\over\Nc}\ {\Dvc\over \Dv}
&,\ for\ \twef\gg 1,\ \thef\ll 1,\ and\ \tau_{0,13}\gsim 3,
\label{xx29}\\
\simeq& \rlap{$1$}
\phantom{x_r^\epsilon\, k_A\,\tau_{0,12}^\epsilon\ {N\over\Nc}\ {\Dvc\over \Dv}}
&,\ for\ \twef\gg 1\ and\ \thef\gg 1 .
\label{xx30}
\end{eqnarray}
The $k_A$ and $\epsilon$ are those for $\tau_0\gsim 3$.  Only in the last case is the line ratio 
completely independent of the abundance ratio.  When $\tau_{0,13}\gsim 3$, $I(\cO)/I(\CO)\propto 
x_r^\epsilon$ which is the same proportionality that the X-factor has in the case when the X(CO) 
is not spatially varying (see first paragraph of Section~\ref{ssec38}).  This represents a very 
important contrast with the case of uniform gas: {\it even when both $\COoneit$\/} and {\it 
$\cOoneit$ are optically thick on the central sightlines of the clumps, the observed 
$\cOoneit/\COoneit$ line ratio does\/} not {\it reach an asymptotic value of unity,\/} unless 
$\thef\gg 1$.  And when both $\twef\ll 1$ and $\thef\ll 1$ for $\tau_{0,13}\gsim 3$, the line ratio 
saturates near an asymptotic value that depends on the abundance ratio and the clump fluffiness. 

Expression~(\ref{xx21}) and its limiting forms, (\ref{xx25a}) to (\ref{xx30}) inclusive, can be
applied to the $\cOone$ and $\COone$ data of the Orion$\,$A and B molecular clouds \citep{W06}.
The line ratio $I(\cO)/I(\CO)$ is plotted against $\twef$ in the panels of Figure~\ref{fig05}.
The $\twef$ was found from solving equation~(\ref{bi04}) for $\tef$.  The $\Jnu$ was determined
from the observed $\Ia/\Ib$ ratio and the one- and two-component models described in \citet{W06}.
The uncertainties in $\tef$, $\sigma(\tef)$, were estimated from  
\begin{equation}
\sigma(\tef) = {\sigma(\eta_{ff})\over 1\ -\ \eta_{ff}} \qquad ,
\label{xx31}
\end{equation}
where $\eta_{ff}\equiv\Tr/\Jnu$.  Any attempt to model the points in Figure~\ref{fig05} must
consider the large scatter.  The points with $\tef\lsim 0.3$ and $I(\cO)/I(\CO)\gsim 0.3$, for
instance, have large vertical scatter due to their large vertical error bars (which are really
twice as large as shown, see figure caption).   Even if the vertical scatter for these points
is not entirely real, the high $I(\cO)/I(\CO)$ ratios still seem to be likely --- i.e., 
$I(\cO)/I(\CO)\sim 0.6$.  This combination of high $I(\cO)/I(\CO)$ and low $\tef$ is difficult 
to explain, and is best represented by expression~(\ref{xx27}).  For $x_r = 1/60$, a line ratio 
of $I(\cO)/I(\CO) = 0.6$ requires $\epsilon=0.12$, roughly consistent with clumps that are hard 
spheres.  Hence these points could be explained by gas with a low filling factor (and, accordingly, 
low $\tef$) and with clumps approximating hard spheres. 

On the other hand, the vertical scatter is probably real in the quasi-horizontal band of points
that extends from $\tef\simeq 0$ to $\tef > 1$ for $I(\cO)/I(\CO)\lsim 0.3$ to 0.5.  Accordingly,
no {\it single\/} theoretical curve can adequately model this band; models that have two curves
each are plotted in the panels of Figure~\ref{fig05}.   Before the two-curve models are discussed, 
the one-curve models are described.

\subsubsection{Crude Modeling of $I(\cO)/I(\CO)$ versus $\twef$\label{sssec331}}

The data points in Figure~\ref{fig05} are those positions in the Orion$\,$A and B molecular clouds
for which the peak $\Tr(\COone) > 3\,\sigma$, and $\Ia$, $\Ib$, and $\Ic$ are all $> 5\,\sigma$ 
for a total of 372 points \citep[see][]{W06}.  One set of $\twef$ values were derived from one-component
models, hereafter $\twef^{(1)}$, and another from two-component models, hereafter $\twef^{(2)}$.  These 
models were necessary for estimating the source function, $\Jnu$, from the 140$\um$ and 240$\um$ DIRBE 
data \citep[see details in][]{W06}.  While both panels of Figure~\ref{fig05} show roughly the same 
basic trend of $I({\cO})/I(\CO)$ with $\twef$, the models fit the $I({\cO})/I(\CO)$ versus $\twef^{(1)}$
plot much better (i.e., upper panel) than the $I({\cO})/I(\CO)$ versus $\twef^{(2)}$ (i.e., lower panel). 
This is due to the unrealistically small errors in $\twef^{(2)}$.  But these errors are only the formal
errors.  While the two-component models for the Orion clouds are better than the one-component models in 
many ways, they do not seem to adequately estimate the source function, $\Jnu$.  This source function 
depends on a model parameter, $c_0$, that is not well constrained (unless combined with another parameter).
Therefore, when systematic effects are considered, the effective errors in $\twef^{(2)}$ would be much
larger.   This must be borne in mind when considering the models described below. 

One of the simplest models to consider is that where each model curve is characterized only by a 
value of $a$, where $a$ is defined in expression~(\ref{xx22}).   This kind of model represents a 
picture in which the clump properties are constant (i.e. fixed $\tau_0$, $\Dvc$, $\Nc$) and $\twef$ 
and $\thef$ increase only because the average number of clumps per sightline within each clump 
velocity width is increasing (i.e., increasing ${N\over\Nc}{\Dvc\over\Dv}$), thereby increasing 
the beam-averaged column density $N$.  In the limit where $\twef\to 0$ and $a\twef\to 0$, 
(\ref{xx21}) becomes
\begin{equation}
{I({\cO})\over I(\CO)}\to a\ +\ 0.5\,a\,(1-a)\,\twef
\label{xx32}
\end{equation}
If $a$ is constant, then the line ratio would have a linear dependence on $\twef$ with slope
$0.5a(1-a)$ and intercept $a$.  It turns out that (\ref{xx32}) is very good approximation 
{\it even when\/} the limit $\twef\to 0$ is {\it not\/} valid.  For example, (\ref{xx32})
yields a result that is within 2\% of that of expression~(\ref{xx21}) even for $\twef=3$, when 
$a=0.3$.  A sample of clumpy gas with identical clumps, but varying $\twef$ would have a clear 
signature in a diagram like Figure~\ref{fig05}: a straight line extending to the right from the 
intercept on the vertical axis with a positive slope, and there would be a well-defined 
relationship between the slope and the intercept.  Figure~\ref{fig05} shows that, for the Orion
clouds at least, such a simple signature is not obvious.  

Nevertheless, models with one curve were attempted using the orthogonal regression method 
described in \citet{W06}.  Models with two curves were also tried.  {\it Neither the one-curve
nor the two-curve models were successful.\/} The two-curve models produced fits with reduced 
chi-square values of $\chi_\nu^2= 13.9$ and 24.4 for the $\twef^{(1)}$ and the $\twef^{(2)}$, 
respectively.  These are factors of roughly 3 better than for the corresponding one-curve models.   
The recovered $a$ values for the $\twef^{(1)}$ are 0.13 and 0.24 from the two-curve models.  
These are the same for the $\twef^{(2)}$.

Models that are slightly more complicated are those that assume constant $\Atw/\twef$.  The model 
curves here are characterized by two parameters: the $\Atw/\twef$ ratio, $b$, and the clump fluffiness,
$\epsilon$.  This kind of model represents a picture in which the average number of clumps per 
sightline within each clump velocity width is constant (i.e., constant ${N\over\Nc}{\Dvc\over\Dv}$), 
but $\tau_{0,12}$ and $\tau_{0,13}$ increase, thereby increasing $\twef$ and $\thef$.  Combining
equation~(\ref{bi09a}) with the constant $b$ yields,
\begin{equation}
\tau_{0,12} = b^{1/\epsilon}\ k_A^{-1/\epsilon}\ \twef^{1/\epsilon} \qquad ,
\label{xx33}
\end{equation}
where $k_A$ and $\epsilon$ are those for $\tau_{0,12}\gg 3$.  Substituting (\ref{xx24}) into 
(\ref{xx33}) gives
\begin{eqnarray}
a &=& b_\tau\,\twef^{(1-\epsilon)/\epsilon} \qquad ,
\label{xx34}\\
\noalign{\noindent where\medskip}
b_\tau &\equiv& x_r\ k_A^{-1/\epsilon}\ b^{1-\epsilon\over\epsilon}\qquad .
\label{xx35}
\end{eqnarray}
Substituting (\ref{xx34}) into (\ref{xx21}) results in
\begin{eqnarray}
{I({\cO})\over I(\CO)}&\simeq& {1\ -\ exp(-b_\tau\twef^{1/\epsilon})\over 1\ -\ exp(-\twef)} \qquad .
\label{xx36}\\
\noalign{\noindent The above expression results in $\thef$ growing very rapidly for small $\epsilon$.
This growth cannot continue unchecked and must slow when $a$ reaches its saturation value, given by
expression~(\ref{xx25}).  Consequently, in this limit,\medskip}
{I({\cO})\over I(\CO)}&\simeq& {1\ -\ exp(-x_r^\epsilon\twef)\over 1\ -\ exp(-\twef)} \qquad .
\label{xx37}
\end{eqnarray}
Each model curve combines (\ref{xx36}) and (\ref{xx37}): the minimum of the two $I({\cO})/I(\CO)$ 
values for each $\twef$ is used.  As a result, the $I({\cO})/I(\CO)$ versus $\twef$ curves are from 
expression~(\ref{xx36}) for lower $\twef$ and expression~(\ref{xx37}) for higher $\twef$.  This 
produces an abrupt transition at some intermediate $\twef$ value (see curves in Figure~\ref{fig05}).
This abrupt transition is not real and only appears in these model curves because the smooth 
transition of $\At$ from optically thin to optically thick behavior (i.e., for $1\lsim\tau_0\lsim 3$)
has not been considered here.  Nevertheless, these intermediate $\tau_0$ values represent only a 
small portion of the curves.  Accordingly, the abrupt transitions in the curves only introduce small 
inaccuracies into the final results. 

Models with one curve and models with two curves were fitted, each fit optimizing $b_\tau$ and $\epsilon$. 
The best fits of the one-curve models had reduced chi-square values that were factors of 2.1 to 2.5 worse 
than those of the two-curve models.  The data points in Figure~\ref{fig05} are not entirely independent, 
so the effective number of degrees of freedom is about a factor of 9 lower than the number of points 
\citep[see discussion of this in][]{W06}, or about 40.  Therefore, the F-test states that the two-curve 
models are better than the one-curve models at a 97.5\% confidence level.  The resultant parameter and 
reduced chi-square values are given in Figure~\ref{fig05}.   Even if the two-curve models are superior 
to the one-curve models, the fits of the former to the data are not impressive.  Nonetheless, the fit to 
the $\twef^{(1)}$ is not too bad ($\chi_\nu^2\simeq 4$) considering the vertical spread in the points, even 
in the quasi-horizontal band.   The fit to the $\twef^{(2)}$ is considerably worse, but then this is due to 
the unrealistically small horizontal error bars.  These two-curve models (where each curve uses two 
parameters) are better than the previous simpler two-curve models (where each curve uses one parameter) at 
the 95\% confidence level or better, according to the F-test.  For $\twef$ beyond some threshold, the 
constant $a$ models (with one parameter per curve) superficially resemble the constant $\Atw/\twef$ models 
(with two parameters per curve), because the latter have constant $a$ (i.e., $a=x_r^\epsilon$) in this 
$\twef$ range.  Consequently, the latter being significantly better than the former is because the latter 
models have a rapid non-linear rise at smaller $\twef$ where the majority of the points lie.   
 
The systematic uncertainties in the resultant parameter values were found by adopting systematic errors
of 30\% for $I({\cO})/I(\CO)$, $\eta_{ff}$, and $x_r$.  Specifically, 0.7 and 1.3 were each multiplied by
one of these quantities and the models were refit, a total of 6 refits.  The minimum and maximum $b_\tau$
and $\epsilon$ were chosen from these refits as measures of the uncertainties.  The uncertainty in $b_\tau$
is about a factor of 2 and for $\epsilon$ about 30\%.  Expression~(\ref{xx35}) and $\epsilon\sim 0.3$ imply
that the systematic uncertainty in $b$ is roughly a factor of $\sqrt{2}$ or about 40\%.  Given that $1/b$ 
is in reality ${N\over\Nc}{\Dvc\over\Dv}$, this latter also has a systematic uncertainty of about 40\%. 

With these systematic uncertainties in mind, the $\epsilon$ values given in Figure~\ref{fig05} for the
upper panel are consistent with those in the lower panel.  The $b_\tau$ values given in Figure~\ref{fig05}
are significantly different between the two panels.  Nonetheless, the changes in $\epsilon$ from one panel 
to the other conspire with the $b_\tau$ values to produce $b$ values that are much closer together (in a 
logarithmic sense).  Solving (\ref{xx35}) for $b$, adopting $k_A=1.5$, and using the $b_\tau$ values in 
Figure~\ref{fig05} results in
\begin{eqnarray}
{N\over\Nc}{\Dvc\over\Dv} =& 1\times 10^{-2}\ and\ \ 6\times 10^{-2} &,\ for\ \twef^{(1)},
\label{xx38}\\
                          =& 2\times 10^{-2}\ and\ \ 8\times 10^{-2} &,\ for\ \twef^{(2)}.
\label{xx39}
\end{eqnarray}
Considering that the effective spatial resolution of the observations was slightly less than 1$^\circ$
(or a linear resolution of 8$\,$pc for the adopted distance of 450$\,$pc), these numbers are reasonable.  
The fluffiness values are more succintly summarized as roughly the following: $\epsilon\simeq 0.3\pm 0.1$.  
This range is consistent with that found in Section~\ref{ssec36}.
 
One concern raised by the modeling here is that many positions seem to have non-negligible optical depths 
in the $\cOone$ line.  The modeling of physical conditions by \citet{W06} required $\cOone$ to be optically 
thin.  Examination of the upper panel of Figure~\ref{fig05} suggests that the majority of points are not 
towards optically thick sightlines in $\cOone$.  Most of the points are associated with the curve with the 
lower $b_\tau$ value and, hence, lower $\Atw$ values.  The majority of these points are at lower $\twef$ 
than occurs for the abrupt transition at $\twef = 0.34$.   It is easy to compute that $\tau_{0,13}\simeq 2$ 
at this transition.  Accordingly, the majority of points have $\tau_{0,13}<2$.  Given that $\tau_{0,12}
\propto\twef^{1/\epsilon}$ (cf. equation~\ref{xx33}), the $\tau_{0,13}$ is very strong function of $\twef$ 
and drops rapidly with $\twef$ decreasing below the transition point.  Therefore, the majority of points do 
not represent sightlines optically thick in $\cOone$, although some important minority of points may indeed 
represent such sightlines.  

In summary, the modeling of the $I(\cO)/I(\CO)$ versus $\twef$ diagram did not produce exceptionally
good fits, but does strongly favor one scenario over another.  Specifically, it favors the case of
growing clump optical depth as the column density increases from position to position, rather than the
case of a growing number of clumps per sightline.  The model fits suggest that the average number of 
clumps per sightline per clump velocity width is a $few\times 10^{-2}$ and the fluffiness is $0.3\pm 0.1$,
consistent with that expected for the observed $X_f$ value (see Section~\ref{ssec36}).

\section{Discussion\label{sec4}}

The Orion clouds have been used as a test bed for the ideas developed here.   More general implications are 
examined in this section.

\subsection{Comparisons with Previous Work\label{ssec41}}

Much of the previous work on the X-factor was based in one way or another on DSS86.  Two examples of 
this are \citet{Sakamoto96} and \citet{Weiss01}, which examined the relationship between $X_f$ and 
physical parameters --- i.e., $\nH$ and $\Tk$.   The former used LVG models to explore the behaviour
the X-factor under non-LTE conditions.  The latter used observational data to estimate the dependence
of $X_f$ on density and temperature.  

\citet{Sakamoto96} has the advantage that it deals with non-LTE conditions, but has the disadvantage 
that it does not consider radiative transfer through a clumpy cloud.  Obviously, the advantage and 
disadvantage of the current paper complement those of \citet{Sakamoto96}.   Accordingly, a treatment 
that combines non-LTE conditions with radiative transfer in a clumpy medium would be desirable, but is 
beyond the scope of the current work. 

Any attempt to at least roughly estimate the properties of such a combined treatment must compensate for 
some minor errors found in \citet{Sakamoto96}. 
\bupskip
\begin{itemize}
\bupskip
\item[$\bullet$] The expression for the virial mass, equation~(5) of that paper, is missing a factor of 
$1\over2$.  This can be verified for the uniform density case by comparing with the derivation given in 
Section~\ref{ssec22} of the current paper.  This can also be verified for the $\rho(r)\propto r^{-1}$ 
case by comparing with the expressions given in \citet{Maloney90}.  Both of these comparisons confirm 
the absence of the factor $1\over 2$ in equation~(5) of \citet{Sakamoto96}.  So equations using the 
parameter $k_3$ must use $k_3\over 2$ in place of $k_3$.  Consequently, the numerical values of $X_f$
computed from equation~(7) of that paper must be corrected downward by a factor of $\sqrt{2}$.
\bupskip
\item[$\bullet$] The claim that $X_f$ is independent of $\Tk$ in the optically thin (i.e. $\tau(\COone)
\ll 1$), thermalized limit is incorrect.  This claim and its explanation are given in point~\#2 (which 
contains expression~11) of \citet{Sakamoto96}.  Section~\ref{ssec32} shows that such a claim is false.  
A very simple physical argument makes it obvious that $X_f$ {\it must\/} depend on $\Tk$ in the
optically thin, LTE limit: as $\Tk$ increases in the high-$\Tk$, LTE limit the CO molecules populate 
more and more of the upper-$J$ levels, thereby decreasing the column density of CO in the $J=1$ level.  
If the $\Jone$ line is optically thin, then $I(CO)$ decreases for fixed total $N(H_2)$, resulting in
$X_f$ increasing with increasing $\Tk$ (cf. Section~\ref{ssec32}).  The panels of Figure~3 of
\citet{Sakamoto96} do indeed show that $X_f$ varies by less than factor of about 2 (from smallest to
largest) over the range $\Tk = 10$ to $100\unit K$ for CO abundances less than about $10^{-5}$. 
However, contrary to point~\#2 of \citet{Sakamoto96}, even $\nH$ as high as $10^4\unit cm^{-3}$ does 
{\it not\/} thermalize the $\COone$ line when it is optically thin --- LTE still does {\it not\/}
apply.  Given that the critical density of the $\COone$ line is $\sim 3\times 10^3\ H_2\unit cm^{-3}$,
it is natural to assume that densities corresponding to the lower two panels of Figure~3 of that paper
--- i.e., $3\times 10^3\ H_2\unit cm^{-3}$ and $1\times 10^4\ H_2\unit cm^{-3}$ --- are sufficient to
bring the $\Jone$ line of CO close to LTE.  The problem is that critical densities are normally 
computed as though a given atom or molecule is just a two-level system.  For the CO molecule, there
are many rotational levels {\it and\/} the radiative decay rates increase very rapidly --- roughly
like $J^3$ --- as one moves up the rotational ladder.  When the CO~$\Jone$ line is optically thin,
radiative trapping effects are minimized and the CO molecules pile up mostly in the $J=0$ and $J=1$
states, often resulting in superthermal $J=1/J=0$ population ratios (i.e., the excitation temperature
of the $\Jone$ transition is greater than $\Tk$) or even population inversions, because of the higher
levels with large radiative decay rates.   Therefore, non-LTE effects play a role in the behavior of
the $\Jone$ line even at densities around $3\times 10^3\ H_2\unit cm^{-3}$ (provided that the $\Jone$ 
line is optically thin).  For the specific cases of the densities considered in Figure~3 of 
\citet{Sakamoto96} and for CO abundances $\lsim 10^{-5}$, the increasing $\Tk$ populates the $J=1$ level 
from the $J=0$ level at an increasing rate, but also {\it de}populates the $J=1$ level to the higher-$J$ 
levels at an increasing and roughly comparable rate.  Accordingly, the column density of CO molecules in the 
$J=1$ level stays roughly constant.  Given that the CO~$\Jone$ line is optically thin for these abundances,
the observed line brightness is directly proportional to the column density in the $J=1$ (and {\it not\/}
the $J=0$) state, resulting in roughly constant line brightness and, ergo, roughly constant $X_f$.
Therefore, expression~(11) of \citet{Sakamoto96}, which states that $X_f\propto X(CO)^{-1}$ with no 
dependence on $\Tk$ nor $\nH$, is approximately correct.   And again, this is only in this particular 
{\it non}-LTE case.   In the LTE limit when CO~$\Jone$ is optically thin, the dependence of $X_f$ on $\Tk$ 
is as given in Section~\ref{ssec32}.
\bupskip
\item[$\bullet$] The statement that the observed spectral profiles of the CO~$\Jone$ line place an upper 
limit on its opacity is erroneous, and is at odds with the clumpy cloud picture presented in Figure~1 of
\citet{Sakamoto96}.  That paper states that the $\COone$ line cannot be very optically thick or its
profiles would be flat-topped.  For a given volume of gas with more or less uniform density, that would 
be true; increasing the optical depth would saturate the line core before it would do so in the line wings.  
But this is {\it not\/} necessarily the case in a clumpy cloud.  The optical depth of the $\COone$ line
within each clump can increase without bound and, as long as the filling factor within each velocity
interval within the line profile is sufficiently small, the observed line profile can still be Gaussian 
(e.g., see equation~\ref{bi07} of the current paper).   This point is important for avoiding unrealistically 
low upper limits on the optical depth of the $\COone$ line.
\end{itemize}
\bupskip 
With these caveats in mind, a relatively {\it simple} treatment of clumpy clouds that includes both 
radiative transfer and non-LTE effects can be developed.  This will be particularly useful for developing 
insights into the emission of high-dipole molecules like CS, whose transitions are farther from LTE than 
those of CO. 

In the meantime, a rough estimate of the behavior of $X_f$ in the non-LTE limit is 
possible from combining the results of \citet{Sakamoto96} with the current work.  For example, in the
high-density, metal-rich case \citet{Sakamoto96} finds that $X_f\propto\nH^{1/2}\ \Tk^{-1}$.  In the
current paper, this can correspond to any of the cases in Sections~\ref{ssec33}, \ref{ssec34}, or
\ref{ssec35}.   All of these cases show a weaker dependence of $X_f$ on $\nH$ and on $\Tk$.  The 
side-on Gaussian cylinder, for instance, requires roughly the following: $X_f\propto C_T\,\Tk^{-0.6}\,
\nH^{0.4}$.   Therefore, as has been noted previously, the radiative transfer through a clumpy medium 
has the effect of softening the dependence of the X-factor on density and kinetic temperature.  Two 
other cases are discussed in that paper.   The case of intermediate density (i.e. $\sim 3\times 10^3
\ H_2\unit cm^{-3}$) and low metallicity shows no dependence of $X_f$ on $\nH$ nor $\Tk$.  Including
radiative transfer through a clumpy medium is unlikely to change this.  The case of low density 
(i.e. $\lsim 3\times 10^2\ H_2\unit cm^{-3}$) and low metallicity has the following behavior: 
$X_f\propto\nH^{-1}\ \Tk^{-1/2}\ X(CO)^{-1}$.   Combining this with radiative transfer in clump medium
will yield $X_f\propto\nH^{-\delta_1}\ \Tk^{-\delta_2}\ X(CO)^{-1}$, where $\delta_1 < 1$ and $\delta_2
< {1\over 2}$.  Very crude guesses would be that $\delta_1\sim 0.6$ and $\delta_2\sim 0.4$.  Something 
else worth considering is that in low-metallicity clouds, such as those in irregular galaxies, the CO 
abundance {\it within the CO-emitting region within each cloud} is close that in our Galaxy \citep[see]
[and references therein]{Israel97a, Israel00}.   Consequently, the behavior of the X-factor in such 
clouds may be closer to that described in Sections~\ref{sec3} and \ref{ssec37} than assuming optically
thin, non-LTE $\COone$ emission. 

\citet{Weiss01} observationally tested the dependence of the X-factor on density and temperature. They 
observed the central region of the galaxy M$\,$82 in the lower-$J$ lines of $\CO$, $\cO$, and $\Co$
to constrain estimates of the molecular gas density and temperature.  Using their estimates of $\NH$,
$\nH$, and $\Tk$, they plotted $\NH/I(CO)$ versus $\nH^{1/2}\ \Tk^{-1}$ for the observed positions and 
found a roughly linear relationship (see their Figure~11).  Determining the densities and kinetic
temperatures from line ratios with sufficient accuracy is fraught with difficulties;  \citet{Weiss01}
do not include the error bars in this plot and it is easy to show that the total horizontal uncertainty
of each plotted point (i.e. from smallest to largest value for each point) is roughly 1/3 the horizontal 
extent of the entire figure.  Accordingly, a large range of dependences on temperature and density are
{\it not\/} ruled out.  In general, such observational tests of the behavior of the X-factor are very
uncertain.



\subsection{The Dependence of ${\bf X_f}$ on CO Abundance\label{ssec42}}

The molecular gas in low-metallicity irregular galaxies has very low CO abundance over large volumes and 
has roughly the Galactic abundance level in small internal regions within the molecular clouds \citep[see]
[and references therein]{Israel97a, Israel00}.  Based on this picture, Section~\ref{ssec37} discussed the 
dependence of the X-factor on the CO abundance and found that $X_f\propto X(CO)^{-\theta}$, where $\theta$ 
can be anywhere from 0.5 to 1.0.  \citet{Israel00} examined the dependence of $X_f$ on the metallicity in
a sample of Magellanic irregular galaxies and had the result that $X_f\propto X(O)^{-2.5}$.  If the CO
abundance is largely determined by the O abundance, then $X_f\propto X(CO)^{-2.5}$.  If, however, the
CO abundance is largely determined by the C abundance and, given that $X(C)\propto X(O)^{1.7}$ 
\citep[][and references therein]{Israel00}, $X_f\propto X(CO)^{-1.5}$.  Instead, the CO abundance
might be strongly affected by the photodissociation rate and would be inversely proportional to that rate.
Given that the photodissociation rate is proportional to $X(O)^{-3}$ \citep[][and references therein]
{Israel97a}, this results in $X_f\propto X(CO)^{-0.8}$.  Hence, there are three possibilities:
\begin{eqnarray}
X_f &\propto& X(CO)^{-2.5}\qquad ,
\label{d1}\\
\noalign{\noindent or}
X_f &\propto& X(CO)^{-1.7}\qquad ,
\label{d2}\\
\noalign{\noindent or}
X_f &\propto& X(CO)^{-0.8}\qquad . 
\label{d3}
\end{eqnarray}
Expressions~(\ref{d1}) and (\ref{d2}) represent dependences that are steeper than one would expect 
even in the optically thin case (i.e., $X_f\propto X(CO)^{-1.0}$).   Having such a steep dependence
would require that other physical parameters like density and kinetic temperature also depend on
the CO abundance.  When the volume-averaged $X(CO)$ is small, the interstellar radiation field that
reaches the CO-containing gas is strong, resulting in high gas temperatures.   This would produce a 
smaller X-factor at small $X(CO)$ (based on the expressions of Section~\ref{sec3}) and the dependence 
of $X_f$ on $X(CO)$ would be less steep.  On the other hand, the strong interstellar radiation field 
would preferentially photodissociate the low-density molecular gas, leaving the high-density gas.  The 
low X(CO) would be associated with higher densities and higher densities mean higher $X_f$ (see 
Section~\ref{sec3}).  Higher $X_f$ at low $X(CO)$ means a steeper dependence of $X_f$ on $X(CO)$.  
Consequently, the effects of the higher temperature would compete against the effects of higher density, 
making it uncertain if such steep relations are possible or even likely.   The most likely dependence 
is that of expression~(\ref{d3}).  And this is consistent with the range found in Section~\ref{ssec37}.

\subsection{The Relationship between ${\bf X_f}$ and $\cOonebf/\COonebf$\label{ssec43}}

That a relationship exists between the X-factor and the $\cOone/\COone$ line ratio is not new and 
has been used to determine whether the X-factor varies spatially.  For example, \citet{Rickard85} and
\citet{Paglione01} observed the $\Jone$ lines of $\CO$ and $\cO$ in external galaxies and found that
the $I(\cO)/I(\CO)$ ratio was significantly different in the nuclear regions from that in the disks.   
This is evidence that the X-factor in galactic nuclei is also different from that in the disks.   Based 
on the current work, how do changes in $I(\cO)/I(\CO)$ relate to changes in $X_f$?  This can be assessed
by using an equation analogous to (\ref{xx6}) for $\cO$.  Equation~(\ref{xx6}) assumes that $\COone$ is 
optically thin (i.e., $\tau_{0,12}\lsim 1$).   Since $\cOone$ is likely to be optically thin on the scales
of many parsecs, the appropriate expression is  
\begin{equation}
X_f = C_T\,\,k_\tau^{-1}\,\hbox{$\Tk^{\gamma -1}$}\ x_r^{-1}\ {I({\cO})\over I({\CO})}\quad .
\label{d4}
\end{equation}
The $X_f$ and $k_\tau$ are those for $\CO$ and not for $\cO$; accordingly the $I({\cO})/I({\CO})$ and 
$x_r^{-1}$ are necessary to correct for those, respectively.  Notice that when $\COone$ is optically thin 
as well ($\tau_{0,12}\lsim 1$), the line ratio is $x_r$ (see equation~\ref{xx25a}) and equation~(\ref{d4}) 
reduces to (\ref{xx6}), as required.  When $\tau_{0,12} > 3$, equation~(\ref{xx26}) for $I({\cO})/I({\CO})$ 
applies; when this is substituted into expression~(\ref{d4}), equation~(\ref{bi29}) for $X_f$ is recovered, 
also as required.  Therefore, the X-factor is proportional to the $\cOone/\COone$ line ratio.

However, the situation is not quite that simple.  Given that the X-factor is also proportional to
$\Tk^{\gamma -1}\ x_r^{-1}$, determining spatial variations in $X_f$ from one observed position to another 
can be obscured by unknown spatial variations in $\Tk$ and $x_r$.  For example, $X_f$ may be varying spatially 
while $I({\cO})/I({\CO})$ remains constant and vice versa.   One way to estimate the dependence between $X_f$ 
and $I({\cO})/I({\CO})$ when there are also spatial variations in $\Tk$ is to assume some kind of correlation 
between the $\Tk$ and $\nH$.  For example, ${\Tk}\propto{\nH}^w$.  One of the limiting forms of 
equation~(\ref{xx21}) in Section~\ref{ssec39} --- i.e., equations~(\ref{xx25a}) to (\ref{xx29}) --- is useful 
here: the most appropriate of these for the bulk of molecular clouds is equation~(\ref{xx26}).   Using that 
expression along with that of (\ref{bi22x}) and substituting into (\ref{d4}) results in the following possible 
dependence: 
\begin{equation}
X_f\propto \left[{I({\cO})\over I({\CO})}\right]^u\qquad\phantom{for\ {\Tk}= 10\ to\ 20\, K},
\label{d5}
\end{equation}
where
\begin{equation}
u\equiv {{1\over2}(1-\epsilon)\ +\ w(\gamma\epsilon-1.32)\over ({1\over2}-\gamma w)
(1-\epsilon)}\qquad for\ {\Tk}= 10\ to\ 20\, K.
\label{d6}
\end{equation}
In the high-$\Tk$ limit, the ``1.32'' is replaced with ``1.00''.  Of particular interest is pressure 
equilibrium, where $w=-1$.  For $\epsilon = 0.2$ to 0.4 (see Section~\ref{sssec331}), $u=0.8$ to 0.7 when 
$\Tk=10$ to 20$\,$K.  In the high-$\Tk$ limit, the corresponding range in $u$ is 0.5 to 0.3.  Hence, if 
there are spatial variations of the molecular gas kinetic temperature from one observed position to another, 
and pressure equilibrium applies, the apparent correlation between the X-factor and the $\cOone/\COone$ line 
ratio could be weak. 

There are, of course, additional complications.  One is that equation~(\ref{d4}) assumes LTE; non-LTE effects 
might be important in some sources and this would require an explicit density dependence in the relationship 
between $X_f$ and $I({\cO})/I({\CO})$.  Another complication is that the $\cOone/\COone$ ratio may be saturated
(see equation~\ref{xx27}) toward some positions in some sources.  (See Section~\ref{ssec44} for more details.)

Nonetheless, if a large number of positions are sampled in a source where the kinetic temperature
is roughly constant, such as in a galactic disk on large scales, then there could be an approximately 
linear correlation between the X-factor and the $\cOone/\COone$ line ratio.

\subsection{Determining $\epsilon$ from $\cOonebf/\COonebf$\label{ssec44}}

One way to constrain estimates of $X_f$ is to model the data in the $I({\cO})/I({\CO})$ versus $\twef$ 
plot, as was done in Section~\ref{sssec331}, thereby estimating $\epsilon$ and possibly other parameters.  
However, this method has the considerable drawback that it requires some measure of the source function.   
\citet{W06} did this for the Orion clouds by using the far-IR continuum data.   The Orion clouds are 
10$^\circ$ to 20$^\circ$ out of the Galactic plane, meaning that the origin of the continuum emission was 
unambiguous.   Due to the lack of velocity information, continuum data is not very useful for isolating 
the emission of individual clouds along lines of sight through the Galactic plane.  Therefore, determination 
of the source function of such clouds with continuum data would be very difficult, if possible at all.

Another simpler approach is to use the histogram of $I({\cO})/I({\CO})$ values.  
One such histogram for the Orion clouds
is shown in Figure~\ref{fig07}.    It shows a prominent peak
between $\cOone/\COone$ line ratios of about 0.1 and 0.3 with a maximum at about 0.2.  The left edge of 
this peak roughly corresponds to the self-shielding limit for molecules --- i.e., $A_v\simeq 1\, mag$. 
The $\cOone/\COone$ line ratio approximately correlates with the gas column density.  For the Orion
clouds, $I({\cO})/I({\CO})=0.1$ corresponds to total hydrogen-nuclei column density, $\Ngas$, of
0.7 to $1.5\times 10^{21}\unit cm^{-2}$ or about $A_v=0.5$ to 1.0 magnitudes.  (Note that this range
applies regardless of whether the column densities were derived from one- or two-component models and 
regardless of imposed restrictions on the signal-to-noise ratio; cf. \citet{W06}.)  The right edge 
of the peak can be interpreted by inspecting the panels of Figure~\ref{fig05}; the value of $I({\cO})/I({\CO})$
corresponding to the right edge of the histogram peak corresponds to where the model curves are linear 
--- i.e., where the $\cOone/\COone$ ratio saturates.  According to equation~(\ref{xx27}), this is 
$x_r^\epsilon$.  If $r_e$ is the $\cOone/\COone$ ratio at the right edge of the peak, then $\epsilon$
is given by
\begin{equation}
\epsilon = {ln(r_e)\over ln(x_r)}\qquad .
\label{d7}
\end{equation}
If the $\cO/\CO$ abundance ratio, $x_r$, is the same in the Orion clouds as in the inner
Galactic disk, then the fluffiness, $\epsilon$, of the molecular cloud clumps are also the same.  For
$x_r=1/60$, $\epsilon=0.29$.  This is consistent with model results of Section~\ref{sssec331}.  However, 
the observations of \citet{Langer90} show that $x_r$ in the inner galaxy systematic varies with 
Galactocentric radius by a factor of about 2 from the center to the solar circle.  Because of the 
logarithmic dependence on $x_r$ the affect on $\epsilon$ is relatively small: $\epsilon = 0.29$ to 0.35
from the solar circle to the inner Galaxy.  And these values are still consistent with the model results
of Section~\ref{sssec331}, even though those results only apply to the Orion clouds. Whether the X-factor
is the same in the Orion clouds as in the inner Galaxy also depends on the typical densities and temperatures
in the molecular cloud clumps in the two.  If they are the same on average, then Figure~\ref{fig04} provides
a crude estimate of the X-factor: $X_f\simeq 1.2$ to 2.2$\, X_{20}$ depending on the curve and $\epsilon$ value 
(i.e. 0.29 or 0.35) used. 

Of course, not all positions have $\cOone/\COone$ ratios between 0.1 and 0.3.  A few are much larger, up 
to 0.95.  Line ratios this high can only occur if either $\epsilon$ is smaller than the value(s) that occurs 
for the positions in the peak, as discussed in Section~\ref{ssec39}, or if $\thef$ is large.  It is possible 
that there is a distribution of $\epsilon$ values in clouds.  This distribution would probably have to be 
sharply peaked to permit the peak in the histogram to have reasonably defined right edge (e.g., the peak from 
its maximum down to some poorly defined zero level on the right side has $\cO/\CO$ line ratios from about 0.2
to about 0.45).  To have a rough guess at what such a distribution might look like, it could be {\it assumed\/} 
that {\it all\/} the observed positions have saturated $\cOone/\COone$ ratios.  Then equation~(\ref{d7}) with 
all the observed line ratios used in place of $r_e$ will give a histogram that would contain such a distribution 
within its boundaries.  Adopting symmetry about the maximum of the $\epsilon$ distribution as a working assumption 
implies that there is an excess in the number of positions with high $\epsilon$ values ($\epsilon>0.3$), which 
corresponds to an excess in the number of positions with low $\cOone/\COone$ ratios ($<0.3$).  This excess 
supports the interpretation that these low line ratios are {\it un}saturated as expected, and do not represent 
part of this hypothetical distribution of $\epsilon$ values. 

Comprehending exactly what limits the range of $\epsilon$ values in a molecular cloud, cloud complex, or
galaxy is key to understanding the X-factor.  It is also key to comprehending the narrow range of $\cOone/\COone$ 
ratios.  Therefore, understanding the narrow range of $\cOone/\COone$ ratios is key to understanding the X-factor.
These two are linked.  Even if the formulation for the X-factor proposed in this paper were incorrect, even if it
were not possible to clearly define a quantity like $\epsilon$ in real molecular clouds, the connection between 
the restricted range of $\cOone/\COone$ ratios and the X-factor would exist independently.

\subsection{Clump/Interclump\label{ssec45}}

Real molecular clouds (like Orion and Rosette) have about ${3\over 4}$ of their mass and ${1\over 10}$ of their 
volume in the form of clumps, and the remainder in an interclump medium \citep[see, e.g.,][]{Ostriker99, Williams95, 
Bally87}.  Adopting $\theta_H$ as the former quantity and $f_H$ as the latter \citep[as done in][]{Ostriker99}, then
$r_M = \theta_H^{-1}\, -\, 1={1\over 3}$ and $r_\sigma = [(1-f_H)/f_H]^{1\over 3}\simeq 2$.  For $r_M<1$, we need 
$r_\sigma\gg r_M^{1\over 3}$ for the $r_\sigma$ large limit to apply; since $2\gg 0.7$ is more or less valid, $X_f$ 
would be increased by about $(1\ +\ r_M)^{0.5(1-\epsilon)}$, which is an increase of less than 15\%.  Therefore, if 
most molecular clouds are like the examples given here, then the interclump medium will have a small effect on the 
X-factor.  This is especially true if this interclump gas is mostly atomic \citep[e.g., see][and references therein]
{Williams95}.

\subsection{Radial Density Profiles\label{ssec46}}

The density dependence on radius in dark clouds or in the clumps (or cores) of clouds is often $\rho\propto 
r^{-\alpha}$, where $\alpha$ is 2 or nearly so \citep[e.g.,][]{Harvey01, Tachihara00, Lada99, Alves98, 
Henriksen97, Williams95}.  As discussed in Appendices~\ref{appb} and \ref{appc} (see Sections~\ref{appb4}
and \ref{apppc31}) and Section~\ref{ssec35}, a $\rho\propto r^{-2}$ dependence requires specifying an outer 
radius, $r_1$, to keep the total mass finite and an inner radius to the $r^{-2}$ region, $r_0$, to keep the 
density finite.  Also, the ratio $r_1/r_0$ determines the fluffiness, $\epsilon$, of the clump.  And estimates 
of $\epsilon$ can be constrained by observations $X_f$ or of the $\cOone/\COone$ intensity ratio (cf. 
Sections~\ref{ssec39} and \ref{sssec331}).  Accordingly, the observations can constrain the $r_1/r_0$ ratio 
(assuming that the bulk of the $\COone$ emission originates in structures with $\rho\propto r^{-2}$).  
The 
histogram of $I(\cOone)/I(\COone)$ values suggests that $\epsilon\simeq 0.3$, the same as that found 
in the modeling in Section~\ref{sssec331} for the Orion clouds.  If the range of $\epsilon$ values found from 
that modeling are relevant to much of the Galactic disk, then $\epsilon = 0.2$ to 0.4. 
If the clumps are 
spherical, then $r_1/r_0$ would be in the range 2 to 9.  If the clumps are cylindrical filaments viewed side-on, 
then $r_1/r_0$ would be from 4 to 42.  (End-on cylindrical filaments would have a much more restricted range of
$r_1/r_0$, but these need not be considered here.  It is much more likely to view high-aspect cylinders close 
to side-on than close to end-on.  Hence, the side-on cylinders would probably dominate the emission from most
clouds.)

Observations suggest $r_1/r_0$ ratios that are consistent with the $\epsilon$ values specified above. 
Specifically, $r_1/r_0$ is found to be from about 4 to 15 \citep[e.g.,][]{Harvey01, Tachihara00, Lada99, 
Alves98, Henriksen97, Williams95}.  It must be borne in mind that many of these observed $r_1/r_0$ values
are merely lower limits due to limitations in spatial resolution, which place upper limits on $r_0$ ---
the inner radius of the $\rho\propto r^{-2}$ region within each clump.

\subsection{Molecular Clouds in Some Specific Regions\label{ssec47}}

\subsubsection{Rosette and Orion\label{sssec471}}

The literature has papers with lists of clump temperatures, masses, and dimensions for certain molecular 
clouds.  Two examples are \citet{Williams95} for the clumps in the Rosette Molecular Cloud (RMC) and 
\citet{Nagahama98} for the clumps in the Orion$\,$A cloud.   With these clump properties, rough estimates 
of $X_f$ can be made. 

For the RMC, Table~2 of \citet{Williams95} provides the necessary information, after adjusting the adopted
$N(H_2)/N(\cO)$ ratio.  \citet{Williams95} adopt $N(H_2)/N(\cO) = 5\times 10^{5}$, 
which may be quite reasonable, especially for the $\rho\,$Oph cloud \citep[see][and references therein]
{Bertoldi92}.   \citet{Frerking82} find an abundance of $8\times 10^{-5}$ for $\CO$ and \citet{Langer90} 
find 1/60 for $X(\cO)/X(\CO)$ in the Galactic disk approximately at the solar circle (and out to Orion).  
These imply $N(H_2)/N(\cO)\simeq 7.8\times 10^5$.  Adopting this abundance ratio implies 55\% higher clump 
masses (see $M_{LTE}$ in their Table~2) and an observationally determined $X_f$ that is also 55\% higher or 
1.7$\, X_{20}$, which is about the value expected for molecular clouds in the Galactic disk.  At the very
least, this higher $N(H_2)/N(\cO)$ ratio is the same as that adopted for the Orion clouds.     

For the RMC clumps, \citet{Williams95} find ${r_1\over r_0} = 5$ (cf. their Figure~22), which may only
be a lower limit due to resolution constraints.  If taken at face value, this means that $\epsilon=0.30$
and $k_A=1.80$.  Section~\ref{appb4} of Appendix~\ref{appb} computes the specific expression for $X_f$ 
for those parameter values --- expression~(\ref{apb39}).  The densities of the clumps were determined
by assuming spherical clumps with the masses (i.e., $M_{LTE}$ increased by 55\%) and radii (i.e., 
$\Delta R$) as listed in Table~2 of \citet{Williams95}.  The densities and temperatures of the clumps 
either inferred from, or listed in, Table~2 of that paper will then give the $X_f$ value corresponding 
to each clump --- that is, the $X_f$ value that would result in a hypothetical cloud filled with clumps 
identical to this one.  The median of these $X_f$ values is then taken as an estimate of the $X_f$ of 
the cloud.  (The mean is not used because it is influenced by extreme outliers.  A better way to do this 
is to develop a formulation that includes clumps with a distribution of densities and temperatures, but 
that is beyond the scope of the current paper.)  Here the $X_f$ values use the density as averaged over
the entire clump volume (i.e. equation~\ref{apb42}) and is more appropriate in this case \citep[see][]
{Williams94, Williams95}.  The $X_f$ values found for the RMC range from about 2 to 10$\, X_{20}$ with a 
median of 3.4$\, X_{20}$.  This median is double that determined observationally (and adjusted to the new 
$N(H_2)/N(\cO)$ ratio): 1.7$\, X_{20}$.   This suggests that the formulation is moderately successful, for 
the RMC at least, because it estimates the X-factor to within a factor of 2. 

It may be possible, however, to improve the agreement between the model $X_f$ and that observed.  At least some 
of the temperatures listed in Table~2 of \citet{Williams95} are probably underestimated.  These temperatures are 
excitation temperatures of the $\Jone$ transition of $\CO$ that were computed from the peak radiation temperature, 
$\Tr$, of that line.  In practical terms, this excitation temperature is about the same as the kinetic temperature, 
given that the $\COone$ line is easily thermalized (and \citet{Williams95} themselves implicitly assumed equivalence 
between the two temperatures when they determined column densities of {\it all\/} the molecular gas and {\it not 
just\/} the column densities in the $J=1$ state of $\cO$.)  This method of determining excitation temperatures or, 
roughly equivalently in this case, kinetic temperatures implicitly assumes that the emitting gas fills the beam 
within each velocity interval of the line profile, especially at the line peak.  Despite the scatter, the inferred 
clump temperatures have a definite positive correlation with the clump diameters; the Spearman rank-order 
correlation test places a 99.998\% confidence level on the correlation (or the significance of the null 
proposition of no correlation is $\sim 2\times 10^{-5}$).  In contrast, in the Orion$\,$A cloud, where the 
kinetic (excitation) temperatures were also estimated using the peak $\Tr$ of the $\COone$ line \citep[see][]
{Nagahama98}, the Spearman test places only a 40\% to 60\% confidence on the correlation (depending on 
whether the clump size used was filament length or filament diameter).  Accordingly, any correlation 
between clump size and inferred clump temperature in Orion$\,$A is either weak or non-existent.  (Although 
it should be mentioned that these temperatures are not entirely reliable either, given that certain values 
are frequently repeated.)  The Orion clumps have linear sizes (i.e., in parsecs) factors of 3 to 5 larger 
than those of the RMC and the Orion clouds are nearly a factor of 4 closer than the RMC. The interpretation 
of why the RMC clumps clearly show a correlation between size and inferred temperature while the Orion clumps 
do not is then fairly straightforward: the smaller clumps in the RMC are not being resolved in the velocity 
interval at the line peak, while the larger clumps are better resolved.  Hence it is likely that at least 
some of the inferred clump temperatures for the RMC are too low.  These inferred temperatures range from about 
5 to 31$\,$K with a median of about 9$\,$K.  For Orion$\,$A, the range is 13 to 37$\,$K with a median of about 
18$\,$K, which is the dominant dust temperature of the Orion clouds \citep[see][]{W96, W06} and even of the 
Galactic plane clouds \citep{Sodroski94}.  Both Orion$\,$A and the RMC are GMCs and should 
have clumps with roughly similar properties; the median temperature of the clumps of the RMC should not be half 
of that of the Orion$\,$A clumps.  In short, there are three reasons why some of the clump temperatures in the 
RMC are probably underestimated:
\bupskip
\bupskip
\begin{enumerate}
\item The temperature versus clump size correlation argues for a spatial resolution effect.
\bupskip
\item Both the RMC and Orion$\,$A are GMCs forming massive stars and their clumps should have similar
      temperatures.
\bupskip
\item An additional reason is that the RMC is close to the Galactic plane (i.e. about 2$^\circ$) and
      the dust temperatures in the plane is about 18$\,$K \citep[e.g., see][]{Sodroski94}.  If the RMC
      is like the Orion clouds in that its dust-gas temperature difference is small \citep{W06}, then
      the gas kinetic temperatures must be roughly double the listed values for some of the clumps. 
\end{enumerate} 
\bupskip
Correcting all the RMC clump temperatures upwards by a factor of 2 would not be appropriate; the temperature-size 
correlation suggests that the correction factor should be largest for the ``coldest,'' smallest clumps and 
progressively decrease towards unity for the ``warmer,'' larger clumps.  

How do these corrections to the temperatures affect the inferred X-factor values?  A simple way to address this
is to raise all the listed temperatures by a factor of 1.5.  This approach is less extreme than doubling all of 
the temperatures and is simpler than finding some prescription for applying different scale factors to the
temperatures of different-sized clumps.  This temperature correction will then affect the observed column densities
and, in turn, the inferred averaged densities by a factor 1.5$^{\gamma-1.32}$ or 1.19. (Again, note that the 
``1.32'' in the exponent becomes ``1.00'' for $\Tk\gg 20\,$K.) The theoretical $X_f$ will then change by a factor 
of 1.5$^{-0.47}\times 1.19^{0.35}$ or 0.88.  The median $X_f$ becomes 2.8$\, X_{20}$.  The observationally inferred 
$X_f$, however, increases by a factor 1.19 to about 2.0$\, X_{20}$.  Consequently, there is still a discrepancy between 
the $X_f$ from the currently proposed formulation and that that would be inferred observationally; the theoretical 
$X_f$ must be corrected downwards by 30\%.

There is a similar but somewhat smaller discrepancy for the Orion clouds.  The Orion clumps are filaments 
\citep{Nagahama98} and are treated here as $\rho\propto r^{-2}$ cylinders.  This type of cylinder was treated
in Section~\ref{apppc31} of Appendix~\ref{appc}, but it was collapsing and magnetized.  Here we assume
the cylinders to be virialized and again adopt the average density to be over the entire volume.  This gives 
$k_v = 2.76\times 10^{-16}$ in $cgs$ units and $k_N = (r_1/r_0)[2-(r_0/r_1)][1+2\,ln(r_1/r_0)]^{-1}$.  Here
$r_1/r_0 = 14$ is adopted because it has the fluffiness of that obtained from the modeling in 
Section~\ref{sssec331} (also see Section~\ref{ssec44}); numerical integration of (\ref{bi02}) and comparison 
with (\ref{bi09a}) results in $\epsilon = 0.30$ and $k_A = 1.76$.  Therefore, 
\begin{equation}
X_f(X_{20}) = 0.57\ C_T\ \Tk^{-0.48}\ n_a^{0.35}
\label{d8}
\end{equation}
The $X_f$ value corresponding to each clump listed in Table~2 of \citet{Nagahama98} was determined from the
temperatures listed in, and the densities inferred from, that table.  The densities were estimated from the
listed filament masses and dimensions.  For the Orion clouds, assuming cylindrical clumps and that the inferred 
densities are over the entire volume will result in $X_f$ values that range from 1.5 to 2.7$\, X_{20}$ 
with a median value of 1.9$\, X_{20}$, which is close to the observed value of about 2.1$\, X_{20}$ \citep[][]{W06}.  
(Note that this $X_f$ applies to the one-component models of \citet{W06}.  The two-component models would imply that 
$X_f\simeq 3.5\, X_{20}$, except that it is uncertain exactly how much mass is in the other colder component --- see 
\citet{W06b}.)   One additional consideration is that at least some of the observed filaments are being viewed partly 
end-on.  If we were to simplistically assume that there were only end-on filaments with the same dimensions, densities, 
and temperatures as those listed in Table~2 of \citet{Nagahama98}, then the median $X_f$ would roughly double, even 
though the $r_1/r_0=14$ cylinders considered here have $\epsilon=0.53$ when viewed end-on.  If we then consider that 
some of the filaments are partly end-on, then the best estimate of the theoretical $X_f$ would be somewhere around 
3$\, X_{20}$.  In fact, given that high-aspect filaments are more likely viewed side-on than end-on, this would
probably be closer to about 2.5$\, X_{20}$.  Accordingly, the derived X-factor must be corrected downward by about 
20-30\% (assuming that $r_1/r_0=14$ is appropriate for the Orion filaments). 

       There are a few points to consider here.  One is that average density defined as being over the whole 
cylindrical volume results in an X-factor that {\it increases\/} slowly with increasing $\epsilon$; this is quite 
different from the behavior seen with the previous definition of average density (see Section~\ref{ssec36} and 
Figure~\ref{fig04}).  Another point is that defining the average density this way results in fewer of the Orion$\,$A 
clumps being virialized; the fraction of clumps with velocity widths within 40\% of the virialized width drops from 
74\% to 51\%.   And the final point is that the derived X-factor values must be corrected downwards by about 20-30\% 
to match those observed.  This may only require a better definition for the average density.   It is also possible 
that simply redefining the average density is not enough to account for the discrepancy between the theoretical and 
observed X-factor values.  There must be some additional consideration to be included in the formulation.  This is 
not surprising given that the current formulation simplistically neglects the effects of magnetic fields and surface 
pressure \citep[e.g., see][]{Bertoldi92, Tilley03}. It is likely that improved understanding of molecular cloud 
physical conditions is also necessary for reducing this discrepancy.  In any event, the derived X-factor is within a 
factor of 2 of that observed.



\subsubsection{The Galactic Center\label{sssec473}}

The molecular gas within the central few hundred parsecs of the Galaxy represents an environment that
is distinct from that of the Galactic disk.  Compared to the disk gas, the molecular gas in the Galactic 
center is denser and hotter by an order of magnitude and very dynamically active \citep[e.g., see][]
{Martin04, RF01, Paglione98, Huettemeister98, Huettemeister93, Bally88, Bally87a, Harris85, Guesten85}.
Consequently, it is no surprise that the X-factor for the Galactic center clouds differs greatly from 
that for the Galactic disk clouds.  Specifically, it is between a factor of a few and more than an order
of magnitude smaller in the Galactic center than in the disk \citep[see, e.g.,][]{Dahmen98, Oka98,
Sodroski95}.   Smaller X-factors may be common in the central few hundred parsecs of spiral galaxies
\citep[e.g.,][]{Rickard85, Israel88, W93, Regan00, Paglione01}\citep[see also][and references therein]
{Dahmen98}.

How do we account for these low X-factors?  Given that tidal forces may be appreciable in the Galactic center, 
the outer layers of clumps or clouds may be sheared off, producing a substantial interclump or intercloud 
medium \citep{Stark89}.  As mentioned in Section~\ref{ssec38a}, if interclump gas dominates the $\COone$ 
emission and the gas mass, then $X_f$ decreases because of the low average density in this gas.  As stated 
in the previous paragraph, there is evidence for dense, hot molecular gas in the Galactic center region, but 
does this apply to the bulk of the CO-emitting gas in the Galactic center?  If, for example, this CO-emitting 
gas is an order of magnitude hotter and an order of magnitude {\it less\/} dense than that in the Galactic disk, 
then the low $X_f$ might be explained.  However, the evidence for this is less than compelling.  \citet{Dahmen98}
find high $I(\COone)/I(\Coone)$ ratios that are sometimes nearly as high as the $\CO/\Co$ abundance ratio. 
They argue that the higher the $I(\COone)/I(\Coone)$ ratio, the lower the gas density, claiming densities as low 
as $\nH\sim 10^2\unit cm^{-3}$, even though a simpler explanation is that the higher ratio only really implies 
lower optical depths in the $\Jone$ lines of $\CO$ and $\cO$.  These authors used observed molecular gas velocity 
gradients (or velocity widths per unit size) to impose constraints on the model velocity gradient.  These observed 
gradients only really apply to the scale of entire clouds, and applying them to the model gradients is no more 
valid than applying such constraints to gas volume density.  Such cloud-scale quantities are only rough lower 
limits to the relevant quantities to be used in non-LTE radiative codes, such as the LVG code.  Assuming that the 
lowest and highest observed cloud-scale velocity gradients represent both lower and upper limits on the velocity 
gradients to be used in the LVG code is not correct and imposes an artificial connection between optical depth and 
density --- both rise and fall together.   Thus when the $I(\COone)/I(\Coone)$ ratio approaches the abundance ratio,
the optical depths of both $\COone$ and $\Coone$ drop and, because of the imposed constraint on the velocity 
gradient, the density drops too.  Thus imposing constraints on the velocity gradients in this way leads to the 
assertion that $I(\COone)/I(\Coone)$ being close to $X(\CO)/X(\Co)$ cannot occur in LTE --- which is incorrect.  
Densities of $\nH\sim 10^2\unit cm^{-3}$ may indeed be present in the CO-emitting gas in the Galactic centre region, 
but it cannot be verified with this line ratio alone; additional information is needed (and something more than 
{\it large}-scale velocity gradients).

Nonetheless, partly explaining the low X-factors in the Galactic center region is still possible.  The
large observed values of the $I(\COone)/I(\Coone)$ ratio \citep{Dahmen98}, for example, suggest that
$\COone$ is optically thin or nearly so.  From Table~\ref{tbl-1} we see that the optically thin case
has $X_f$ an order of magnitude or more lower than the Hard Sphere case, this latter having roughly the 
$X_f$ for Galactic disk clouds when $\Tk = 20\,$K and $\nH = 2000\unit cm^{-3}$.  One problem with this
is that there is evidence that the molecular gas could have high kinetic temperatures, e.g. $\Tk\gsim 100\,$K
\citep[e.g.,][]{Martin04, RF01, Huettemeister98, Huettemeister93, Guesten85, Harris85}.  If $\Tk$ is 100$\,$K,
then $X_f$ {\it rises\/} to about 0.5$\, X_{20}$ (see equation~\ref{xx6} in Section~\ref{ssec32}), or a factor 
of $\sim 4$ lower than that for the Galactic disk clouds.   To have more than an order of magnitude lower $X_f$,
an additional effect is necessary.  The Galactic center's clouds are very dynamically active and have line
widths factors of 5 to 10 larger than found in Galactic disk clouds \citep[see, e.g.,][]{Bally87a, Bally88}.
If this also means that the velocities of the Galactic center gas are factors of 5 to 10 beyond virialization,
then the effective $k_v$ of this gas is larger than the virialized value by the same factors.  Given that
$X_f\propto k_v^{\epsilon - 1}$, $X_f$ would be reduced by factors between 1 and 5 to 10, depending on
$\epsilon$.  If the $\COone$ line truly is optically thin, then $\epsilon = 1$ and the velocity structure
is irrelevant (i.e., only {\it in}directly relevant because the velocity structure partly determines the optical 
depth) for determining $X_f$.  Not all the observed $I(\COone)/I(\Coone)$ ratios are high enough to indicate
optically thin $\COone$, so an intermediate $\epsilon$ of 0.5 might be more appropriate.  But this would only
result in an X-factor that is only about a factor of 6 lower than the disk value.  Finding X-factors that are
more than an order of magnitude smaller \citep[see][]{Dahmen98} in molecular gas that is warm and optically
thin (or nearly so) is difficult to explain.  Heightening the abundance of $\CO$ by a factor of a few might
do this, but it would lead to other difficulties, such as explaining how $\COone$ could be so close to being
optically thin (as suggested by the high $I(\COone)/I(\Coone)$ ratios).  The best way to reduce $X_f$ seems
to be assuming that the gas behaves like fluffy clumps, something like the Squared Lorentzian Sphere.  If the
gas has a temperature of about 100$\,$K and is still marginally optically thick, then it is easy to obtain
$X_f\simeq 0.4\, X_{20}$ --- a factor of about 5 smaller than in the Galactic disk.  If we also consider that 
the gas velocity widths are an order of magnitude larger than required by virialization, then this reduces the 
X-factor another factor of nearly 3 and it can be an order of magnitude smaller than that in the disk.

In short, it is possible to explain X-factor values that are an order of magnitude smaller than is found in
Galactic disk molecular clouds, but the really low values --- nearly two orders of magnitude smaller 
\citep[see][]{Dahmen98} --- are not so easily accounted for.



\subsubsection{Irregular Galaxies\label{sssec474}}

As discussed in Sections~\ref{ssec37} and \ref{ssec42}, the unusually high X-factors found in molecular
clouds in irregular galaxies are largely due to the low abundances of $\CO$ (or low $X(\CO)$) in these 
galaxies.  As listed in \citet{Israel97}, the $X_f$ values found in molecular clouds of irregular galaxies 
can range from a factor of a few to about 100 higher than that found in the Galactic disk molecular clouds. 
As described in Section~\ref{ssec37}, the abundance $X(\CO)$ is an average over the volume of the molecular
gas; there are CO-emitting regions with a Galactic abundance of CO surrounded by envelopes of CO-deficient 
molecular gas \citep[see][and references therein]{Israel97, Israel00}.  Consequently, a map of some molecular 
gas tracer that can show the molecular gas not traced by CO can be compared with CO maps to test the correlation 
between the observed $X_f$ and the inferred volume ratio of molecular gas to CO-emitting gas.  \citet{Madden97}, 
for example, map the low-metallicity irregular galaxy IC$\,$10 in the [C$\,$II] 158$\um$ line and find that
this emission is largely associated with molecular gas in photodissociation regions (PDRs).  Their 
Figure~5 shows the contours of the [C$\,$II] 158$\um$-line emission with superposed contours of CO
emission.  Depending on exactly how the relative areas are estimated, the ratio of projected areas
of molecular gas to CO-emitting gas is about 10 to 20.  Assuming that the volume goes like (area)$^{1.5}$,
then these correspond to volume ratios of about 30 to 90.  Accordingly, the volume-averaged $X(\CO)$ 
would be roughly the Galactic disk value divided by these volume ratios.   Given that the X-factor goes
like $X_f\propto X(\CO)^{-0.8}$ (see Section~\ref{ssec42}), $X_f$ would be factors of about 15 to 40
larger than found in the Galactic disk; \citet{Madden97} find that this factor to be about 100.
Therefore, using only very crude estimates of the volume ratio of molecular gas to CO-emitting gas,
the X-factor is estimated to within a factor of a few.  If the molecular gas in IC$\,$10 is very roughly
representative of that in other irregular galaxies, then it is easy to understand the high X-factors in 
these galaxies.  If $X_f/X_{fG}$ is the ratio of the observed X-factor to the ``standard'' value found
in Galactic disk clouds, and if $A_r$ is the ratio of projected areas of molecular gas to CO-emitting
gas, then $X_f/X_{fG}\simeq A_r^{1.2}$ is expected.  Deviations from this dependence would be extremely
interesting, providing clues to differences in physical conditions (e.g., $\epsilon$, $\nH$, $\Tk$)
between those in the molecular clouds of irregular galaxies and those in Galactic disk clouds. 


\section{Summary and Conclusions\label{sec5}}

An improved formulation for the X-factor is proposed that combines virialization of the gas with radiative 
transfer in a clumpy medium.  The statement that the velocity-integrated radiation temperature of the 
$\COone$ line, $I(\CO)$, ``counts'' optically thick clumps is quantified using the formalism of MSH84 for 
line emission in a clumpy cloud.  Adopting the simplifying assumptions of thermalized $\COone$ line emission 
and isothermal gas, an effective optical depth, $\tef$, is defined as the product of the clump filling factor 
within each velocity interval, ${N\over\Nc}{\Dvc\over\Dv}$, and the clump effective optical depth (or effective 
optically thick area), $\At$, as a function of the optical depth, $\tau_0$, on the clump's central sightline.  
The $\At$ is well approximated (to within about 13-26\%) as a power law in $\tau_0$ with power-law index, 
$\epsilon$, called the clump ``fluffiness,'' and has values between zero and unity.   While the $\COone$ line 
is optically thick within each clump (i.e., high $\tau_0$), it is optically thin ``to the clumps'' (i.e., low 
$\tef$).  Thus the dependence of $I(CO)$ on $\tef$ is linear, resulting in an X-factor that depends only on the 
properties of the clumps rather than having a direct dependence on the entire cloud.  Assuming virialization of 
the clumps yields an expression for the X-factor whose dependence on physical parameters like density and 
temperature are ``softened'' by power-law indices of less than unity that depend on the fluffiness parameter, 
$\epsilon$.  The X-factor provides estimates of gas column density because each sightline within the beam has 
optically thin gas within certain narrow velocity ranges.  Determining column density from the optically thin 
gas is straightforward and parameters like $\epsilon$ then allow extrapolation of the column density of the 
optically thin gas to that of all the gas.  Implicit in this formulation is the assumption that fluffiness is, 
on average, constant from one beam to the next.  This is less true for density and temperature for which the 
X-factor, $X_f$, may have a weaker dependence. 

The proposed formulation addresses the problems of the explanation proposed by DSS86:
\bupskip
\bupskip
\begin{enumerate}
\item {\it Treatment of radiative transfer.\/}  The dependence on fluffiness, $\epsilon$, represents a
radiative transfer parameter of the clumps.  The optically thin case results by simply setting $\epsilon 
= 1$: dependence on virialization disappears and the expression for column density in terms of $I(CO)$
in the optically thin case remains.  In a hypothetical completely optically thick case with flat-topped
line profiles, $\epsilon\to 0$ and the radiative transfer vanishes, leaving only virialization (as in
DSS86).  
\bupskip
\item {\it Reduced sensitivity to T$_{_K}$ and n(H$_2$).\/}  The effect of the fluffiness reduces the
dependence of $X_f$ on density and temperature from $X_f\propto{\nH^{0.5}\over\Tk}$ to roughly
$X_f\propto\left({\nH\over\Tk}\right)^{0.3}$ in the high-$\Tk$ limit.  Thus variations of
an order of magnitude in either $n$ or $\Tk$ would allow $\rm X$ to vary by less
than a factor of 2.
\bupskip
\item {\it Virialization of entire clouds is {\rm un}necessary.\/}  The densities required to
give reasonable values of $X_f$ are consistent with those found in cloud clumps (i.e. $\sim 10^3\, H_2
\unit cm^{-3}$).  Thus virialization of clumps, rather than of entire clouds, is consistent with the
observed values of $X_f$.  And even virialization of clumps is not strictly required; only a relationship 
between clump velocity width and column density similar to that of virialization can still yield
reasonable values of the X-factor.  The underlying physics is now at the level of cloud clumps,
implying that the X-factor can probe sub-cloud structure. 
\bupskip 
\item {\it Stronger dependence of peak $T_{_{R}}$ on N(H$_2$) than of $\Delta$v on N(H$_2$) is now 
explained.\/}  The peak $\Tr$ depends linearly on the filling factor within each velocity interval,
${N\over\Nc}{\Dvc\over\Dv}$, thereby accounting for its dependence on $N(H_2)$.  The $\Dv$, on the
other hand, is the observed line width and is not necessarily directly related to the beam-averaged
column density, $N(H_2)$.  If virialization is 
important at the level of the clumps, then the velocity width-column density relation is between 
that of the clump velocity width, $\Dvc$, and clump column density, $\Nc$.
\end{enumerate}
\bupskip

X-factor values were computed for both spherical clumps and cylindrical clumps (i.e. filaments) of 
densities $2\times 10^2$, $2\times 10^3$, and $2\times 10^4\, H_2\unit cm^{-3}$ and kinetic temperatures 
10 and 20$\,$K and different internal density variations.  The clumps of average density $2\times 10^3\, 
H_2\unit cm^{-3}$ that reproduce the standard observed $X_f$ value of about 2$\, X_{20}$ for the Galactic 
disk clouds within a factor of 2 are the Hard Sphere (or uniform-density sphere), the Gaussian Sphere, 
and the Gaussian Filament (see Table~\ref{tbl-1}).  Aside from the clumps listed in Table~\ref{tbl-1}, 
spherical clumps and filaments with average densities of $\sim 10^3\, H_2\unit cm^{-3}$ and an $r^{-2}$ 
density variation can also produce X-factors within a factor of 2 of the standard value.  Testing these
clump types reveals a potentially strong inverse dependence of $X_f$ on $\epsilon$, as shown in 
Figure~\ref{fig04}.  Increasing the fluffiness has the advantage of weakening the dependence of the
X-factor on density and temperature, but the disadvantage of decreasing $X_f$ to values appreciably below 
the standard value.  This strong dependence of $X_f$ on $\epsilon$ is related to how the average density
is defined.  For clump types with no clearly defined outer radius (e.g., Gaussian or Squared Lorentzian),
the average density was defined as that within a spherical or cylindrical (depending on clump geometry)
diameter equal to the FWHM of the projected surface density distribution.  For clumps with an $r^{-2}$ 
density variation, an outer radius must be defined to keep the clump mass finite.  The average density 
can be defined as done previously, but also can be defined over the whole clump volume.  The latter 
definition can result in a weakly rising $X_f$ as a function of $\epsilon$. 

The proposed formulation also suggests a specific dependence of the X-factor, $X_f$, on the $\CO$ 
abundance, $X(\CO)$.  In the molecular clouds in irregular galaxies, the CO abundance within the
CO-emitting regions is about the same as that in Galactic disk clouds, but the average over the
entire molecular cloud volume is much lower \citep[see][and references therein]{Israel97, Israel00}.
If $X(\CO)$ is this volume-averaged abundance, then the formulation predicts $X_f\propto X(\CO)^{0.7\
to\ 0.8}$ (for $\epsilon\simeq 0.3$).  This is consistent with observations (see Section~\ref{ssec42}).

This formulation has implications for the interpretation of spectral line ratios, especially for the
$I(\cOone)/I(\COone)$ ratio.  Modeling the plot of this ratio against the effective optical depth, $\tef$,
--- determined from the peak radiation temperature of the $\COone$ line normalized to the source function (in 
temperature units) --- for the Orion clouds provides crude estimates of the fluffiness: $\epsilon\simeq 
0.3\pm 0.1$.  Histograms of $\cOone/\COone$ line strength ratios can also provide estimates of $\epsilon$,
and have the very important advantage that no estimates of the source function are necessary.  
Such a histogram from observations of the Orion clouds shows a peak that
extends from $I(\cOone)/I(\COone)\simeq 0.1$ to 0.3.
These limits for the Orion clouds (and possibly
for other Galactic disk clouds) correspond to the minimum column density for self-shielding against the
interstellar radiation field (i.e. $A_v\simeq 1\, mag$) at the low end and to the saturation of the
$\cOone/\COone$ line ratio at the high end.  The value of this ratio at saturation is determined from the 
dominant value of $\epsilon$ (i.e. $\sim 0.3$) within the clouds' substructures.  Consequently, a narrow 
range in $\epsilon$ can simultaneously account for the limited range of $\cOone/\COone$ line ratios and for 
a relatively constant X-factor.  In any event, it is no surprise that $\cOone/\COone$ line ratio is related 
to the X-factor.

Observations of the $\cOone/\COone$ line ratio have been used, for example, to infer differences in the 
X-factor between the molecular clouds of the nucleus of a spiral galaxy and that of its disk clouds 
\citep[e.g.,][]{Rickard85, Paglione01}.  The formulation finds a linear relationship between $X_f$ and 
$I(\cOone)/I(\COone)$, provided that $\Tk$ and $X(\cO)/X(\CO)$ (or $x_r$) are constant.  If these quantities 
vary spatially, then $X_f$ can have a dependence on $I(\cOone)/I(\COone)$ that is weaker than linear.

The proposed formulation and observed $I(\cOone)/I(\COone)$ ratios, or observed X-factor values, can 
constrain estimates of properties of substructures within molecular clouds, often by imposing limits on 
estimates of the fluffiness, $\epsilon$.  Many clumps or small clouds have an $r^{-2}$ density (or $\rho$) 
dependence.  If the clump outer radius is, $r_1$, and the inner radius of the $\rho\propto r^{-2}$ region 
within the clump is $r_0$, then the ratio $r_1/r_0$ is constrained by the limits on $\epsilon$.  Given that 
$\epsilon\simeq 0.2$ to 0.4, and assuming that $\rho\propto r^{-2}$ structures are the dominant source of 
the $\COone$ emission, spherical clumps would have $r_1/r_0 = 2$ to 9 and cylindrical clumps would have 
$r_1/r_0 = 4$ to 42.  Observations are apparently consistent with these limits, where $r_1/r_0$ is about 
4 to 15 \citep[e.g.,][]{Harvey01, Tachihara00, Lada99, Alves98, Henriksen97, Williams95}, although spatial 
resolution limitations often mean that the observed numbers are merely lower limits.

The properties of real clumps in real molecular clouds can be used to estimate the X-factor within these
clouds and then be compared with the observationally determined X-factor.  Applying this to the Orion$\,$A
cloud \citep{Nagahama98} and the Rosette Molecular Cloud or RMC \citep{Williams95} yields estimates of the
X-factor that are within a factor of 2 of the observed values.  While this is acceptable as a start, reducing
this discrepancy will require improving the formulation.  Simply changing the definition of the average 
density can reduce the discrepancy, but it seems likely that something else is missing from the current 
formulation.  
     
A future, improved formulation for the X-factor must address the following shortcomings of the current
formulation:
\bupskip
\bupskip
\begin{itemize}
\item[A)] It is not entirely clear why the fluffiness seems to be constant, or at least sharply peaked, 
	  at one value.  This value seems to be about $\epsilon\simeq 0.3$.
\bupskip
\item[B)] The resultant X-factor values are too dependent on the precise definition of the average
	  density.  A closer to optimal way of defining such a density must be found.
\bupskip
\item[C)] The desired insensitivity of the X-factor to density and temperature often comes at the
	  price of too low an X-factor value.  Again, this is affected by the definition of average
	  density.
\bupskip
\item[D)] There is a potentially strong dependence on the fluffiness, $\epsilon$.  This dependence 
	  can be weakened by the appropriate choice for the definition of average clump density.
\bupskip
\item[E)] The current formulation does not consider clumps with a spectrum of properties, such as
	  distributions of densities and temperatures.  This will introduce other parameters in
	  addition to $\epsilon$, $\Tk$, and $\nH$ for determining the X-factor. 
\bupskip
\item[F)] Additional physical effects must be considered.  These would include non-LTE effects and the 
	  effects of magnetic fields, surface pressure, and turbulence. 
\bupskip
\item[G)] Observational determinations of parameters such as $\epsilon$ can be complicated by clumps
	  having a spectrum of properties, such as ranges in densities, temperatures, optical depths,
	  etc.  The presence of interclump gas might complicate this as well. 
\end{itemize}

Despite these shortcomings, the currently proposed formulation represents the first major improvement
in understanding the X-factor since \citet{Dickman86} (i.e., DSS86), because it includes radiative 
transfer.  Previous explanations of the X-factor involved counting optically thick clumps (or entire 
clouds) and a relationship between the gas column density and the velocity width of the CO$\,\Jone$ 
spectal line \citep[e.g., DSS86][]{Israel88, Evans99}.  But applying the DSS86 approach directly to
cloud clumps often overestimates the X-factor.   At first glance, applying radiative transfer
to an optically thick line is apparently pointless.  However, portions of the gas are not optically
thick, permitting estimates of the mass of the gas when radiative transfer is considered.  And 
including radiative transfer does indeed result in reasonable estimates of the X-factor even when
applying the formulation to the level of individual clumps. 

Determination of the X-factor on scales of many parsecs can constrain the average properties of the
molecular gas at scales of just a few parsecs.  Future formulations may refine the X-factor into a 
potent probe of molecular cloud structure.

\acknowledgments This work was supported by CONACyT grant \#211290-5-0008PE 
to W.~F.~W. at {\it INAOE.\/} I am very grateful to W.~T.~Reach
for his comments and support.
I thank T.~A.~D.~Paglione, G.~MacLeod, 
E.~Vazquez Semadeni, F.~P.~Israel and others for stimulating and useful discussions. 
The author is grateful to R.~Maddalena and T.~Dame, who 
supplied the map of the peak $\COone$ line strengths and provided important 
calibration information.




\appendix

\section{The Effective Optical Depth of a Single Clump\label{appa}}

MSH84 derive the effective optical depth of a clump, $\At$, which they call the 
effective optically thick area, for the case of a clump with a Gaussian spatial 
variation of its optical depth.  This appendix gives a quick derivation of $\At$
for a clump with a more general optical depth variation.  MSH84 derive the 
effective optical depth on a line of sight through a clumpy cloud:
\begin{equation}
\tef(v_z) = {\Ncs\over\sqrt{2\pi}\;\Dv}\int dv\, exp\left(-{v^2\over 2\,\Dv^2}\right)
\int dx\int dy \left\{1-exp\left[-\tau(x,y)\, exp\left(-{(v-v_z)^2\over 2\,\Dvc^2}\right)
\right]\right\}\quad .
\label{apa01}
\end{equation}
Notice that different notation from that of MSH84 is used here.  The \Ncs\ is the 
number of clumps per unit projected area of the cloud and corresponds to the $N$
of MSH84.  The velocity widths here are rms velocity widths, whereas those of
MSH84 are ratios of the Gaussian line profile areas to their amplitudes.  This 
results in extra factors of $\sqrt{2\pi}$.  The spectral line velocity width was 
represented by $\sigma$ in MSH84 and is given by $\sqrt{2\pi}\Dv$ here.  The
velocity width of a single clump was $v_o$ in MSH84 and is $\sqrt{2\pi}\Dvc$ here.
The velocity $v_z$ is the bulk velocity of a clump along the sightline and
the velocity, $v$, is the velocity of a given element of gas along the sightline.
Following MSH84, the assumption of the line width being much greater than the
velocity width of a single clump, i.e. $\Dv\gg\Dvc$, is adopted.  This means that
$v$ cannot deviate too far from $v_z$ without making the integrand very small. At 
the same time, the first exponential factor of the integrand does not change much
from $exp(-v_z^2/(2\,\Dv^2))$, because $\Dv$ is so much larger than $\Dvc$.  
Consequently, expression (\ref{apa01}) simplifies:
\begin{equation}
\tef(v_z) = {\Ncs\over\sqrt{2\pi}\;\Dv}\, exp\left(-{v_z^2\over 2\,\Dv^2}\right)
\int dv'\int dx\int dy \left\{1-exp\left[-\tau(x,y)\, exp\left(-{v'^2\over 2\,\Dvc^2}
\right)\right]\right\}\quad ,
\label{apa02}
\end{equation}
where $v'$ replaced $v-v_z$.  Now the integrals are only over the clump velocity 
width and the clump's projected surface area.  Therefore, the integrals, when
properly normalized, must give the effective optical depth of a single clump,
$\At$.  Normalizing for the velocity width requires dividing by $\sqrt{2\pi}\Dvc$.
Normalizing for the effective projected area of the clump requires dividing by
$\aef$ (see equation \ref{bi03}).  This results in equation~(\ref{bi02}), as
desired.
 
Equation~(\ref{apa02}) then becomes
\begin{equation}
\tef(v_z) = {\aef\Ncs\,\Dvc\over\Dv}\,\At\, 
exp\left(-{\hbox{$v_z$}^2\over 2\,\Dv^2}\right)\quad .
\label{apa03}
\end{equation}
Given that $\aef\Ncs=N/\Nc$, equation~(\ref{apa03}) becomes equation~(\ref{bi01}),
also as desired.

Equation~(\ref{bi01}) can be used as the starting point instead.  It represents
a logical relationship between the effective optical depth through the cloud
and that for a single clump; its derivation is trivial.  Comparing 
equation~(\ref{bi01}) with equation~(\ref{apa02}) then gives equation~(\ref{bi02}) 
for $\At$.

The reader may have noticed that the definition of $\At$ is lacking a certain
geometric correction.  For example, in the $\tau_0\ll 1$ limit, $\At\simeq\tau_0$.
And yet $\At$, which is an appropriately averaged $\tau$ over the projected 
surface of the clump, cannot simply be the $\tau$ through the center of the
clump, i.e. $\tau_0$.  It must be less than this value, at least for those
clumps where $\tau_0$ is highest on the central sightline.  Consider, for
example, the hard sphere, which means a sphere with clearly defined edges
and uniform internal density.  (See MSH84 for their treatment of hard spheres.)
Since we are only considering LTE and uniform kinetic temperatures, and that these 
clumps have uniform internal density, the optical depth on any given sightline 
through the clump is proportional to the path length through the clump on that
sightline.  The average path length through a hard sphere is its volume divided
by its projected area or ${2\over 3} d$, where $d$ is the sphere's diameter. 
The average optical depth of such a sphere in the optically thin limit is then
${2\over 3}\tau_0$ rather than simply $\tau_0$.  However, the clump optical
depth, $\Nc$, as defined in the text is also used as the column density on
the central sightline through the clump.  This means that $N/\Nc$ is not 
exactly the number of clumps in the beam as stated in the text: a slight 
correction factor is needed.  Nevertheless, this correction factor cancels 
that needed for $\At$, thereby yielding the correct result for $\tef$.  Another
way to argue this is to consider the factors $\At/\Nc$ and $N\Dvc/\Dv$.  The
former factor is the average optical depth per unit column density for a clump, 
since both $\At$ and $\Nc$ are referred to the central sightline.  The latter
factor is the beam-averaged column density within velocity interval $\Dvc$
at the line central velocity, $v_z=0$.  Therefore, the product of the the two
factors must give the optical depth at the line central velocity, and this is
equation~(\ref{bi01}) evaluated at $v_z=0$, as desired.  

\bigskip
\bigskip

\section{The X-Factor for a Small Beam on a Uniform-Density Cloud\label{appab}}

Here the $N(H_2)/I(CO)$ conversion factor is estimated for the case of a beam much smaller
than the source, where that source is a uniform-density cloud.  The velocity-integrated 
radiation temperature of the $\COone$ line, $I(CO)$, in LTE in a cloud of uniform kinetic 
temperature is
\begin{equation}
I(CO) = \Jnu\ \int_{line}\, dv\ [1\ -\ exp(-\tau(v))]\qquad,
\label{ab01}
\end{equation}
where the integral is over the spectral line profile.  The optical depth velocity profile is 
\begin{equation}
\tau(v) = \tau_0\, exp\left(-{v^2\over2\,\Dv^2}\right)\qquad .
\label{ab02}
\end{equation}
Numerically integrating (\ref{ab01}) shows that the integral can be approximated by
\begin{equation}
\int_{line}\, dv\ [1\ -\ exp(-\tau(v))]\simeq\sqrt{2\pi}\,\Dv\, k_A\, \tau_0^\epsilon\qquad ,
\label{ab03}
\end{equation}
where $k_A = 1.3$ and $\epsilon=0.14$.  If we now assume a uniform density sphere and virialization, then 
the derivation of $\Dv$ of Section~\ref{ssec22} is relevant.  It is also assumed that the velocity width 
on whatever sightline through the cloud reflects the virialization of the entire cloud and does not have 
appreciable contributions from other types of motion.  Repeating the derivation of Section~\ref{ssec24}
again yields equation~(\ref{bi29}).  This means that this equation and its variants (see Appendices~\ref{appb}
and \ref{appc}) are more general than only treating radiative transfer through a clumpy medium.  All that is 
required is that $I(CO)\propto\tau_0^\epsilon$ and $\Dv\propto (N\,L)^{0.5}$, where $N$ and $L$ are the column 
density and pathlength, respectively, on the sightline through the cloud.  In fact, the density and kinetic 
temperature that appear in expression~(\ref{bi29}) must be replaced, in the current treatment only, by the 
corresponding quantities that are appropriately averaged over the entire cloud, rather than for just an 
individual clump.

\bigskip
\bigskip

\section{Treatment of Spherically Symmetric Clumps\label{appb}}

Here we assume that the mass density $\rho$ is simply a function of the radius, 
$r$, within each spherical clump: $\rho = \rho(r)$.   This is similarly true 
for the number density, $n(r)$.  If the $x$-$y$ plane is perpendicular to the 
observer's sightline and $z$ is measured along the sightline, then $n(r)$ is 
$n(\sqrt{p^2+z^2})$, where $p\equiv\sqrt{x^2 + y^2}$ is the projected radius as 
seen by the observer.  The clump column density as a function of projected radius, 
$N_c(p)$, is given by
\begin{equation}
N_c(p) = \int_{-\infty}^{+\infty}\back\back dz\ n(\sqrt{p^2+z^2}) \quad .
\label{apb01}
\end{equation}
The mass within radius $r$, $M(r)$, is given by
\begin{equation}
M_c(r) = 4\pi \int_0^r\back d\rs\ \rs^2\,\rho(\rs) \quad .
\label{apb02}
\end{equation}
A useful quantity is the average density $\rb$.  This can be defined in a 
number of ways, but is chosen here to be in terms of the 
half-width-at-half-maximum projected radius of the clump, $\ph$, defined such 
that $N_c(\ph)\equiv 0.5\, N_c(p=0)$.  Then $\rb$ is 
\begin{equation}
\rb \equiv {M_c(r=\ph)\over {4\pi\over 3}\ph^3} \quad .
\label{apb03}
\end{equation}
The self-potential due to gravity is
\begin{equation}
W = -16\pi^2 G\int_0^\infty\back d\rs\ \rs^2 \rho(\rs) \int_0^\rs\back d\rs'
\ {\rs'^2\over\rs}\,\rho(\rs') \quad .
\label{apb04}
\end{equation}
If $N_c$ is defined as the column density through the clump center --- i.e., 
$N_c\equiv N_c(p=0)$ --- and $M_c$ is defined as the total clump gas mass
--- i.e., $M_c\equiv M_c(r=\infty)$, then we can write these and $W$ as
\begin{eqnarray}
N_c &=& 2\, k_N\ \nb\ \ph \quad ,
\label{apb05}\\
M_c &=& k_M\ \rb\ \ph^3 \quad ,
\label{apb06}\\
W &=& - k_W\ {G M_c^2\over\ph} \quad .
\label{apb07}
\end{eqnarray}
The constants $k_N$, $k_M$, and $k_W$ are given by the precise functional form of $\rho(r)$.
Specifically, these constants are determined by comparing equations~(\ref{apb01}),
(\ref{apb02}), and (\ref{apb04}) with (\ref{apb05}), (\ref{apb06}), and (\ref{apb07}),
respectively. 

The simple form of the Virial theorem applied in Section~\ref{ssec22} and the
full-width-at-half-maximum, $\Dh\equiv 2\ph$, are now used to derive $\Dvc$:
\begin{eqnarray}
\Dvc &=& k_v\ \nb^{1\over 2}\ \Dh \quad ,
\label{apb08}\\
\noalign{\noindent where}
k_v &\equiv& \left[{1\over 12}\ k_W\ k_M\  \mu\mh G\right]^{1\over 2}\quad ,
\label{apb09}
\end{eqnarray}
where the expressions~(\ref{apb06}) and (\ref{apb07}) and $\rb =\mu\,\mh\,\nb$ was also used 
(also see Section~\ref{ssec22}). 

In addition to the above equations, we need the appropriate equation for $\tau_0$.  Starting
with equation~(\ref{bi22}) and plugging in (\ref{apb08}) and (\ref{apb05}) yields,
\begin{equation}
\tau_0 = {k_\tau k_N\over k_v\sqrt{2\pi}}\, \nb^{0.5}\Tk^{-\gamma}\quad ,
\label{apb10}
\end{equation}
which is a more general form of (\ref{bi22x}).   Equation~(\ref{bi22x}) assumed a sphere
of uniform density, for which $k_N=1$, and (\ref{apb10}) reduces to (\ref{bi22x}). 

Now we can combine the expressions developed here with (\ref{bi25}), (\ref{bi26}), and 
(\ref{bi09a}) to obtain,
\begin{equation}
X_f = (2\pi)^{{1\over 2}(\epsilon - 1)}\, C_T\, k_A^{-1}\,k_\tau^{-\epsilon}
\,k_v^{\epsilon - 1}\, k_N^{1-\epsilon}\,\hbox{$\Tk^{\gamma\epsilon - 1}$}
\,\nb^{{1\over 2}(1-\epsilon)}
\quad .
\label{apb11}
\end{equation}
This is a more general version of equation~(\ref{bi29}), which assumed a uniform-density
sphere.  In fact, equation~(\ref{apb11}) is more general than for just spherical clumps: 
as long as $\Dvc$, $N_c(0)$, $\tau_0$, and $\At$ can be represented by expressions~(\ref{apb08}),
(\ref{apb05}), (\ref{apb10}), and (\ref{bi09a}), respectively, the expression for the X-factor
can have the specific form given by (\ref{apb11}). 

A few examples of spherically symmetric clumps are examined in the following subsections.

\subsection{Uniform-Density Sphere\label{appb1}}

This case is also called the ``Hard Sphere'' in Section~\ref{sec3}, the tables, and in MSH84.  
Because the hard sphere has a well-defined edge with a well-defined radius, $R$, this 
$R$ is used in place of $\ph$.  The density, $\rho(r)$, is constant and equal to both $\rho_c$ and 
$\rb$. Similarly $n(r)\equiv n_c$ and $n(r)\equiv\nb$ for all $r$ from zero up to $R$.  Accordingly, 
it is trivial to show that $k_M = {4\pi\over 3}$, $k_W = {3\over 5}$, and $k_N = 1$. It then follows 
that expressions~(\ref{bi14}), (\ref{bi22x}), and (\ref{bi29}) result from (\ref{apb09}), (\ref{apb10}), 
and (\ref{apb11}), respectively.  The central sightline optical depth, $\tau_0$, is then,
\begin{equation}
\tau_0 = 1.18\times 10^2\, n_c^{0.5}\,\Tk^{-1.75}\quad ,
\label{apb12oo}
\end{equation}
for $\Tk=10$ to 20$\,$K.   These values along with the $k_A$ and $\epsilon$ values given in 
Subsection~\ref{ssec32} yield,
\begin{equation}
X_f(X_{20}) = 0.488\, C_T\,\hbox{$\Tk^{-0.76}$}\,\nb^{0.43}
\quad ,
\label{apb12o}
\end{equation}
where this X-factor is in units of $X_{20}$ or $10^{20}\, H_2\ molecules\cdot\cKkms$.

Specific numerical results of these calculations are listed in Tables~\ref{tbl-1} through
\ref{tbl-5}.

\subsection{Gaussian Sphere\label{appb2}}

The optical depth profile, $\tau(p)$, for the Gaussian sphere is given by (~\ref{bi09}).
For an isothermal clump in LTE, the volume density, $\rho(r)$, that would give this $\tau(p)$ 
is Gaussian:
\begin{equation}
\rho(r) = \rho_0\, exp\left(-{r^2\over\sigma_r^2}\right)\quad ,
\label{apb12}
\end{equation}
where $\rho_0$ is the density at the center of the sphere and $\sigma_r$ is the (1/e)-folding
radius for the density variation.  The analogous expression exists for $n(r)$ and placing this
in (\ref{apb01}) results in
\begin{equation}
N_c(p) = n_0\,\sqrt{\pi}\, exp\left(-{p^2\over\sigma_r^2}\right)\quad .
\label{apb13}
\end{equation}
so that $N_c = n_0\, \sqrt{\pi}\, \sigma_r$.  From (\ref{apb13}) we see that $\sigma_r = 
\ph/\sqrt{ln\,2}$.  The mass as a function of $r$ is
\begin{eqnarray}
{M_c(r)\over M_c(\infty)} &=& erf\left({r\over\sigma_r}\right) 
- {2r\over\sigma_r\sqrt{\pi}}\,exp\left(-{r^2\over\sigma_r^2}\right) \quad ,
\label{apb14}\\
\noalign{\noindent where,}
M_c(\infty) &=& \pi^{3\over 2}\, \rho_0\, \sigma_r^3 \quad .
\label{apb15}
\end{eqnarray}
The $erf$ is the Gaussian error function.  Replacing $r$ with $\ph$ in (\ref{apb14})
yields $M_c(\ph)/M_c(\infty) = 0.2878$.  Placing this into (\ref{apb03}) gives us
$\rb = 0.6630\,\rho_0$.  Accordingly, $k_N = 1.606$, $k_M = 14.55$.   Using 
expressions~(\ref{apb04}) and (\ref{apb07}) gives $k_W = \sqrt{(ln\,2)/2\,\pi} = 0.3321$, 
which in turn gives $k_v = 3.420\times 10^{16} cgs$.  The central sightline optical depth, 
$\tau_0$, is accordingly,
\begin{equation}
\tau_0 = 1.38\times 10^2\, \nb^{0.5}\,\Tk^{-1.75}\quad ,
\label{apb16o}
\end{equation}
for $\Tk=10$ to 20$\,$K. These together with the $k_A$ and $\epsilon$ values given in 
Subsection~\ref{ssec33} yield,
\begin{equation}
X_f(X_{20}) = 0.204\, C_T\,\hbox{$\Tk^{-0.38}$}\,\nb^{0.32}
\quad .
\label{apb16}
\end{equation}

Specific numerical results of these calculations are listed in Tables~\ref{tbl-1} through
\ref{tbl-5}.

\subsection{Squared-Lorentzian Sphere\label{appb3}}

This case was discussed in MSH84.  The optical depth profile is given by
\begin{equation}
\tau(p) = \tau_0(0)\,\left[1+\left({p^2\over r_0^2}\right)\right]^{-2}\quad ,
\label{apb17}
\end{equation}
where the $r_0$ is the projected radius at which $\tau$ falls to $0.25\,\tau_0$.
It also gives the effective area, where $\aef = \pi\, r_0^2$.  For an isothermal
clump in LTE, it follows that
\begin{equation}
N_c(p) = N_c(0)\,\left[1+\left({p^2\over r_0^2}\right)\right]^{-2}\quad .
\label{apb18}
\end{equation}
It is easy to show that $r_0 = (\sqrt{2} - 1)^{-0.5}\,\ph$.   The volume density,
$\rho(r)$, that corresponds to the surface density of equation~(\ref{apb18}) is
\begin{equation}
\rho(r) = \rho_0\,\left[1+\left({r^2\over r_0^2}\right)\right]^{-2.5}\quad .
\label{apb19}
\end{equation}
This is easily demonstrated by substituting the expression analogous to (\ref{apb19})
for $n_c(r)$ into (\ref{apb01}).  Doing so gives,
\begin{equation}
N_c = {4\over 3}\, n_0\, r_0\quad .
\label{apb20}
\end{equation}
Substituting (\ref{apb19}) into (\ref{apb02}) results in,
\begin{eqnarray}
{M_c(r)\over M_c(\infty)} &=& \left[\left({r^2\over r_0^2}\right)+1\right]^{-1.5} \quad ,
\label{apb21}\\
\noalign{\noindent where,}
M_c(\infty) &=& {4\pi\over 3}\, \rho_0\, r_0^3 \quad .
\label{apb22}
\end{eqnarray}
Applying (\ref{apb03}) yields $\rb = 2^{-0.75}\,\rho_0$.  Consequently, 
\begin{eqnarray}
k_N &=& {4\over 3}\,\left(2-\sqrt{2}\right)^{-0.5}
\label{apb23}\\
\noalign{\medskip}
    &=& 1.742
\nonumber\\
\noalign{\noindent and}
k_M &=& {4\pi\over 3}\,\left(2^{0.75}\right)\,\left(\sqrt{2} - 1\right)^{-1.5}
\label{apb24}\\
\noalign{\medskip}
    &=& 26.43 \quad .
\nonumber
\end{eqnarray}
Plugging (\ref{apb19}) into (\ref{apb04}) and doing some work yields
\begin{eqnarray}
k_W &=& {3\pi\over 32}\left(\sqrt{2}-1\right)^{0.5}
\label{apb25}\\
\noalign{\medskip}
    &=& 0.1896 \quad .
\nonumber
\end{eqnarray}
From these the following is obtained:
\begin{eqnarray}
k_v &=& \left[{\pi^2\over 96 \left(\sqrt{2}-1\right)}\, \left(2^{0.75}\right)
\,\mu\,\mh\, G\right]^{0.5}
\label{apb26}\\
\noalign{\medskip}
    &=& 3.481\times 10^{-16} cgs \quad .
\nonumber
\end{eqnarray}

The above results give the central sightline optical depth: 
\begin{equation}
\tau_0 = 1.46\times 10^2\, \nb^{0.5}\,\Tk^{-1.75}\quad ,
\label{apb27o}
\end{equation}
for $\Tk=10$ to 20$\,$K.  All of the above results along with the $k_A$ and $\epsilon$ 
values given in Subsection~\ref{ssec33} give
\begin{equation}
X_f(X_{20}) = 0.0756\, C_T\,\hbox{$\Tk^{-0.003}$}\,\nb^{0.22}
\quad .
\label{apb27}
\end{equation}

Specific numerical results of these calculations are listed in Tables~\ref{tbl-1} through
\ref{tbl-5}.

\subsection{${\bf\rho\propto r^{-2}}$ Sphere\label{appb4}}

The adopted sphere density, $\rho(r)$, is flat for $r\leq r_0$ and falls like $r^{-2}$ outside of 
this and out to $r=r_1$:
\begin{eqnarray}
\rho(r) &=& \rho_0 \qquad\qquad\ for\ r\leq r_0
\label{apb28}\\
        &=&\rho_0 \left({r_0\over r}\right)^2 \quad for\ r_0\leq r\leq r_1\qquad .
\label{apb29}
\end{eqnarray}
It is easy to show that the mass is given by
\begin{equation}
M_c(r_1) = {4\pi\over 3}\, \rho_0\, r_0^3 \left(3{r_1\over r_0}\ -\ 2\right)\qquad .
\label{apb30}
\end{equation}
Notice that the mass, $M_c(r_1)$, would diverge if $r_1\to\infty$, so a maximum radius, $r_1$
must be specified.  The usual quantities (e.g., for $k_v$ and $k_N$) must now be determined in
terms of $M_c(r_1)$ instead of $M_c(\infty)$.  Employing expressions~(\ref{apb01}) and the number 
density analogs of (\ref{apb28}) and (\ref{apb29}) yields
\begin{eqnarray}
N_c(p) &=& 2 n_0\sqrt{r_0^2 - p^2}\ + {2\ n_0\, r_0^2\over p}\,\left[arccos\left({p\over r_1}\right)
 - arccos\left({p\over r_0}\right)\right]\ for\ p\leq r_0
\label{apb31}\\
       &=& \pi\,\ n_0\, r_0^2\, p^{-1}\, arccos\left({p\over r_1}\right)\qquad\qquad\qquad
       \qquad\qquad\qquad for\ r_0\leq p\leq r_1 ,
\label{apb32}
\end{eqnarray}
where $N_c(0) = 2\, n_0\, r_0\, (2\ -\ {r_0\over r_1})$. 

Determining the self-potential, $W$, and applying the simple form of the Virial theorem leads to
\begin{equation}
\Dvc^2 = 4\,\pi\, G\,\rho_0\, r_0^2\ \ {\strut {r_1\over r_0}\ -\ {2\over 3}
	\ ln\left({r_1\over r_0}\right)\ -\ {14\over 15}\over \strut 3{r_1\over r_0}\ -\ 2}\qquad .
\label{apb33}
\end{equation}
For simplicity, ${r_1\over r_0} \gg 1$ is assumed and
\begin{equation}
\Dvc = \left[{4\,\pi\over 3}\, G\,\mu\,\mh\right]^{0.5}\ n_0^{0.5}\ r_0\qquad ,
\label{apb34}
\end{equation}
where $\rho_0 = n_0\,\mu\,\mh$ was used.  It can be shown that 
\begin{equation}
r_0 = {2\over\pi}\ \ph\qquad 
\label{apb35}
\end{equation}
and 
\begin{eqnarray}
\rho_0 &=& {\pi^3\over 4\,(3\pi\ -\ 4)}\ \ \rb\qquad , 
\label{apb36}\\
\noalign{\medskip}
&=& 1.429\ \rb
\nonumber
\end{eqnarray}
Expressions~(\ref{apb35}) and (\ref{apb36}) again assume that ${r_1\over r_0}\gg 1$.  With
these expressions in hand, expressions for $k_N$ and $k_v$ follow:
\begin{eqnarray}
k_N &=& {\pi^2\over 3\pi\ -\ 4}\qquad , 
\label{apb37}\\
\noalign{\medskip}
&=& 1.819 
\nonumber
\end{eqnarray}
and 
\begin{eqnarray}
k_v &=& \left[{\pi^2\over 3\,(3\pi\ -\ 4)}\, G\,\mu\,\mh\right]^{0.5}\qquad , 
\label{apb38}\\
\noalign{\medskip}
&=& 4.20\times 10^{-16}\ cgs\qquad . 
\nonumber
\end{eqnarray}

Computing $X_f$ requires specifying $r_1/r_0$.  The particular example discussed in Section~\ref{sssec471}
is of the Rosette Molecular Cloud, where \citet{Williams95} find ${r_1\over r_0} = 5$ for many of the
clumps.  Simple numerical integration yields $\epsilon = 0.30$ and $k_A = 1.80$ for this value of 
${r_1\over r_0}$.  Therefore, 
\begin{equation}
X_f(X_{20}) = 0.22\ C_T\ \Tk^{-0.47}\ \nb^{0.35}
\label{apb39}
\end{equation}

Because the $r^{-2}$ sphere must have an outer radius, the average density can also be defined as
over the entire volume --- i.e., $n_a$.  Doing this changes $k_N$ and $k_v$:
\begin{equation}
k_N = 2\, {r_1\over r_0}\qquad , 
\label{apb40}
\end{equation}
and 
\begin{eqnarray}
k_v &=& \left[{\pi\over 9}\, G\,\mu\,\mh\right]^{0.5}\qquad , 
\label{apb41}\\
\noalign{\medskip}
&=& 3.18\times 10^{-16}\ cgs\qquad . 
\nonumber
\end{eqnarray}
For the clumps of the Rosette Molecular Cloud, again using the numbers of the previous
paragraph, we have,
\begin{equation}
X_f(X_{20}) = 0.88\ C_T\ \Tk^{-0.47}\ n_a^{0.35}
\label{apb42}
\end{equation}

\section{Treatment of Cylindrically Symmetric Clumps\label{appc}}

\subsection{Viewed Perpendicularly to the Axis of Symmetry: The Side-on Case\label{appc1}}

Here the mass density $\rho$ is adopted to be a function of the radius from an axis 
of symmetry, $r$, within each cylindrical clump: $\rho = \rho(r)$.   This is also 
true for the number density, $n(r)$.  Again, the $x$-$y$ plane is chosen to be 
perpendicular to the observer's sightline and $z$ is measured along the sightline.
The $x$-axis is selected to be along the cylinder's symmetry axis and is also
perpendicular to the observer's sightline.  The $y$-axis is perpendicular to both the
observer's sightline and the symmetry axis; the $y$ is the projected distance from
the cylinder's central axis.  The radius, $r$, is then $y^2+z^2$.  The cylinder's length
is $h$.  The clump column density as a function of projected distance from the central 
axis, $N_c(y)$, is given by
\begin{equation}
N_c(y) = \int_{-\infty}^{+\infty}\back\back dz\ n(\sqrt{y^2+z^2}) \quad .
\label{apc01}
\end{equation}
The mass within radius $r$, $M(r)$, is given by
\begin{equation}
M_c(r) = 2\pi\ h \int_0^r\back d\rs\ \rs\,\rho(\rs) \quad .
\label{apc02}
\end{equation}
The half-width-at-half-maximum distance from the symmetry axis, $\yh$, is defined by 
$N_c(\yh)\equiv 0.5\, N_c(y=0)$.  The average density, $\rb$, is defined analogously
to that of expression~(\ref{apb03}):
\begin{equation}
\rb = {M_c(r=\yh)\over \pi\,\yh^2\, h} \quad .
\label{apc03}
\end{equation}
The self-potential due to gravity is
\begin{equation}
W = - k_F\ {G M_c^2\over h} \quad ,
\label{apc04}
\end{equation}
where $k_F = 1$ in the limit $h/\Dh\gg 1$ (and $\Dh = 2\yh$).  Equation~(\ref{apc04})
in this limit is found from solving the Poisson equation in cylindrical coordinates. 
The parameter $k_F$ is a correction factor for when $h/\Dh\gg 1$ is false; a crude 
numerical analysis suggests that for $h/\Dh\ge 1$, $k_F < 1.4$.  A similar value is
found from a very crude analytical approach.  If the cylinder is of uniform density
with radius, $R$, and has length, $h = 2R$, then the cylinder very roughly approximates
a uniform density sphere of radius, $R$.  From Appendix~\ref{appb1}, we know that 
$k_W = 0.6$.  Placing $h=2R$ into (\ref{apc04}) and comparing with the uniform sphere,
we find that $k_F\simeq 1.2$.  In any event, the final dependence of the relevant 
quantities on $k_F$ will be weak and setting this to unity will be sufficient (see
below).  

Many parameters have similar or identical expressions to those in Appendix~\ref{appb}.
The parameter $k_N$ is defined similarly to that in equation~(\ref{apb05}), but with 
$\yh$ in place of $\ph$.   The $k_M$ is defined in terms of a cylindrical variation of 
equation~(\ref{apb06}):
\begin{equation}
M_c \equiv k_M\ \rb\ h\ \yh^2 \quad .
\label{apc05}\\
\end{equation}
The parameter $k_W$ does not apply in this case because $W$ is not dependent on $r$, except 
for the weak implicit $r$-dependence of $k_F$.  The expression for $\Dvc$ is the same as
(\ref{apb08}) and the expression for $k_v$ is almost the same as (\ref{apb09}):
\begin{equation}
k_v  = \left[{1\over 12}\ k_F\ k_M\  \mu\,\mh\, G\right]^{1\over 2}\quad .
\label{apc06}
\end{equation}
The equations for $\tau_0$ and $X_f$ are still (\ref{apb10}) and (\ref{apb11}). 
Given that $X_f\propto k_v^{\epsilon-1}$ and that $k_v\propto k_F^{0.5}$, $X_f\propto 
k_F^{0.5(\epsilon-1)}$.   Even for a value of $\epsilon$ as low as that in a hard 
sphere, the dependence of $X_f$ on $k_F$ is weak: $X_f\propto k_F^{0.43}$.  Since 
$k_F$ is between 1.0 and 1.4, the effect on $X_f$ is less than 16\%.

\subsubsection{Gaussian Cylinder\label{apppc1}}

The volume density, $\rho(r)$, is given by,
\begin{equation}
\rho(r) = \rho_0\, exp\left(-{r^2\over\sigma_r^2}\right)\quad .
\label{apc07}
\end{equation}
This is apparently identical to (\ref{apb12}), except that the $r$ and $\sigma_r$ here are 
defined as distances from the central axis rather than a central point.  From (\ref{apc07}),
it is easy to show that,
\begin{eqnarray}
N_c(y) &=& N_c(0)\ exp\left(-{y^2\over\sigma_r^2}\right) \quad ,
\label{apc08}\\
\noalign{\noindent where,}
N_c(0) &=& \sqrt{\pi}\ n_0\ \sigma_r \quad ,
\label{apc09}\\
\noalign{\noindent and,}
M_c(r) &=& M_c(\infty)\left[1\ -\ exp\left(-{r^2\over\sigma_r^2}\right)\right] \quad ,
\label{apc10}\\
\noalign{\noindent where,}
M_c(\infty) &=& \pi\ \rho_0\ h\ \sigma_r^2 \quad .
\label{apc11}
\end{eqnarray}
From the above expressions, we see that $\yh =\sigma_r\,\sqrt{ln\, 2}$, $\rb = \rho_0/(2\ ln\,2)$,
$k_N=\sqrt{\pi\ ln\,2}$, $k_M = 2\pi$, yielding,
\begin{eqnarray}
k_v &=& \left[{\pi\over 6}\, k_F\,\mu\,\mh\, G\right]^{0.5} \quad ,
\label{apc12}\\
\noalign{\medskip\noindent and adopting $k_F = 1$,}
    &=& 3.899\times 10^{-16} cgs \quad .
\nonumber
\end{eqnarray}
The above results give the central sightline optical depth: 
\begin{equation}
\tau_0 = 1.10\times 10^2\, \nb^{0.5}\,\Tk^{-1.75}\quad ,
\label{apc13o}
\end{equation}
for $\Tk=10$ to 20$\,$K.  These results along with the $k_A$ and $\epsilon$ values given in 
Subsection~\ref{ssec34} give
\begin{equation}
X_f(X_{20}) = 0.309\, C_T\,\hbox{$\Tk^{-0.56}$}\,\nb^{0.37}
\quad .
\label{apc13}
\end{equation}

The numerical results of the above calculations are listed in Tables~\ref{tbl-1} through
\ref{tbl-5}.

\subsection{Viewed Along the Axis of Symmetry: The End-on Case\label{appc2}}

The mass density is again of the form $\rho = \rho(r)$, but with the sightline along
the axis of symmetry.  The $x$-$y$ plane is still chosen to be perpendicular to the 
observer's sightline and $z$ is measured along the sightline and along the cylinder's
symmetry axis.  The radius, $r$, is again the distance from the central axis, but this 
is now $x^2+y^2$.  The cylinder's length again is $h$.  The clump column density as a 
function of projected distance from the central axis, $N_c(r)$, is given by
\begin{equation}
N_c(r) = n_c(r)\ h \quad ,
\label{apc14}
\end{equation}
which is because $n_c(r)$ has no $z$-dependence. The HWHM radius, $\rh$, is then found
from
\begin{equation}
n_c(\rh) \equiv 0.5\ n_c(0)\quad .
\label{apc15}
\end{equation}
The expressions for $M_c(r)$, $k_M$, $\rb$, and $k_N$ are analogous to those for the side-on 
case --- i.e., (\ref{apc02}), (\ref{apc05}), (\ref{apc03}), and (\ref{apb05}) --- but with $\rh$ 
in place of $\yh$ or $\ph$.  Combining this last with (\ref{apc14}) yields,
\begin{equation}
k_N =  {h\over\Dh} {n_c(0)\over\nb} \quad ,
\label{apc16}
\end{equation}
where $\Dh\equiv 2\,\rh$.  The above states that $k_N\propto h/\Dh$; i.e., $k_N$ is proportional
to the cylinder aspect ratio. 

Other expressions are identical to those mentioned previously: $k_v$ is still given by
(\ref{apc06}), $\Dvc$ by (\ref{apb08}), $\tau_0$ by (\ref{apb10}), and $X_f$ by (\ref{apb11}).
The dependence of $X_f$ on $k_F$ is the same as in the side-on case.  The dependence of $X_f$
on $k_N$ implies there is now a dependence on the cylinder aspect ratio: 
$X_f\propto (h/\Dh)^{(1-\epsilon)}$.

\subsubsection{Gaussian Cylinder\label{apppc2}}

The Gaussian cylinder has been treated in the side-on case, so there are similarities in
this end-on case.  The relationship between $\rh$ and $\sigma_r$ is the same as that between
$\yh$ and $\sigma_r$.  The relationship between $\rb$ and $\rho_0$ is unchanged. $k_M$ and
$k_v$ are also unchanged.  One important change is the expression for $k_N$:
\begin{equation}
k_N =  2 (ln\, 2){h\over\Dh} \quad .
\label{apc17}
\end{equation}
These results give $\tau_0$: 
\begin{equation}
\tau_0 = 1.04\times 10^2\, \nb^{0.5}\,\Tk^{-1.75}\ \left({h\over\Dh}\right)\quad ,
\label{apc18o}
\end{equation}
for $\Tk=10$ to 20$\,$K.  Another difference is that the end-on Gaussian cylinder looks like 
the Gaussian sphere. Using the $k_A$ and $\epsilon$ for the Gaussian sphere results in,
\begin{equation}
X_f(X_{20}) = 0.171\, C_T\,\hbox{$\Tk^{-0.38}$}\,\nb^{0.32}\ \left({h\over\Dh}\right)^{0.64}
\quad .
\label{apc18}
\end{equation}

The numerical results of the above calculations are listed in Tables~\ref{tbl-1} through
\ref{tbl-5}.

\subsection{Collapsing, Magnetized Filament\label{appc3}}

\citet{Tilley03} examined the case of a constant toroidal flux-to-mass ratio in a collapsing
cylindrical cloud.  Here we estimate $X_f$ for the side-on case only.  Their equation~(35) 
can be rearranged to give the velocity width:
\begin{equation}
{\Dvc} = 2.89\times 10^{-16}\ n_0^{0.5}\ \lambda_{frag}\ k_{max}\ \ cgs\quad ,
\label{apc19}
\end{equation}
where $\lambda_{frag}$ is the fragmentation wavelength and $k_{max}$ is a dimensionless
wavenumber \citep[see][]{Tilley03}.  They find that the density goes like $r^{-2}$ for a 
strong magnetic field and $r^{-4}$ for a weak magnetic field.  So, assuming a roughly 
constant density inside radius, $r_0$,
\begin{eqnarray}
\rho(r) &=& \rho_0 \qquad\qquad\ for\ r\leq r_0
\label{apc20}\\
        &=&\rho_0 \left({r_0\over r}\right)^\alpha \quad for\ r\geq r_0 ,
\label{apc21}
\end{eqnarray}
where $\alpha = 2$ or 4.  Combining equation~(\ref{apc01}) with the number density
analogs of (\ref{apc20}) and (\ref{apc21}),
\begin{eqnarray}
N_c(y) &=& 2\ n_0\sqrt{r_0^2 - y^2}\ +\ 2\ n_0\, r_0^\alpha\,\int\limits_{\sqrt{r_0^2-y^2}}^\infty
\back\back dz \left(z^2+y^2\right)^{-\alpha/2}\quad for\ y\leq r_0
\label{apc22}\\
       &=& 2\ n_0\, r_0^\alpha\,\int\limits_0^\infty dz \left(z^2+y^2\right)^{-\alpha/2} \qquad 
       \qquad\qquad\qquad\qquad  for\ y\geq r_0.
\label{apc23}
\end{eqnarray}
Combining equation~(\ref{apc02}) with (\ref{apc22}) and (\ref{apc23}) results in
\begin{equation}
M_c(r) = \pi\, h\, r_0^2\, \rho_0\ +\ 2\,\pi\, h\,\rho_0\, r_0^\alpha\,\int_{r_0}^r d\rs\ \rs^{(1-\alpha)}
\quad .
\label{apc24}
\end{equation}
Now we will examine the $\alpha = 2$ and $\alpha = 4$ cases separately.

\subsubsection{$\alpha = 2$\label{apppc31}}

For $\alpha =2$ the mass diverges when $r\to\infty$, so we must specify an outer radius, $r_1$.
The upper limits in equations~(\ref{apc22}) and (\ref{apc23}) are $r_1$ instead of $\infty$.
But, for simplicity, we will assume that $r_1\gg r_0$.  Equations~(\ref{apc22}) and (\ref{apc23}) 
become
\begin{eqnarray}
N_c(y) &=& 2 n_0\sqrt{r_0^2 - y^2}\ + {2\ n_0\, r_0^2\over y}\,\left[arccos\left({y\over r_1}\right)
 - arccos\left({y\over r_0}\right)\right]\ for\ y\leq r_0
\label{apc25}\\
       &=& \pi\,\ n_0\, r_0^2\, y^{-1}\, arccos\left({y\over r_1}\right)\qquad\qquad\qquad
       \qquad\qquad\qquad for\ r_0\leq y\leq r_1 .
\label{apc26}
\end{eqnarray}
It is easy to show that $N_c(0) = 4\, n_0\, r_0$ and, accordingly, that $\yh = {\pi\over 2}\, r_0$,
where it is assumed that $r_1\gg\yh$.  Equation~(\ref{apc24}) becomes
\begin{equation}
M_c(r) = \pi\, h\, r_0^2\, \rho_0\ \left[1\ +\ 2\ ln\left({r\over r_0}\right)\right] \quad .
\label{apc27}
\end{equation}
The maximum radius, $r_1$, must be specified and the total mass, $M_c(r_1)$, is dependent on the 
ratio of the outer-to-inner radii of the $r^{-2}$ region, $r_1/r_0$.  Expression~(\ref{apc03}) now 
gives us $\rb$:
\begin{eqnarray}
\rb &=& \left({2\over\pi}\right)^2\ \left[1\ +\ 2\ ln\left({\pi\over 2}\right)\right]\ \rho_0 \quad ,
\label{apc28}\\
\noalign{\bigskip}
    &=& 0.771 \rho_0
\nonumber\\
\noalign{\noindent With this information in hand (and assuming $r_0/r_1$ to be small) it is easy to 
demonstrate that\bigskip}
k_N &=& {\pi\over 1\ +\ 2\ ln\left({\pi\over 2}\right)} \quad 
\label{apc29}\\
\noalign{\bigskip}
    &=& 1.651\qquad.
\nonumber\\
\noalign{\noindent We can also determine $k_M$, but it is only useful for estimating $\Dvc$ for 
virialized gas.  Also, it is normally defined in terms of $M_c(\infty)$.  Nevertheless, for 
completeness, it is given here for $M_c(r_1)$:}
k_M &=& \pi\ {\ 1\ +\ 2\ ln\left({r_1\over r_0}\right)\over 1\ +\ 2\ ln\left({\pi\over 2}\right)} \quad 
\label{apc30}
\end{eqnarray}
Expression~(\ref{apc19}) is rewritten as
\begin{eqnarray}
{\Dvc} &=& k_v\ \nb^{0.5}\ \lambda_{frag}\ k_{max} \quad ,
\label{apc31}\\
\noalign{\noindent where,}
k_v &=& 3.294\times 10^{-16}\ cgs
\nonumber
\end{eqnarray}
Starting with equation~(\ref{bi22}) and plugging in (\ref{apc31}) and $N_c(0)$ expressed
in terms of $k_N$ yields,
\begin{equation}
\tau_0 = {k_\tau\, k_N\over k_v\, k_{max}\,\sqrt{2\pi}}\ \, \nb^{0.5}\,\Tk^{-\gamma}\ 
\left({\Dh\over\lambda_{frag}}\right)\quad ,
\label{apc32}
\end{equation}
where, again, $\Dh = 2\yh$.

Now we can combine the expressions developed here with (\ref{bi25}), (\ref{bi26}), and 
(\ref{bi09a}) to obtain,
\begin{equation}
X_f = (2\pi)^{{1\over 2}(\epsilon - 1)}\, C_T\, k_A^{-1}\,k_\tau^{-\epsilon}
\,k_v^{\epsilon - 1}\, k_N^{1-\epsilon}\,\hbox{$\Tk^{\gamma\epsilon - 1}$}
\,\nb^{{1\over 2}(1-\epsilon)}\ \left({\Dh\over\lambda_{frag}\ k_{max}}\right)^{1-\epsilon}
\quad .
\label{apc33}
\end{equation}
Except for the extra factor of $[\Dh/(\lambda_{frag}\ k_{max})]^{1-\epsilon}$, this is 
identical to the more-or-less general expression for $X_f$, i.e., equation~(\ref{apb11}). 
Evaluating (\ref{apc33}) numerically depends on the values of $k_A$ and $\epsilon$, which
depend on $r_1/r_0$.  Here we try two values for this ratio: $r_1/r_0 = 10$ and 
$r_1/r_0 = 1000$.  For $r_1/r_0 = 10$, $k_A = 1.74$ and $\epsilon = 0.27$ and,
\begin{eqnarray}
X_f(X_{20}) &=& 0.97\ C_T\,\hbox{$\Tk^{-0.53}$}\,\nb^{0.37}
\ \left({\Dh\over\lambda_{frag}}\right)^{0.74} \quad .
\label{apc34}\\
\noalign{\noindent For $r_1/r_0 = 1000$, $k_A = 1.46$, $\epsilon= 0.69$ and,} 
X_f(X_{20}) &=& 0.0729\ C_T\,\hbox{$\Tk^{+0.21}$}\,\nb^{0.16}
\ \left({\Dh\over\lambda_{frag}}\right)^{0.31}
\quad ,
\label{apc35}
\end{eqnarray}
in which $k_{max}=0.2$ was adopted for both cases \citep[see][]{Tilley03}.

\subsubsection{$\alpha = 4$\label{apppc32}}

Equations~(\ref{apc22}) and (\ref{apc23}) become
\begin{eqnarray}
N_c(y) &=& 2\ n_0\sqrt{r_0^2 - y^2}\ +\ ...
\nonumber\\
&&\ +\ n_0\, r_0^4\, y^{-3}\left[arcsin\left({y\over r_0}\right)\ -
\ \left({y\over r_0}\right)\, \left(1\, -
\, {y^2\over r_0^2}\right)^{1\over 2}\right]
\quad for\ y\leq r_0
\label{apc36}\\
\noalign{\bigskip}
       &=& {\pi\over 2}\,\ n_0\, r_0^4\, y^{-3} 
       \ \qquad\qquad\qquad\qquad\qquad\qquad\qquad\qquad\qquad 
       for\ y\geq r_0 .
\label{apc37}
\end{eqnarray}
It is easy to show that $N_c(0) = {8\over 3}\, n_0\, r_0$ and, accordingly, that 
$\yh = ({3\pi\over 8})^{1\over 3}\, r_0$.  Equation~(\ref{apc24}) becomes
\begin{eqnarray}
M_c(r) &=& M_c(\infty)\left(1\ -\ {r^2\over 2\, r_0^2}\right) \quad ,
\label{apc38}\\
\noalign{\noindent where,}
M_c(\infty) &=& 2\, \pi\ h\ r_0^2\ \rho_0 \quad .
\label{apc39}
\end{eqnarray}
Unlike the $\alpha = 2$ case, the $\alpha = 4$ case has finite mass, even for
$r/r_0\to\infty$. Expression~(\ref{apc03}) gives us $\rb$:
\begin{eqnarray}
\rb &=& \left({8\over 3\,\pi}\right)^{2\over 3}\ 
\left[1\ -\ {1\over 2}\ \left({8\over 3\,\pi}\right)^{2\over 3}\right]\ \rho_0 \quad ,
\label{apc40}\\
\noalign{\bigskip}
    &=& 0.495 \rho_0
\nonumber\\
\noalign{\noindent It is now easy to demonstrate that\bigskip}
k_N &=& {2\,\pi\over \left(3\,\pi\right)^{2\over 3}\ -\ 2} \quad 
\label{apc41}\\
\noalign{\bigskip}
    &=& 2.552
\nonumber\\
\noalign{\noindent For completeness, $k_M$ is also given:}
k_M &=& {2\,\pi\over 1\ -\ {1\over 2}\left({8\over 3\,\pi}\right)^{2\over 3}} \quad 
\label{apc42}
\end{eqnarray}
With $k_N$ known, we have $\Dvc$ given by (\ref{apc31}) and 
\begin{equation}
k_v = 4.114\times 10^{-16}\ cgs \quad .
\nonumber
\end{equation}

The equation for $X_f$ is (\ref{apc33}).  Numerically integrating (\ref{bi02}) and
(\ref{bi03}) gives $k_A = 1.15$ and $\epsilon = 0.37$.  Adopting $k_{max}=0.2$ and 
substituting the values for $k_A$ and $\epsilon$ into (\ref{apc33}) results in
\begin{equation}
X_f(X_{20}) = 0.900\ C_T\,\hbox{$\Tk^{-0.36}$}\,\nb^{0.32}
\ \left({\Dh\over\lambda_{frag}}\right)^{0.64} \quad .
\label{apc43}
\end{equation}




\clearpage


\begin{figure}
\epsscale{0.64}
\plotone{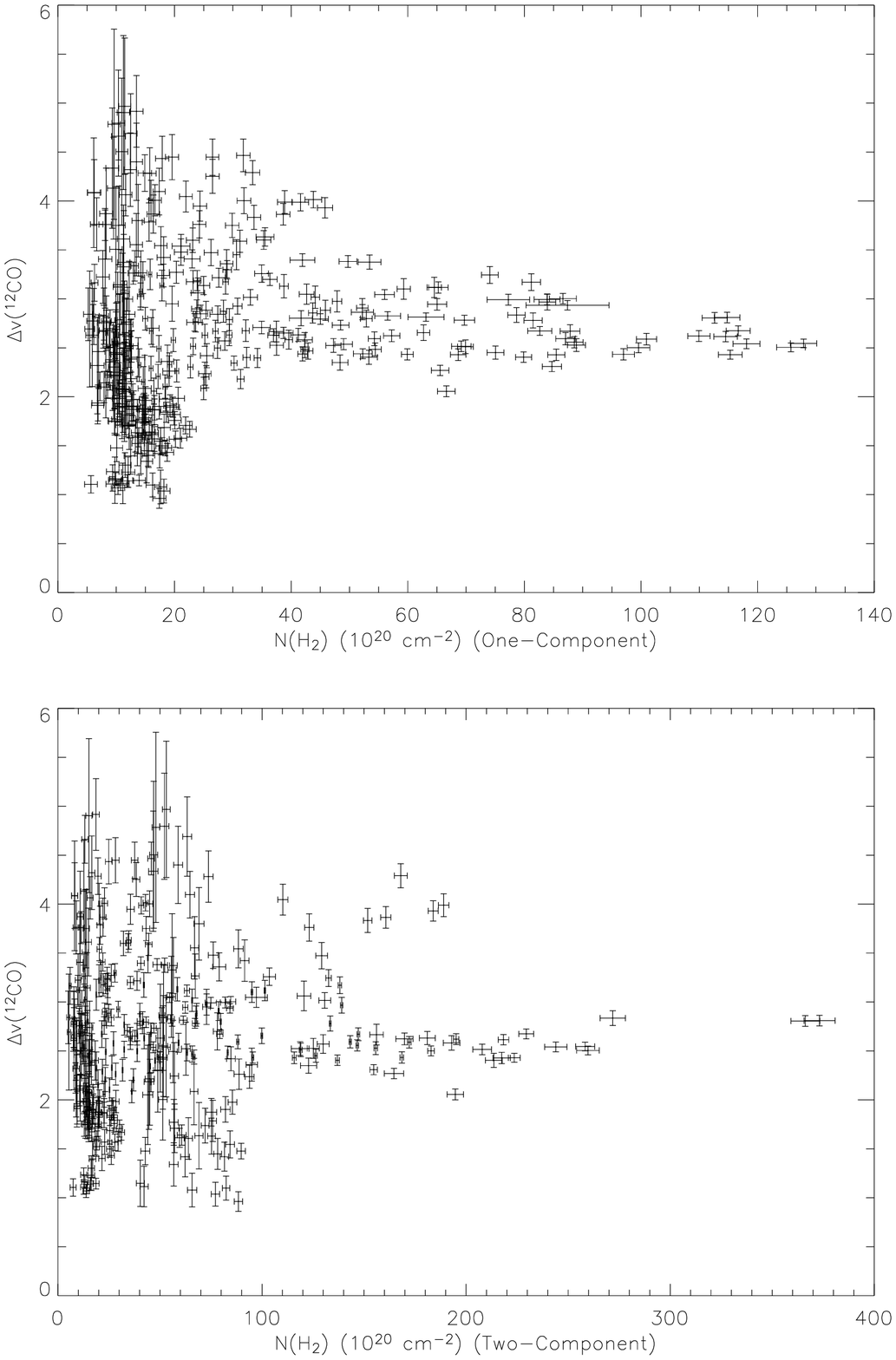}
\caption{The panels above show the velocity widths, $\Delta v(\CO)$, of the 
$\COone$ line for the Orion$\,$A and B molecular clouds versus the molecular
hydrogen column densities, N(H$_2$).  The velocity widths are effective
velocity widths given by the ratio of the velocity-integrated radiation 
temperature, I(CO), divided by the peak radiation temperature of the line.  
The N(H$_2$) values for the upper panel are those determined from the 
one-component, non-LTE models of \citet{W06}.  The N(H$_2$) values for the
lower panel are the two-component, two-subsample, non-LTE models of 
\citet{W06}.  The sample of points are those for which the intensities
are 5-$\sigma$ or more for the 140$\um$ and 240$\um$ continuum, the
$\COone$ line, and $\cOone$ line.  All these maps were convolved to
1-degree resolution \citep[see][for details]{W06}.
\label{fig01}}
\end{figure}

\clearpage

\begin{figure}
\epsscale{0.75}
\plotone{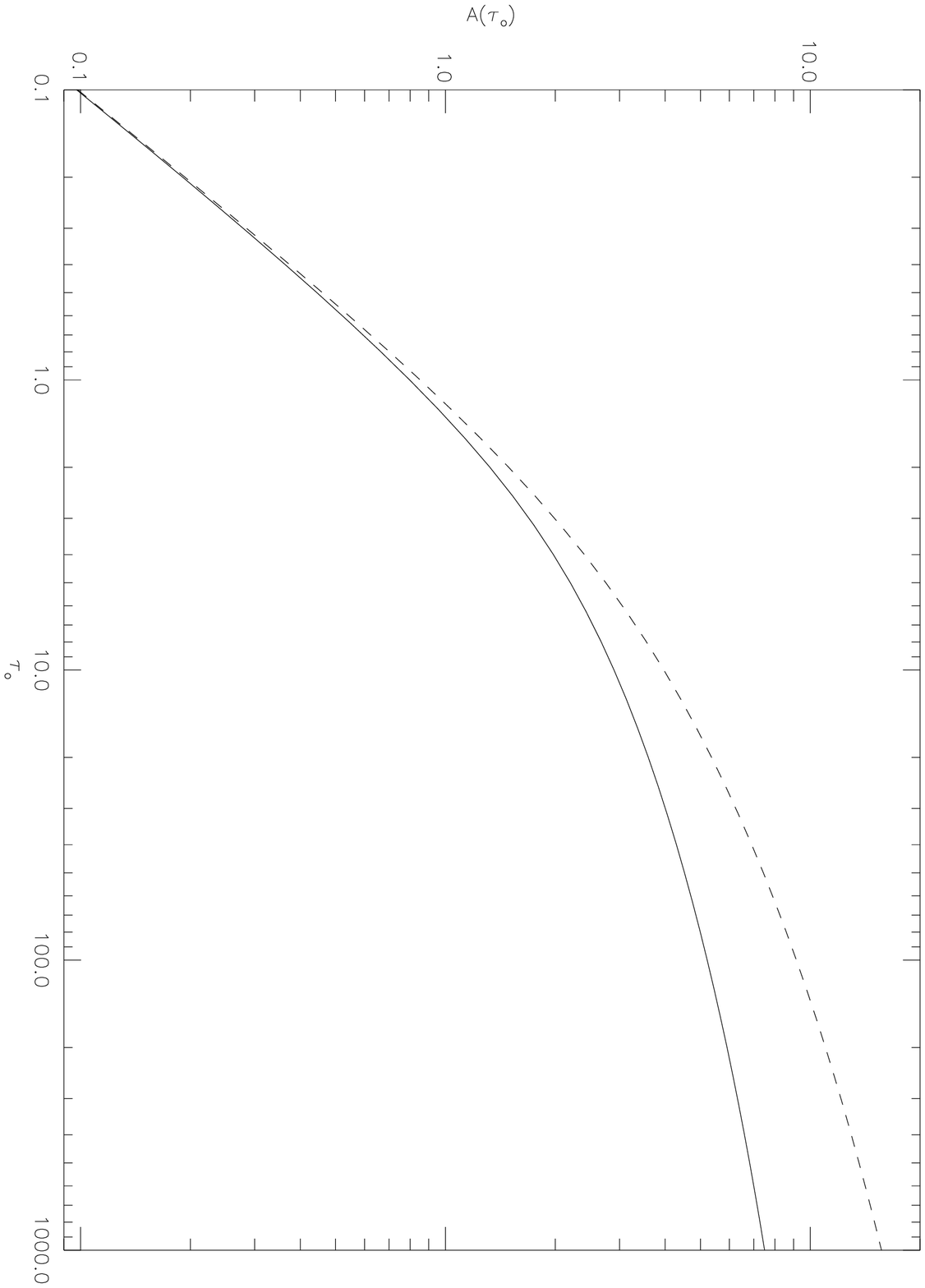}
\caption{The effective optical depth of a clump, $\At$, after averaging over 
its projected area is plotted against the optical depth through the clump
center, $\tau_0$.  The solid curve shows $\At$ versus $\tau_0$ for a cylindrical
clump viewed perpendicularly to the symmetry axis.  The optical depth profile 
across the cylinder, from the central axis towards the edges, is Gaussian.   
The dashed curve shows the corresponding curve for a spherical clump.  The 
optical depth profile from the sphere center towards the edges is also Gaussian. 
The Gaussian spherical clump case was also treated and plotted in MSH84.   
\label{fig02}}
\end{figure}

\clearpage

\begin{figure}
\epsscale{0.66}
\plotone{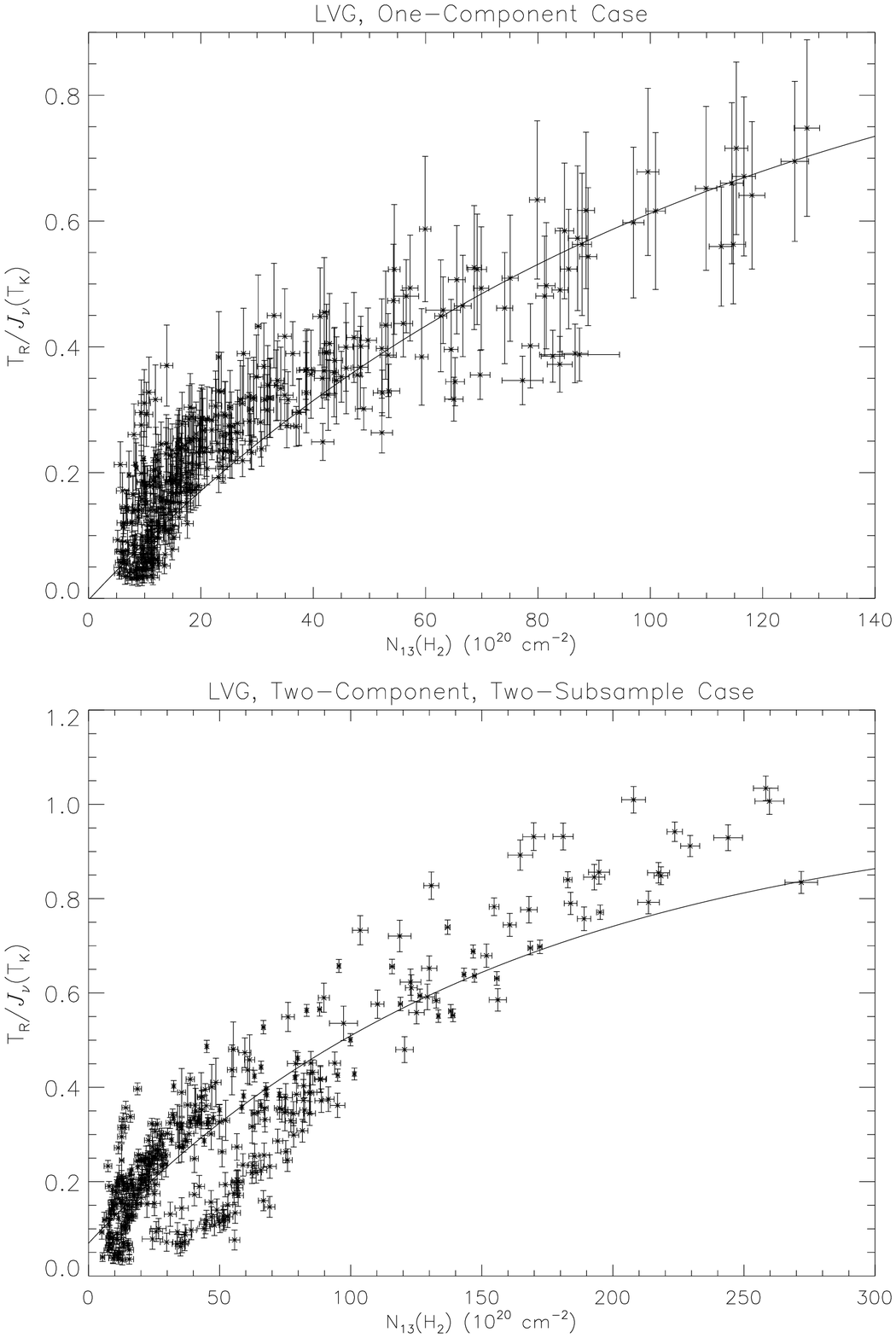}
\caption{Plots of the CO $\Jone$ line radiation temperature, $\Tr$, normalized
to its source function, $\Jnu$, versus the $\cOone$ derived H$_2$ column density.
Both plots are reproduced from \citet{W06}.  The curves are of the form 
$y=1-exp(-a\,x-b)$, where, in the ideal case, $b=0$.  The upper plot is for 
the LVG, one-component models of \citet{W06} and the lower plot is for the
LVG, two-component, two-subsample models of that paper. (The source function for 
the two-component models is the effective source function as defined in \citet{W06}.)  
Given that $x$ is in units of $10^{20}\, H_2\unit cm^{-2}$, $a=(9.5\pm 0.4)\times 
10^{-3}$ and $b=(2.4\pm 5.5)\times 10^{-3}$ for the upper plot and $a=(6.4\pm 0.2)
\times 10^{-3}$ and $b=(7.2\pm 0.8)\times 10^{-2}$ for the lower plot.
\label{fig03}}
\end{figure}

\clearpage

\begin{figure}
\epsscale{0.66}
\plotone{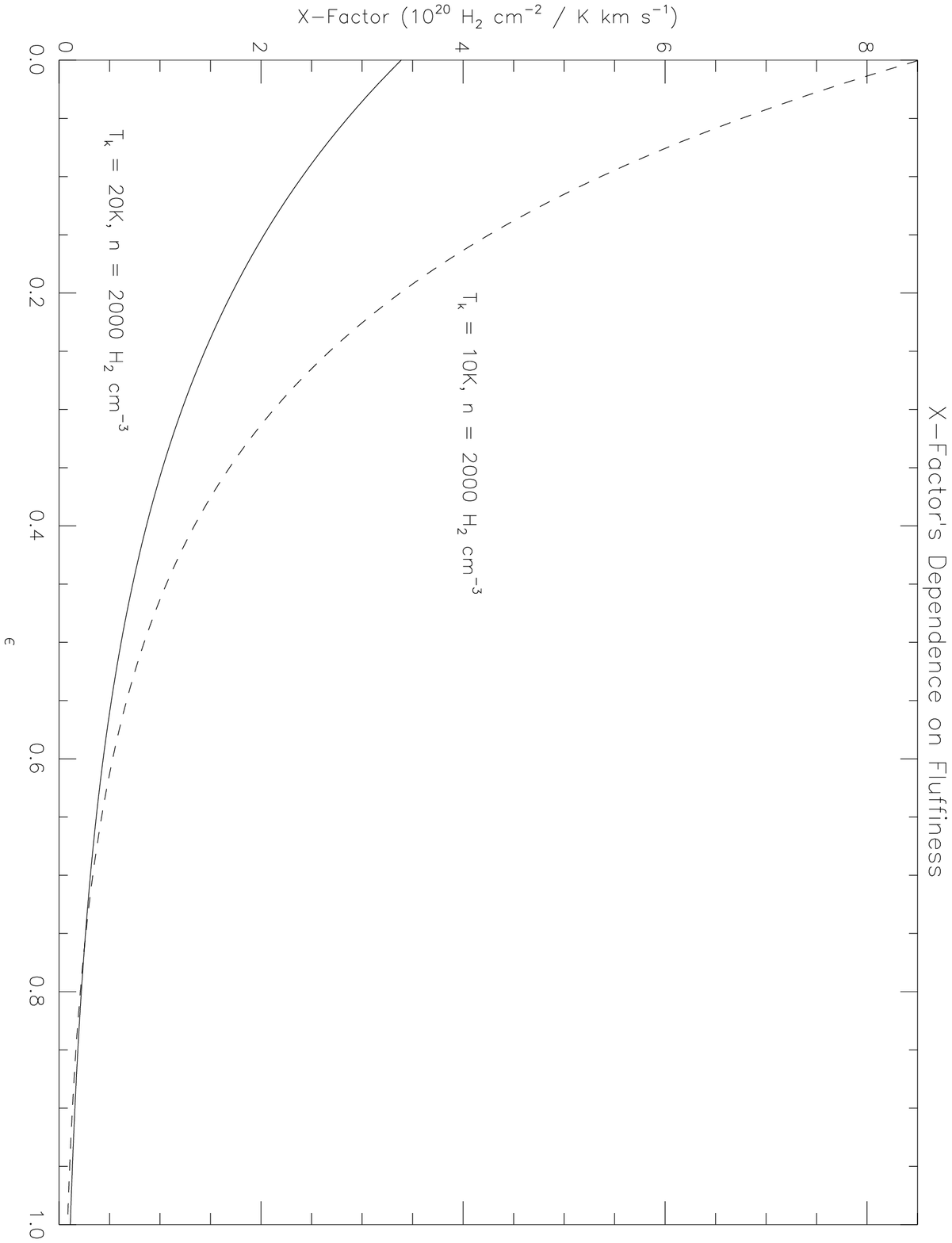}
\caption{Curves that crudely represent the X-factor's dependence on the 
fluffiness parameter, $\epsilon$, are depicted.  The kinetic temperature,
$\Tk$, and the average densities, represented simply by $n$, corresponding
to the curves are shown.  Note that the values of the X-factor for $\epsilon
= 0$ and 1 are too low by about 30\%.  See Section~\ref{ssec36} for details.
It must be emphasized that these curves are dependent on how the ``average''
density is defined (see Section~\ref{sssec471}). 
\label{fig04}}
\end{figure}

\clearpage

\begin{figure}
\epsscale{0.66}
\plotone{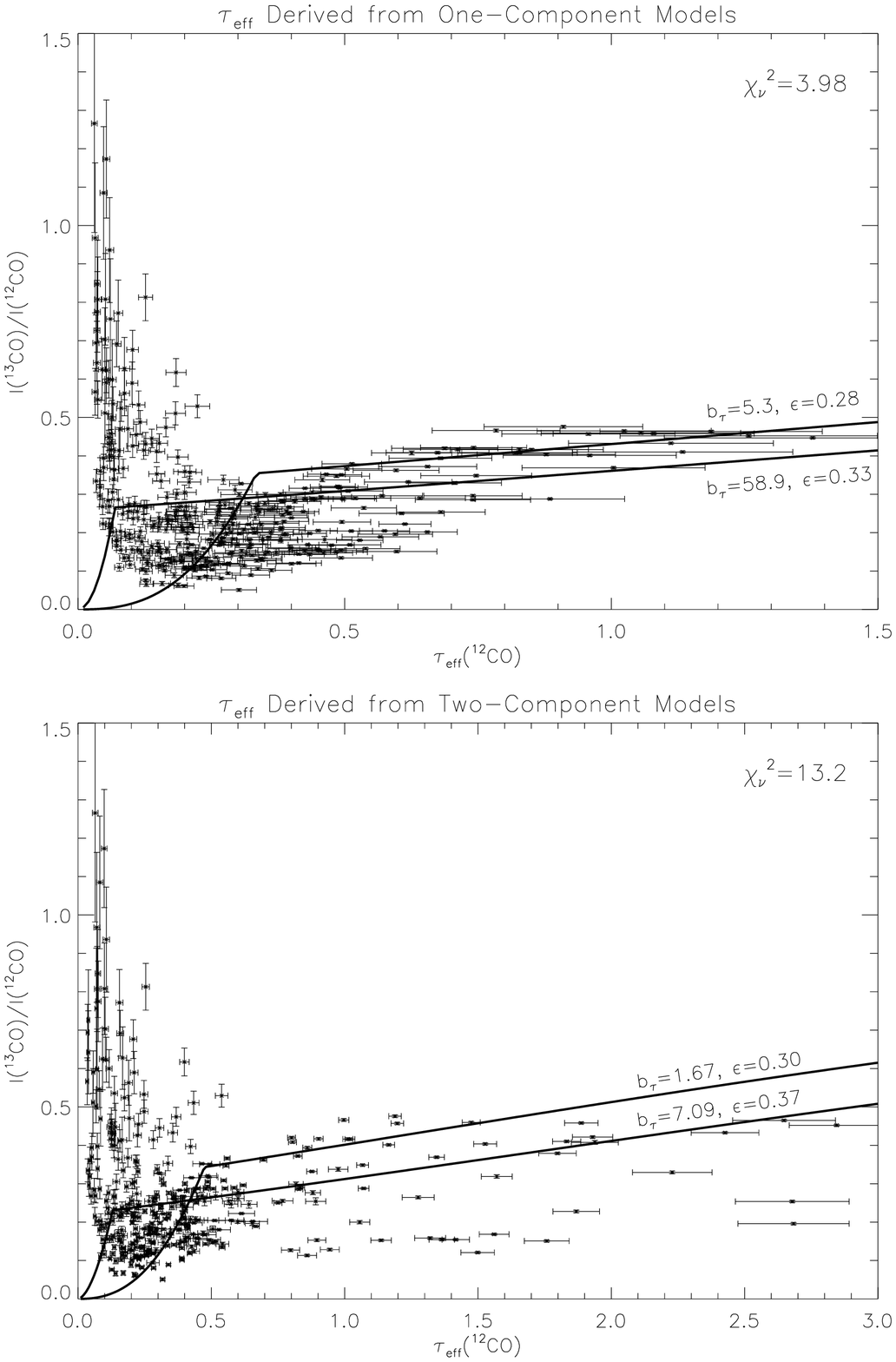}
\caption{The $\cOone/\COone$ line ratio is plotted against the $\tef$ of $\COone$.  
The points represent positions in the Orion$\,$A and B molecular clouds where the 
peak $\Tr(\COone) > 3\,\sigma$, and $\Ia$, $\Ib$, and $\Ic$ are all $> 5\,\sigma$ 
\citep[see][]{W06}.  The solid curves represent model fits that are described in
Section~\ref{ssec39}.  The reduced chi-square of these two-curve fits are given
in the upper-right corner of each panel.  The error bars are 0.5$\,\sigma$ for a 
better view of the distribution of points and of the model curves.  The upper panel 
has the $\tef$ values derived from one-component models and the lower panel has the 
$\tef$ derived from two-component models \citep{W06}.
\label{fig05}}
\end{figure}

\clearpage


\begin{figure}
\epsscale{0.66}
\plotone{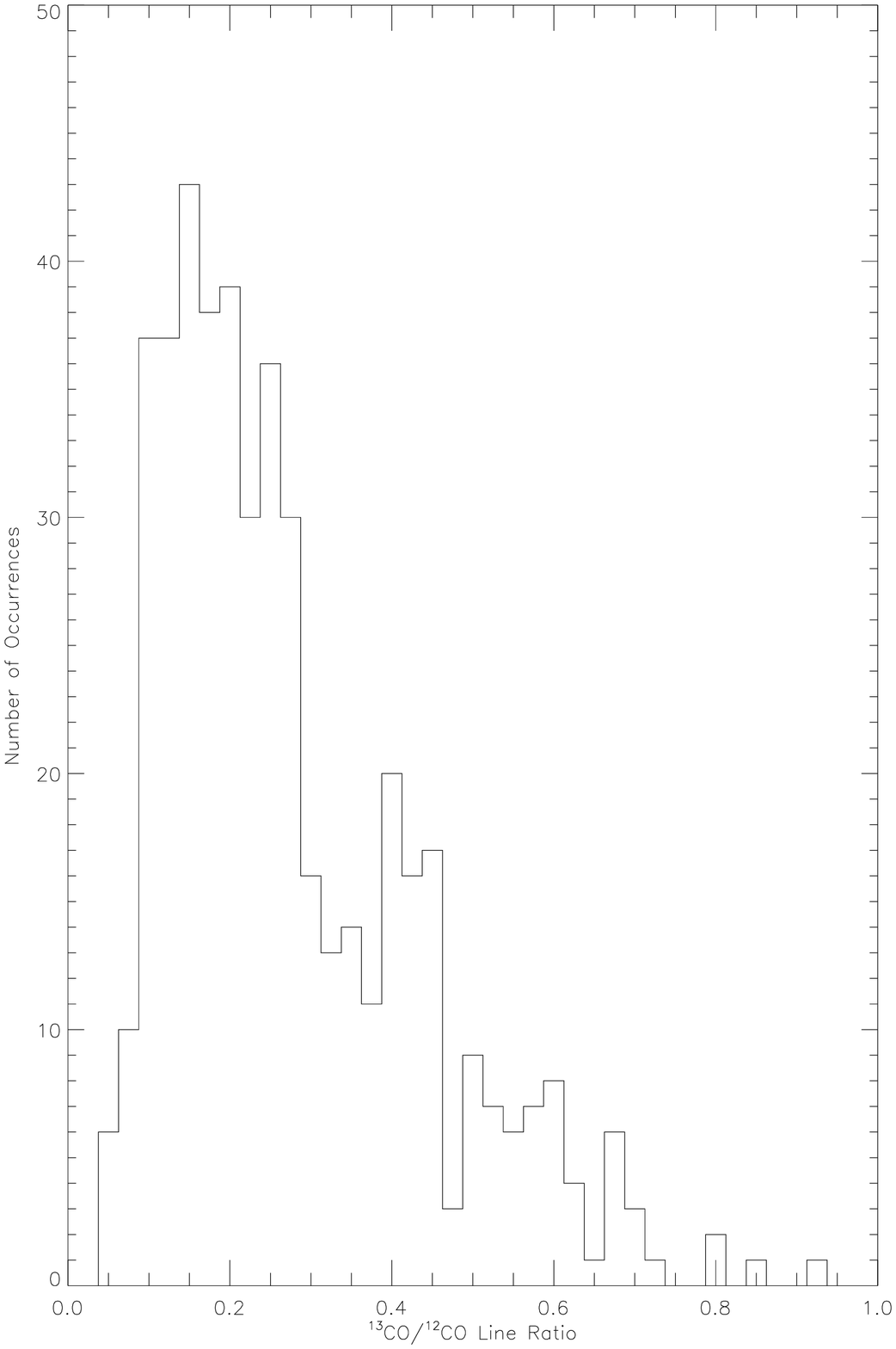}
\caption{The frequencies of values of the $\cOone/\COone$ line ratio in the Orion
clouds is plotted here as a histogram (solid line).  This histogram only includes data 
where $\CO$ and $\cO$ observed line strengths are at or above 5-$\sigma$.
\label{fig07}}
\end{figure}






\clearpage

\begin{deluxetable}{ccccccccc}
\tablecaption{X-Factor Values for Different Model Clump Types\label{tbl-1}}
\tablewidth{0pt}
\tablehead{
\colhead{Type of Clump\tablenotemark{a}} & \colhead{$k_A$} 
& \colhead{\phantom{X}$\epsilon$\phantom{X}} & 
\colhead{$X_f$ $(10^{20} H_2$$\rm\cdot cm^{-2}/K\cdot km\cdot s^{-1})$}
\span\omit\span\omit\span\omit\span\omit\span\omit\\
\noalign{\smallskip}
\noalign{\hbox to 17.73cm{\hglue5.8cm\leaders\hrule\hfill}}
\noalign{\medskip}
\colhead{} & \colhead{} & \colhead{} & \colhead{$\Tk = 10\, K$}\span\omit\span\omit 
& \colhead{$\Tk = 20\, K$}\span\omit\span\omit\\
\noalign{\smallskip}
\noalign{\hbox to 17.73cm{\hglue5.8cm\leaders\hrule\hfill\hglue.2cm\leaders\hrule\hfill}}
\noalign{\medskip}
\colhead{} & \colhead{} & \colhead{} & $\nb\tablenotemark{b}\,=2\times 10^2$ 
& $2\times 10^3$ & $2\times 10^4$ & $\nb\tablenotemark{b}\,=2\times 10^2$ 
& $2\times 10^3$ & $2\times 10^4$\\
\noalign{\bupskip}
}
\startdata
\noalign{\bigskip}
$\tau\gg 1$\tablenotemark{c} & 1\phantom{.5} & 0\phantom{.57} & 3.5\phantom{2} 
& 11\phantom{...12} & 35\phantom{...12} & 1.4\phantom{6} & 4.4\phantom{6} & 14\phantom{...16}\\
$\tau\ll 1$ & 1\phantom{.5}   & 1\phantom{.57}    & 0.12 & 0.12 & 0.12 & 0.16 & 0.16 & 0.16 \\
Hard S.    & 1.7 & 0.14 & 1.3\phantom{2} & 3.4\phantom{2} & 9.1\phantom{2} & 0.59 
& 1.6\phantom{6} & 4.3\phantom{6} \\
Gaussian S. & 1.6 & 0.36 & 0.70 & 1.5\phantom{2} & 3.1\phantom{2} & 0.44 
& 0.91 & 1.9\phantom{6} \\
Sq. Lor. S. & 1.5 & 0.57 & 0.36 & 0.58 & 0.95 & 0.28 & 0.47 & 0.76 \\
G. F. (side-on) & 1.5 & 0.25 & 0.93 & 2.2\phantom{2} & 5.2\phantom{2} & 0.51 
& 1.2\phantom{6} & 2.8\phantom{6} \\
G. F. (end-on)\tablenotemark{d}  & 1.6 & 0.36 & 1.3\phantom{2}  & 2.8\phantom{2} 
& 5.8\phantom{2} & 0.82  & 1.7\phantom{6} & 3.6\phantom{6} \\
\enddata

\tablenotetext{a}{These are the types of model clumps as described in Section 3: Completely 
Optically Thick, Optically Thin, Hard Sphere, Gaussian Sphere, Squared Lorentzian Sphere, 
Gaussian Filament (side-on and end-on).}
\tablenotetext{b}{Clump average density as defined in Section 3 and in the appendices in units
of $H_2\ molecules\cdot\rm cm^{-3}$.}
\tablenotetext{c}{For simplicity, a spherical geometry is adopted for computing $X_f$.}
\tablenotetext{d}{Assuming a length-to-diameter ratio of 3.4.}


\end{deluxetable}

\clearpage

\begin{deluxetable}{cccc}
\tablecaption{Full-Width-at-Half-Maximum Velocity Widths of Individual Model Clumps\label{tbl-2}}
\tablewidth{0pt}
\tablehead{
\colhead{Type of Clump\tablenotemark{a}} & \colhead{$\Dvc$ (FWHM)\tablenotemark{b}\ \ ($\kms$)}
\span\omit\span\omit\\
\noalign{\smallskip}
\noalign{\hbox to 9.35cm{\hglue3.4cm\leaders\hrule\hfill}}
\noalign{\medskip}
\colhead{} & $\nb\tablenotemark{c}\,=2\times 10^2$ & $2\times 10^3$ & $2\times 10^4$ \\
\noalign{\bupskip}
}
\startdata
\noalign{\smallskip}
Hard S.\tablenotemark{d} & 0.46 & 1.4 & 4.6 \\
Gaussian S.\tablenotemark{d} & 0.63 & 2.0 & 6.3 \\
Sq. Lor. S.\tablenotemark{d} & 0.64 & 2.0 & 6.4 \\
G. Filament\tablenotemark{e} & 0.72 & 2.3 & 7.2 \\
\enddata

\tablenotetext{a}{These are some types of model clumps as described in Section 3: Hard Sphere, 
Gaussian Sphere, Squared Lorentzian Sphere, Gaussian Filament.}
\tablenotetext{b}{The earlier computed values have been multiplied by $\sqrt{8\, ln 2}$ to
convert from $rms$ to $FWHM$.} 
\tablenotetext{c}{Clump average density as defined in Section 3 and in the appendices in units
of $H_2\ molecules\cdot\rm cm^{-3}$.}
\tablenotetext{d}{FWHM diameter of 1.8$\,$pc was adopted.}
\tablenotetext{e}{Length of 6.2$\,$pc and FWHM diameter of 1.8$\,$pc were adopted.}


\end{deluxetable}

\clearpage

\begin{deluxetable}{ccccccc}
\tablecaption{Central Sightline Optical Depths, $\tau_0$, of Individual Model Clumps\label{tbl-3}}
\tablewidth{0pt}
\tablehead{
\colhead{Type of Clump\tablenotemark{a}} & \colhead{$\tau_0$}
\span\omit\span\omit\span\omit\span\omit\span\omit\\
\noalign{\smallskip}
\noalign{\hbox to 15.35cm{\hglue3.4cm\leaders\hrule\hfill}}
\noalign{\medskip}
\colhead{} & \colhead{$\Tk = 10\, K$}\span\omit\span\omit 
& \colhead{$\Tk = 20\, K$}\span\omit\span\omit\\
\noalign{\smallskip}
\noalign{\hbox to 15.35cm{\hglue3.4cm\leaders\hrule\hfill\hglue.2cm\leaders\hrule\hfill}}
\noalign{\medskip}
\colhead{} & $\nb\tablenotemark{b}\,=2\times 10^2$ 
& $2\times 10^3$ & $2\times 10^4$ & $\nb\tablenotemark{b}\,=2\times 10^2$ 
& $2\times 10^3$ & $2\times 10^4$\\
\noalign{\bupskip}
}
\startdata
\noalign{\bigskip}
Hard S. & 30 & \phantom{2}94 & 300 & 8.8 & 28 & \phantom{2}88 \\
Gaussian S. & 34 & 109 & 344 
& 10\phantom{...8} & 32 & 100 \\
Sq. Lor. S. & 37 & 120 & 370 & 11\phantom{...8} & 35 & 110 \\
G. F. (side-on) & 28 & \phantom{2}88 & 280 & 8.2 & 26 & \phantom{2}82 \\
G. F. (end-on)\tablenotemark{c} & 90 & 280 & 900 & 27\phantom{...8} & 84 & 270 \\
\enddata

\tablenotetext{a}{These are some types of model clumps as described in Section 3: Hard Sphere, 
Gaussian Sphere, Squared Lorentzian Sphere, Gaussian Filament (side-on and end-on).}
\tablenotetext{b}{Clump average density as defined in Section 3 and in the appendices in units
of $H_2\ molecules\cdot\rm cm^{-3}$.}
\tablenotetext{c}{Assuming a length-to-diameter ratio of 3.4.}


\end{deluxetable}

\clearpage

\begin{deluxetable}{ccccccccc}
\tablecaption{Effective Optical Depths, $\At$, of Individual Model Clumps\label{tbl-4}}
\tablewidth{0pt}
\tablehead{
\colhead{Type of Clump\tablenotemark{a}} & \colhead{$k_A$} 
& \colhead{\phantom{X}$\epsilon$\phantom{X}} & \colhead{$\At$}
\span\omit\span\omit\span\omit\span\omit\span\omit\\
\noalign{\smallskip}
\noalign{\hbox to 17.73cm{\hglue5.8cm\leaders\hrule\hfill}}
\noalign{\medskip}
\colhead{} & \colhead{} & \colhead{} & \colhead{$\Tk = 10\, K$}\span\omit\span\omit 
& \colhead{$\Tk = 20\, K$}\span\omit\span\omit\\
\noalign{\smallskip}
\noalign{\hbox to 17.73cm{\hglue5.8cm\leaders\hrule\hfill\hglue.2cm\leaders\hrule\hfill}}
\noalign{\medskip}
\colhead{} & \colhead{} & \colhead{} & $\nb\tablenotemark{b}\,=2\times 10^2$ 
& $2\times 10^3$ & $2\times 10^4$ & $\nb\tablenotemark{b}\,=2\times 10^2$ 
& $2\times 10^3$ & $2\times 10^4$\\
\noalign{\bupskip}
}
\startdata
\noalign{\bigskip}
Hard S. & 1.7 & 0.14 & 2.7 & 3.2 & 3.8 & 2.3 & 2.7 & 3.2 \\
Gaussian S. & 1.6 & 0.36 & 5.7 & 8.6 & 13\phantom{...7} 
& 3.7 & 5.6 & 8.4 \\
Sq. Lor. S. & 1.5 & 0.57 & 12\phantom{...7} & 23\phantom{...7} 
& 44\phantom{...7} & 6.0 & 12\phantom{...7} & 22\phantom{...7} \\
G. F. (side-on) & 1.5 & 0.25 & 3.4 & 4.6 & 6.1 & 2.5 
& 3.4 & 4.5 \\
G. F. (end-on)\tablenotemark{c} & 1.6 & 0.36 & 8.0 & 12\phantom{...7} 
& 18\phantom{...7} & 5.2 & 7.9 & 12\phantom{...7} \\
\enddata

\tablenotetext{a}{These are some types of model clumps as described in Section 3: Hard Sphere, 
Gaussian Sphere, Squared Lorentzian Sphere, Gaussian Filament (side-on and end-on).}
\tablenotetext{b}{Clump average density as defined in Section 3 and in the appendices in units
of $H_2\ molecules\cdot\rm cm^{-3}$.}
\tablenotetext{c}{Assuming a length-to-diameter ratio of 3.4.}


\end{deluxetable}

\clearpage

\begin{deluxetable}{cccc}
\tablecaption{Masses, $\rm M_c$, of Individual Model Clumps\label{tbl-5}}
\tablewidth{0pt}
\tablehead{
\colhead{Type of Clump\tablenotemark{a}} & \colhead{$\rm M_c$ ($M_\odot$)}
\span\omit\span\omit\\
\noalign{\smallskip}
\noalign{\hbox to 9.98cm{\hglue3.4cm\leaders\hrule\hfill}}
\noalign{\medskip}
\colhead{} & $\nb\tablenotemark{b}\,=2\times 10^2$ & $2\times 10^3$ & $2\times 10^4$ \\
\noalign{\bupskip}
}
\startdata
\noalign{\smallskip}
Hard S.\tablenotemark{c} & $3.9\times 10^1$ & $3.9\times 10^2$ & $3.9\times 10^3$ \\
Gaussian S.\tablenotemark{c} & $1.4\times 10^2$ & $1.4\times 10^3$ & $1.4\times 10^4$ \\
Sq. Lor. S.\tablenotemark{c} & $2.5\times 10^2$ & $2.5\times 10^3$ & $2.5\times 10^4$ \\
G. Filament\tablenotemark{d} & $4.1\times 10^2$ & $4.1\times 10^3$ & $4.1\times 10^4$ \\
\enddata

\tablenotetext{a}{These are some types of model clumps as described in Section~3: Hard Sphere, 
Gaussian Sphere, Squared Lorentzian Sphere, Gaussian Filament.}
\tablenotetext{b}{Clump average density as defined in Section 3 and in the appendices in units
of $H_2\ molecules\cdot\rm cm^{-3}$.}
\tablenotetext{c}{FWHM diameter of 1.8$\,$pc was adopted.}
\tablenotetext{d}{Length of 6.2$\,$pc and FWHM diameter of 1.8$\,$pc were adopted.}


\end{deluxetable}

\end{document}